\pdfoutput=1

\documentclass[a4paper,fleqn]{cas-dc}

\usepackage[numbers]{natbib}

\graphicspath{{./figures/}}
\DeclareGraphicsExtensions{.pdf,.jpg,.png,.xps}
\usepackage[caption=false,font=footnotesize]{subfig}
\usepackage[activate={true,nocompatibility},final,tracking=true,kerning=true,spacing=true,factor=1100,stretch=10,shrink=10]{microtype} 
\usepackage{soul}
\usepackage{algorithm}
\usepackage{algpseudocode}[lines]
\algrenewcommand\algorithmicindent{0.8em}%

\begin{document}
\let\WriteBookmarks\relax
\def\floatpagepagefraction{1}
\def\textpagefraction{.001}

\newcommand{\titlepaper}{Scaling Migrations and Replications of Virtual Network Functions based on Network Traffic Forecasting}

\shorttitle{\titlepaper}
\shortauthors{F. Carpio et~al.}
\title [mode = title]{\titlepaper}

\author{Francisco Carpio}
[
	orcid=0000-0000-0000-0000
]

\address{Institute of Computer and Network Engineering, Technische Universit{\"a}t Braunschweig, 38106 Braunschweig, Germany}

\author{Wolfgang Bziuk}
[
	orcid=0000-0000-0000-0000
]

\author{Admela Jukan}
[
	orcid=0000-0000-0000-0000
]

\cortext[cor1]{Corresponding author}


\begin{abstract}
	Migration and replication of virtual network functions (VNFs) are well-known
	mechanisms to face dynamic resource requests in Internet Service Provider
	(ISP) edge networks. They are not only used to reallocate resources in
	carrier networks, but in case of excessive traffic churns also to offloading
	VNFs to third party cloud providers. We propose to study how traffic
	forecasting can help to reduce the number of required migrations and
	replications when the traffic dynamically changes in the network. We analyze
	and compare three scenarios for the VNF migrations and replications based
	on: (i) the current observed traffic demands only, (ii) specific maximum
	traffic demand value observed in the past, or (iii) predictive traffic
	values. For the prediction of traffic demand values, we use an LSTM model
	which is proven to be one of the most accurate methods in time series
	forecasting problems. Based the traffic prediction model, we then use a
	Mixed-Integer Linear Programming (MILP) model as well as a greedy algorithm
	to solve this optimization problem that considers migrations and
	replications of VNFs. The results show that LSTM-based traffic prediction
	can reduce the number of migrations up to 45\% when there is enough
	available resources to allocate replicas, while less cloud-based offloading
	is required compared to overprovisioning.
\end{abstract}




\maketitle

\section{Introduction}

Internet Service Providers (ISP) recognize Network Function Virtualization (NFV)
as a key concept to reducing capital and operational expenditures. In NFV,
service provisioning is achieved by concatenating Virtual Network Functions
(VNFs) in a specific sequence order, defined as Service Function Chains (SFCs).
The placement of VNFs is a well known problem in the community which can follow
different optimization objectives, such as network load balancing and end-to-end
delay. Once VNFs are deployed in the network, the dynamic traffic demand
patterns require either reallocation or scaling of VNFs to pursuing different
objectives. Moreover, part of the workload may need to be migrated to the cloud
due to, for instance, non-optimal deployments or insufficient resources within
physical servers of the ISP.

The migration and replication of VNFs is a problem widely studied from different
perspectives to date. In all studies, when performing migrations in runtime, it
was shown that the active flows need to be rerouted causing service disruptions.
The use of replications, on the other hand, requires extra server resources, due
to virtualization overhead, and extra network resources, due to state
synchronization tasks. From an ISP-centric point of view, the use of third party
clouds for a possible migration or replication of VNFs has an impact not only on
the performance of the system but also on the monetary costs for the ISP when
using third-party cloud services. For these reasons, accurate prediction of
future resource utilization or traffic demand values is the key for ISP to
better proactively allocate their resources.

We propose to study how traffic forecasting can help us generally reduce the
number of migrations and replications in ISPs, as well as the related placements
in third-party clouds. We formulate the placement problem as a Mixed-Integer
Linear Programming (MILP) model and solve the placement in two phases, the
latter one focused on migrations and replications to be able to better
understand their effects. We analyze and compare three scenarios for the VNF
migrations and replications based on: (i) the current observed traffic demands
only, (ii) specific maximum traffic demand value observed in the past, or (iii)
predictive traffic values. In the latter case, we specifically use LSTM networks
for traffic predictions. The placement model also considers the impact of
migrations on the service delays due to service interruptions and the impact
replications on the network and server resource utilization due to virtual
machine (VM) overhead and synchronization traffic. Since the MILP model cannot
be used as online solution, we propose a greedy algorithm for that purpose and
analyze its performance.

The rest of the paper is organized as follows. Section II presents related work
and our contribution. Section III describes the reference scenario. Section IV
formulates the optimization model. Section V describes the online heuristic
approaches. Section VI analyzes the performance of the model and heuristics and
Section VII concludes the paper.
\section{Related Work and Our Contribution}

\subsection{VNF placement,  migrations and replications}



Significant amount of previous work has focused on the placement of virtual
resources for VNFs \cite{Laghrissi2019}, specially with variants of the joint
optimization placement problem with different objectives. For instance, in
\cite{Tajiki2017}, a resource allocation solution is proposed for optimizing
energy efficiency, while considering delay, network and server utilization.
\cite{Basta2017} proposed models to finding the optimal dimensioning and
resource allocation with latency constraints in mobile networks. \cite{Qu_2017}
studied how to optimize the VNF placement and traffic routing while considering
reliability and end-to-end delays. In \cite{Golkarifard_2021}, the authors
propose to solve a joint decision problem when placing VNFs considering multiple
real-world aspects in order to deal with highly varying traffic requests. Within
the placement problem topic, migration and replications of VNFs are known as
specific sub-problems that need to be solved in the context of resource and
service management.



Regarding migrations, since VNFs are commonly running over VMs, there is the
possibility of migrating VMs entirely \cite{Xia2016} or migrationg only the
internal states of VNFs \cite{Xia2016a} to new VMs. In this regard, while the
interruption and rerouting of active flows is possible
\cite{Gember-Jacobson2014}, there is always a service downtime duration that
will vary depending on the path latencies \cite{Taleb2019}. Some authors, like
in \cite{Cziva2018}, propose a dynamic placement scheduler to minimize the
end-to-end latencies when performing migrations. In \cite{Eramo2017}, a
trade-off was found between the power consumption and QoS degradation to
determine whether a migration is appropriate in order to minimize its negative
impact due to the service interruptions.



On the other hand, replications have been primarily used to provide service
reliability \cite{michael2016, Engelmann2018}, whereby minimization of the
number of required replicas \cite{Ding2017} is one of the main objectives. In
addition, replications need to be studied in the context of reduction of
end-to-end service delays \cite{Yuan2020},  load balancing on the network links
\cite{Carpio2017a} or to load balance the server utilization \cite{Carpio2017b}.
Studies combining both migrations and replications have also been carried out,
e.g., \cite{Huang2018}, where a balancing between the number of migrations and
replications is proposed in order to maximize the network throughput and
minimize the delay. In our previous work \cite{Carpio2018}, we proposed an
optimization method to deriving a trade-off between migrations and replications
while improving server, network load balancing and QoS. Unlike migrations,
replications need to consider the impact on traffic synchronization between
VNFs, which is an important issue that adds considerable traffic overhead in the
network \cite{Alharbi2019}.

\subsection{Traffic forecasting and VNF resource requirement predictions}




While NFV provides network operators more flexibility to instantiate VNFs at
runtime, the dynamic change of network states due to the highly variant traffic
load at the edge requires prediction mechanisms to proactively adapt the
placement of VNFs accordingly. To address this issues, two approaches have been
proposed, one by predicting the resources that VNFs will require based on past
utilization \cite{Mijumbi_2016} while the other one by using traffic forecasting
(predictions) techniques to calculate how much resources the VNFs will need to
serve that traffic correspondingly \cite{Rahman_2018}. In both cases, a more
traditional approach  uses either the statistical analysis of time series, or
machine learning. Examples of the statistical analysis can be found, for
instance, in \cite{Yao_2020} where the authors introduce a mechanism based on
Fourier-Series to determine upcoming demands to perform online VNF scaling. In
\cite{Sun_2016}, the authors also use Fourier-Series with the same purpose but,
in this case, with the objective of reducing blocking probability. A slightly
different approach in this area is proposed in \cite{Tang_2019} where a method
is used based on linear regression to predict traffic and to scale VNFs in order
to improve service availability. Yet another example in \cite{Qu_2020} uses a
fractional Brownian motion (fBm) traffic model to learn traffic parameters in
order to predict time-varying VNF resource demand.

Most of the recent work in this area, however, include machine learning based
methods. In the area of predicting resource requirements, \cite{Mijumbi2017a}
uses Feedforward Neural Networks (FNN) to predict future requirements of VNFs
based on its past utilization and the influence from neighbor VNFs. With a
similar objective, the \cite{Shi2015} uses a Bayesian learning approach to learn
from historical resource usage data from VNFs and predict future resource
reliability. Another example in \cite{Kim_2019} uses an specific type of
Recurrent Neural Network (RNN) which is based on attention and embedding
techniques jointly with Long Short Term Memory (LSTM) model to predict CPU
utilization from VNFs with high accuracy.

For traffic forecasting with ML, \cite{Alawe_2018} uses both RNN and Deep Neural
Networks (DNN) to forecast traffic changes and prove that these methods can
improve delay when provisioning new resources to VNFs as compared to
threshold-based methods. Since one of the main objectives when traffic
predicting is to determine when to scale VNFs, as discussed in
\cite{Subramanya_2019}. Here, it is proposed to use of Multilayer Perceptron
(MLP) to predict the required number of VNFs in relation with the network
traffic to scaling the deployment of VNFs.

\subsection{Our Contribution}

So far, we lack studies on to how traffic prediction can be used to minimize
migrations and replications of VNFs. To this end, we contribute with studying
how traffic forecasting can help on reducing the number of migration and
replication of VNFs by optimizing their placement in a proactive manner. This is
motivated especially by three previously mentioned studies,
\cite{Golkarifard_2021}, \cite{Alawe_2018} and \cite{Subramanya_2019}, that
showed the need to consider highly varying traffic requests when placing VNFs in
5G networks and the role that traffic forecasting plays in placement and scaling
of VNFs. We analyze this problem from an ISP point of view by using a MILP
generic multipath based model comparing three scenarios: (i) when VNFs are
placed only considering current observed traffic demands, (ii) when VNFs are
placed considering the 80\% of the specific maximum traffic demand value and
(iii) when VNFs are placed considering predicted traffic values. For traffic
forecasting, we use an LSTM model which is proven to be one of the most accurate
methods in time series forecasting problems. The placement model also considers
the impact of migration of VNFs have on the service delays due to service
interruptions, considering individual delays per each traffic demand on a
per-path basis, i.e., individually per each path. Regarding replications, we
consider their impact on the network and server resource utilization due to VM
overhead and synchronization traffic used for maintaining states. Additionally,
we propose a greedy algorithm as online solution for the MILP model and we
compare it to basic random- and first-fit approaches. Finally, we contribute to
by showing that traffic prediction can reduce the number of migrations when
enough available resources to allocate replicas, while also reducing the
utilization of the cloud.

\section{Reference Scenario}  \label{sec_3}

We assume that an ISP owns the network infrastructure close to the end users
where it install small groups of servers for the NFV Infrastructure. We also
assume that the ISP uses the cloud as a third party to offload VNFs when, for
instance, its own infrastructure cannot deploy new VNFs. Our model follows a two
phase optimization process, in order to study the impact of migrations and
replications of VNFs have on the ISP network while minimizing the utilization of
the cloud.

\subsection{Optimization Scenarios and assumptions} \label{opt_scenarios}

Since our approach to optimizations is carried out from the point of view of an
ISP who owns the physical server infrastructure, given a certain network
topology with certain number of servers located in network nodes, we assume that
all nodes of that topology have direct links to a third party cloud server. The
specific resulting resource utilization from the links connecting to the cloud
and the cloud servers are not considered in the analysis, but the geographic
location of the cloud servers for service delay is.

The optimization is divided in two phases. During the first one, the model
optimizes by minimizing the placement of VNFs in the cloud, so the ISP network
is as much utilized as possible, and also by minimizing the number of VNF
replicas at certain time step $t$. After that, a second placement is carried out
at time $t + \Delta t$ while considering initial placement of VNFs that took
over during the first phase. In this case, minimizing the migration of VNFs from
the first placement is also added to the objective altogether with the
minimization of replications and cloud VNFs. Since the traffic demands, and,
therefore, the amount of resources allocated by VNFs vary over time, during the
first phase at time $t$, a certain traffic bandwidth is considered which is
different from the one considered during the second phase after $\Delta t$. The
main objective is, therefore, to study how migrations and replications can be
minimized in the network while at the same time also reducing the usage of the
cloud. This is done while comparing three different scenarios when optimizing
during the first phase: i) considering the current observed traffic demands at
time $t$, ii) considering the 80\% of the maximum traffic demand values can have
and iii) considering the predicted traffic demands at time $t + \Delta t$.

For the sake of simplicity, we consider a VNF instance maps 1:1 to a VM where
some server resources are reserved to the VM independently of the processed
traffic. We define the end-to-end service delay, as the sum of propagation delay
(time for the data to travel trough the fiber), processing delay (time for the
VNF to process the data) and service interruption delays caused by migrations.
These delays will be explained in detail in the next section, however, let us
shortly focus on the migration process in order to better understand its impact
on the service delay. We assume a migration occurs when a VNF is reallocated
into a new location and, still, there are active flows being served. So, we omit
here the case of cold migrations. Most of the migration process occurs without
affecting the perceived delay by the end user since, before performing a
migration, a new VNF instance is deployed in a new location and its state is
synchronized with the old instance. However, we consider there is always a short
interruption of the active flows to commute to the new VNF \cite{Taleb2019}. In
this sense, the service delay can be interpreted as a worst case delay. In our
model, we consider a multipath based approach where every SFC can use multiple
paths, whereby each path can exhibit different delays due to different links and
VNFs are traversed. On the other hand, we make use of replications to address
scalability but without introducing delays due to the replication process does
not stop active flows. But, we do consider the synchronization traffic between
replicas in order maintain their states synchronized, as we detail later.

\begin{figure}[!t]
	\centering
	\subfloat[First placement]{\includegraphics[width=0.8\columnwidth]{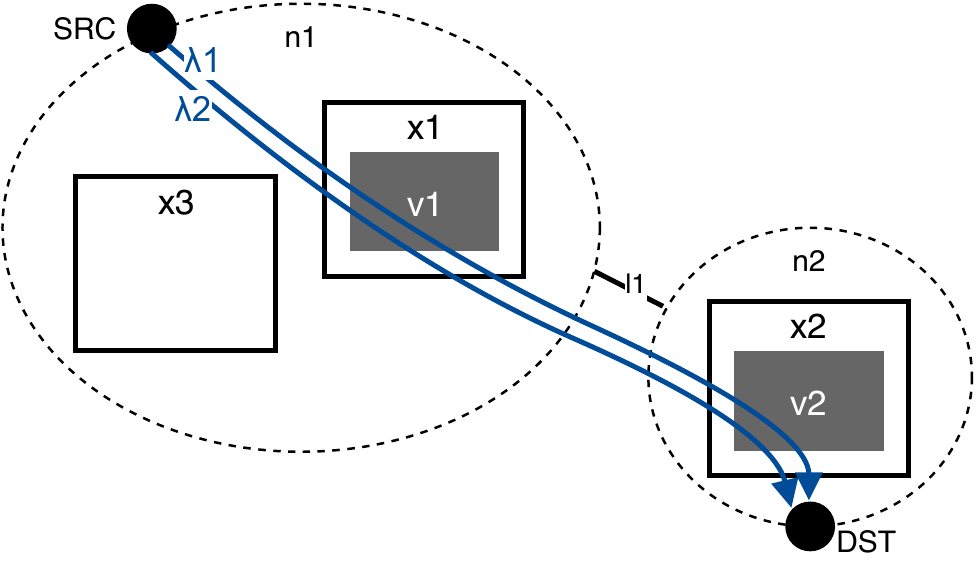}%
		\label{model_init}}
	\hfil
	\subfloat[Migration during second placement]{\includegraphics[width=0.8\columnwidth]{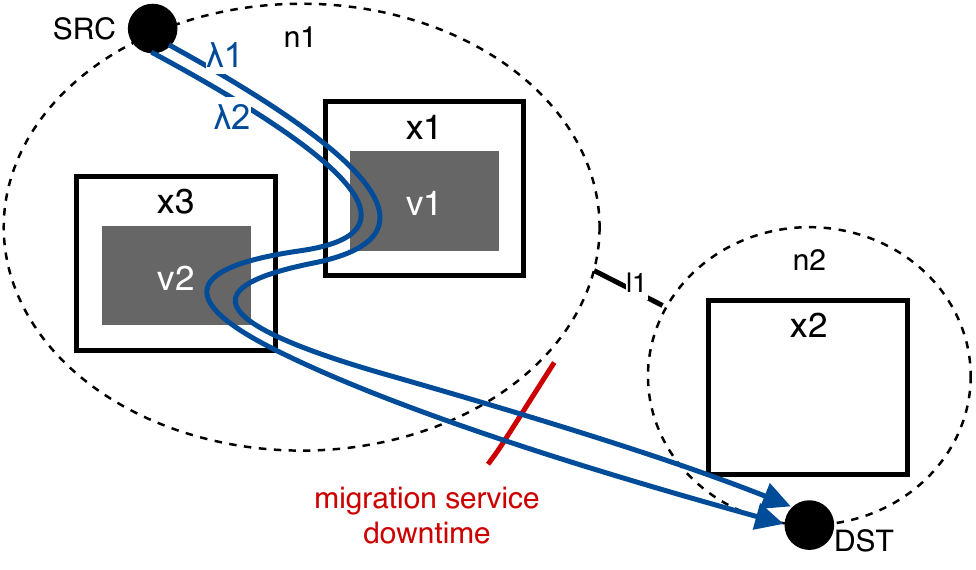}%
		\label{model_mgr}}
	\hfil
	\subfloat[Replication during second placement]{\includegraphics[width=0.8\columnwidth]{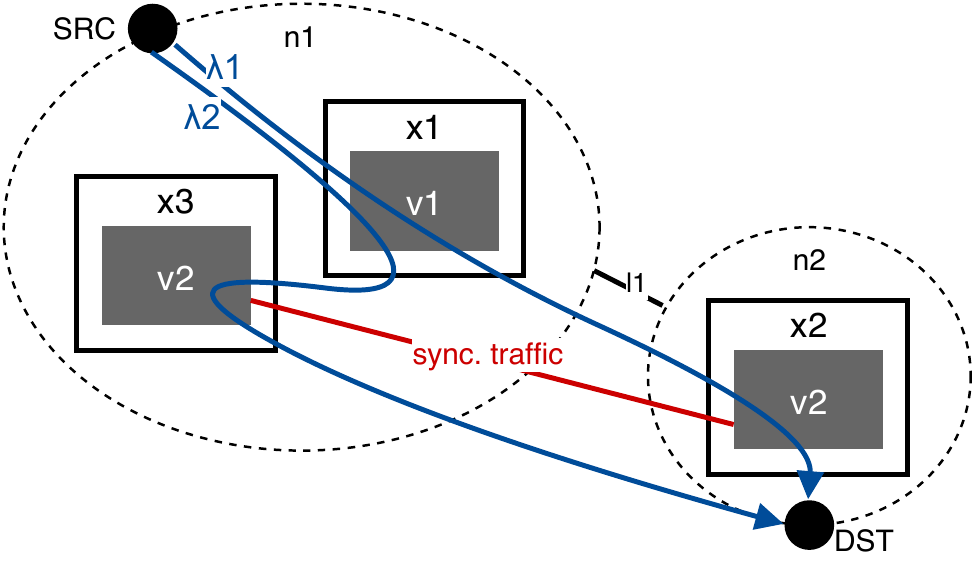}%
		\label{model_rep}}
	\caption{Examples of different possible scenarios for the model}
	\label{models}
\end{figure}

\subsection{Migrations and replications}

To better understand the model, let us now illustrate an example (shown in Fig.
\ref{model_init}) of an SFC is providing service to traffic demands $\lambda1$
and $\lambda2$ with two chained VNFs, $v1$ and $v2$ instantiated in server $x1$,
from node $n1$, and server $x2$ from node $n2$, respectively. Depending on the
functionality, every VNF can be of a different type $t$, however for
simplification in this example, we assume all VNFs are of the same type $t$, so
they all require the same amount resources. The service delay is calculated as
the sum of propagation delays, processing delays and service interruption
delays. As an example, assuming $D_{l}$ is the propagation delay of a link $l$
and $d_{x,v}^{\text{pro}}(\lambda)$ is the processing delay experienced by a
traffic demand $\lambda$ traversing a VNF $v$ on a server $x$, then, the delay
for traffic demand $\lambda_1$ using that specific path $p$ is
$\hat{d}_{p}^{\lambda_1} = D_{l1} + d_{x1,v1}^{\text{pro}}(\lambda1) +
	d_{x2,v2}^{\text{pro}}(\lambda1)$. In this phase, which is taken as the initial
placement for the second phase, we do not consider delays caused by service
interruptions, since there are no migrations yet.

For the second phase, the traffic demands change, so the current VNFs in the
network have the possibility to be either be migrated or replicated. An example
is shown in Fig. \ref{model_mgr}, where VNF $v2$ is migrated from server $x2$ to
server $x3$. From the delay point of view, here because a service interruption
ocurred due to active flows are stopped, a delay will be added and the new
resulting service delay. Another example is shown in Fig. \ref{model_rep}, where
instead of migrating, the VNF $v2$ is replicated into server $x3$ and only
traffic demand $\lambda2$ is routed to the new replica location. In this case,
there is synchronization traffic added between both VNFs $v2$ to maintain their
states synchronized.

\subsection{Traffic demand model and time series forecasting} \label{traffic_model}

\begin{figure}[!t]
	\centering
	\includegraphics[width=0.7\columnwidth]{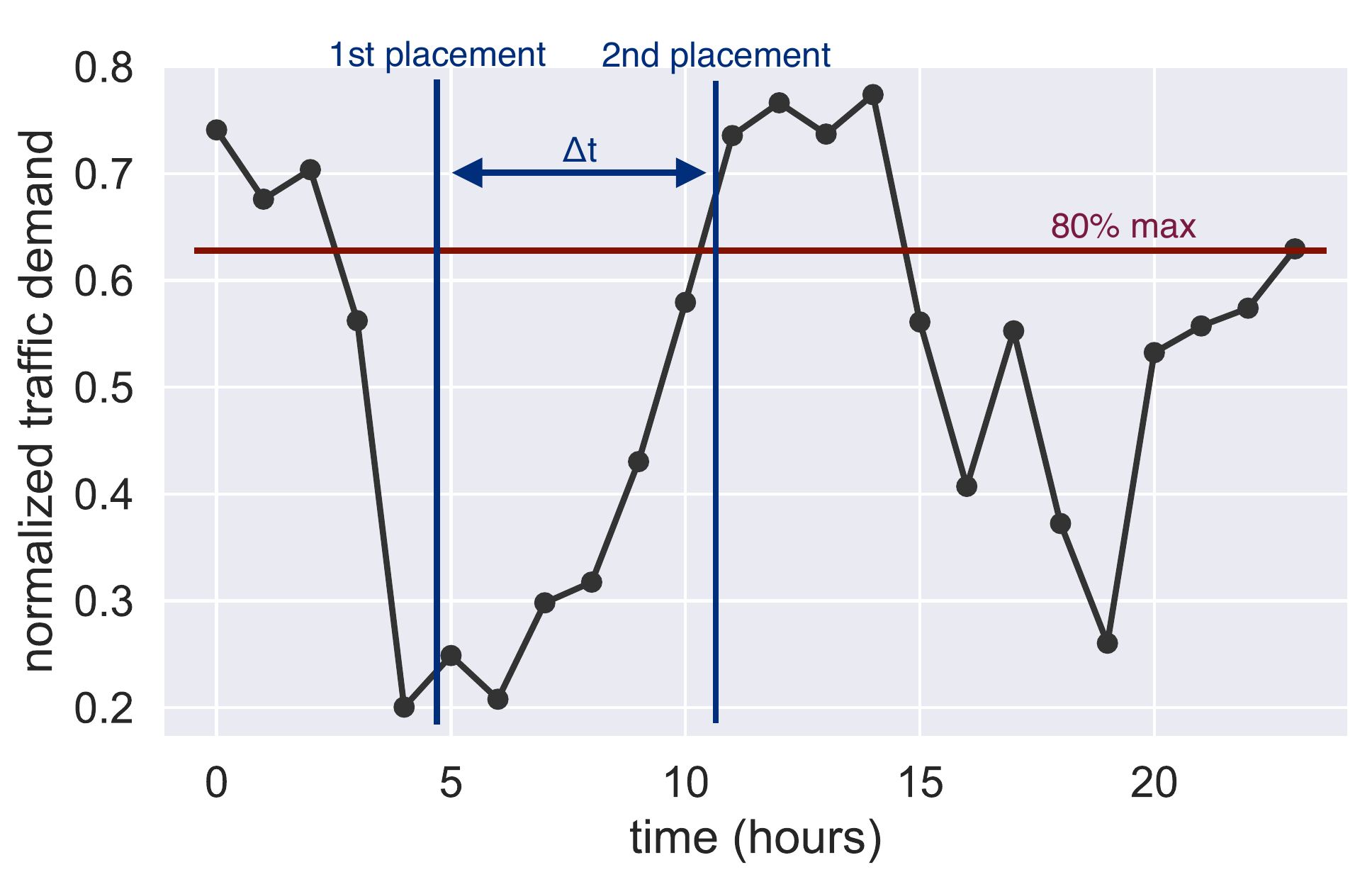}
	\caption{Normalized traffic demand example}
	\label{fig:traffic_gen}
\end{figure}

We assume that every source destination pair of nodes within the ISP network
generates a certain number of traffic demands with specific bandwidth. The
traffic demands data samples are generated using a lognormal distribution with a
time-varying mean and variance, which simulates the behavior of common traffic
patterns in the internet \cite{}. The time-varying mean values are obtained
using superposition of sinusoidal functions, i.e.:
\begin{equation} \label{traffic_equation}
	y(t) = \alpha + \sum_{k=1}^n \beta_k \cdot \sin(\omega_k^t + \phi_k)
\end{equation}
, where $\alpha$ is a constant amplitude, $\beta_k$ and $\phi_k$ are frequency
dependent constants, and $n$ the number of frequency components, in our case
equal to 2. We generate 24 data samples per period simulating one day. An
example of a resulting function is shown in Fig. \ref{fig:traffic_gen}.

In the first scenario, during the first placement the VNFs are allocated based
on the observed traffic at that specific time step. In the second scenario, the
VNFs are allocated assuming the demands values are at the 80\% of the specific
maximum traffic demand value instead of considering the real observed values.
This let us consider this case as the most conservative one, since an
overprovisioning of resources will occur in most of the cases. In the third
scenario, the VNFs are allocated considering the predicted traffic demand values
after $\Delta t$ instead of the observed ones. Then, the resulting placement
from the three scenarios during this first placement is used as initial
condition for the optimization of the second phase, where in all cases only the
the real observed values are considered.

For the last scenario, a time series forecasting problem is modelled where a
certain number of periods $D-1$ are used for training and one period for
evaluation. We specifically use one LSTM network for every traffic demand with
input and output sizes of 1 unit and 8 units in a hidden layer. The model uses
Rectified Linear Unit (ReLU) as the activation function and is fit with Adam
optimizer and optimized using the mean squared error (\emph{mse}) loss function.
The batch size for the model is 4 and the validation data is 10\% of the total.
The number of epochs is not constrained, instead an early stopping function is
used with a minimum delta of 0.001 and a patience of 10 epochs. Specific
parameters are later described during the evaluation of the model.

\newcommand{\nN}{n \in \mathbb{N}}
\newcommand{\mN}{m \in \mathbb{N}}
\newcommand{\nNp}{n \in \mathbb{N}_p}
\newcommand{\m}{m}
\newcommand{\mNp}{\m \in \mathbb{N}_p}
\newcommand{\y}{y}
\newcommand{\xX}{x \in \mathbb{X}}
\newcommand{\xXC}{x \in \mathbb{X}_C}
\newcommand{\yX}{\y \in \mathbb{X}}
\newcommand{\xXn}{x \in \mathbb{X}_n}
\newcommand{\yXm}{\y \in \mathbb{X}_m}
\newcommand{\xXp}{x \in \mathbb{X}_p}
\newcommand{\lL}{\ell \in \mathbb{L}}
\newcommand{\pP}{p \in \mathbb{P}}
\newcommand{\pPs}{p \in \mathbb{P}_s}
\newcommand{\sS}{s \in \mathbb{S}}
\newcommand{\dD}{\lambda \in \Lambda}
\newcommand{\dDs}{\lambda \in \Lambda_s}
\newcommand{\ddDs}{\lambda' \in \Lambda_s}
\newcommand{\vVs}{v \in \mathbb{V}_s}
\newcommand{\tT}{t \in \mathbb{T}}
\newcommand{\yY}{y \in \mathbb{Y}}

\newcommand{\pl}{T_{p}^\ell}
\newcommand{\pnm}{T_{p}^{n, m}}
\newcommand{\loadratio}{\Gamma_{t(v)}^\mathrm{pro}}
\newcommand{\syncload}{\Gamma_{t(v)}^\mathrm{syn}}
\newcommand{\overhead}{\Theta_{t(v)}^s}
\newcommand{\overprovisioning}{\vartheta}
\newcommand{\replicable}{R_{t(v)}}
\newcommand{\Dl}{D_\ell}
\newcommand{\Dsmax}{D_s^{\mathrm{max}}}
\newcommand{\Dsmaxhat}{\hat{D}_{s}^\mathrm{{max}}}
\newcommand{\processtrafficdelay}{D_{t(v)}^\mathrm{proq}}
\newcommand{\processdelay}{D_{t(v)}^\mathrm{prox}}
\newcommand{\maxdelay}{D_{t(v)}^\mathrm{pro,max}}
\newcommand{\minprocessdelay}{D_{t(v)}^\mathrm{pro\_x,min}}
\newcommand{\migrationdelay}{D^\mathrm{dwt}}
\newcommand{\Cx}{C_x^\mathrm{max}}
\newcommand{\Cl}{C_{\ell}^\mathrm{max}}
\newcommand{\maxcapserver}{C_{x, t(v)}^\mathrm{proq,max}}
\newcommand{\energyidle}{E_{i}}
\newcommand{\energyutil}{ \alpha_{u}}
\newcommand{\serverothercosts}{K_x}
\newcommand{\functioncharges}{K_{t(v)}}
\newcommand{\penaltyparam}{\rho}

\newcommand{\xn}{x_n}
\newcommand{\rsp}{z_{p}^s}
\newcommand{\rspd}{z_{p}^{\lambda,s}}
\newcommand{\rspdd}{z_{p}^{\lambda',s}}
\newcommand{\fx}{f_x}
\newcommand{\fxsv}{f_x^{v,s}}
\newcommand{\fysv}{f_\y^{v,s}}
\newcommand{\Fxsv}{F_x^{v,s}}
\newcommand{\fxsvd}{f_{x,\lambda}^{v,s}}
\newcommand{\fxsvdd}{f_{x,\lambda'}^{v,s}}
\newcommand{\fysvd}{ f_{\y, \lambda}^{(v-1),s}}
\newcommand{\gxysv}{ g_{x, \y}^{v,s}}
\newcommand{\hsvp}{h_{p}^{v,s}}
\newcommand{\kl}{k_\ell}
\newcommand{\kx}{k_x}
\newcommand{\ksv}{k_v^s}
\newcommand{\penaltyvar}{q_p^{\lambda, s}}
\newcommand{\penaltyvarA}{\hat{q}_p^{\lambda, s}}
\newcommand{\penaltyvarB}{\hat{\hat{q}}_p^{\lambda, s}}
\newcommand{\penaltyvaraux}{y_p^{\lambda, s}}
\newcommand{\uell}{u_\ell}
\newcommand{\ux}{u_x}
\newcommand{\dsp}{d_p^{s}}
\newcommand{\dsdp}{d_p^{\lambda,s}}
\newcommand{\dsdph}{\hat{d}_p^{\lambda,s}}
\newcommand{\dpro}{d_{x, v, s}^{\mathrm{pro}}}
\newcommand{\dproq}{d_{x, v, s}^{\mathrm{proq}}}
\newcommand{\dprox}{d_{x, v, s}^{\mathrm{prox}}}
\newcommand{\dmgr}{d_{s}^{\mathrm{dwt}}}
\newcommand{\daux}{d_{x, \lambda}^{v,s}}
\newcommand{\trafficaux}{\Lambda_x^{v,s}}

\begin{table}[!h]
	\caption{Parameters and variables notation}
	\label{notation}
	\centering
	\begin{tabular}{>{\centering\arraybackslash}p{0.18\columnwidth} p{0.74\columnwidth}@{}}
		\toprule
		\textbf{Param.}                       & \textbf{Meaning}                                                                                                                                      \\
		\midrule
		$\mathbb{N}$                          & set of nodes: $\mathbb{N} = \{1,...,N\}$, $\nN$.                                                                                                      \\
		$\mathbb{X}$                          & set of servers: $\mathbb{X} = \{1,...,X\}$, $\xX$.                                                                                                    \\
		$\mathbb{L}$                          & set of links: $\mathbb{L} = \{1,...,L\}$, $\lL$.                                                                                                      \\
		$\mathbb{P}$                          & set of admissible paths: $\mathbb{P} = \{1,...,P\}$, $\pP$.                                                                                           \\
		$\mathbb{S}$                          & set of SFCs: $\mathbb{S} = \{1,...,S\}$, $\sS$.                                                                                                       \\
		$\mathbb{T}$                          & set of VNF types: $\mathbb{T} = \{1,...,T\}$, $\tT$.                                                                                                  \\
		$\mathbb{V}_s$                        & ordered set, $\vVs$ is the $v$\textsuperscript{th} VNF in set $\mathbb{V}_s$.                                                                         \\
		$\Lambda$                             & set of traffic demands: $\Lambda = \{1,...,\Lambda\}$, $\dD$.                                                                                         \\
		$\Lambda_s \subseteq \Lambda$         & subset of traffic demands $\dDs$ for SFC $\sS$.                                                                                                       \\
		$\mathbb{N}_p \subseteq \mathbb{N}$   & subset of ordered nodes in path $\pP$.                                                                                                                \\
		$\mathbb{X}_n \subseteq \mathbb{X}$   & subset of servers attached to node $\nN$.                                                                                                             \\
		$\mathbb{X}_p \subseteq \mathbb{X}$   & subset of ordered servers in path $\pP$.                                                                                                              \\
		$\mathbb{X}_C \subseteq \mathbb{X}$   & subset of servers located at the cloud.                                                                                                               \\
		$\mathbb{P}_s \subseteq \mathbb{P}$   & subset of admissible paths $\pPs$ for $\sS$.                                                                                                          \\
		$\pl, \pnm$                           & binary, 1 if path $\pP$ traverses link $\lL$ and 1 if connects node $\nN$ and $\mN$ as source and destination path nodes respectively.                \\
		$\loadratio, \syncload$               & continuous, load ratio of a VNF of type $t \in V_t$ and traffic ratio for synchronization traffic between two VNFs of type $t \in V_t$, respectively. \\
		$\overhead$                           & integer, overhead for VNF $\vVs$ of type $t \in V_t$.                                                                                                 \\
		$\Cl, \Cx$                            & integers, maximum capacity of link $\lL$ and of server $\xX$, respectively.                                                                           \\
		$\maxcapserver$                       & integer, maximum processing capacity that can be assigned by a server $x$ to a VNF of type $t$                                                        \\
		$\Dl$                                 & continuous, propagation delay of link $\lL$.                                                                                                          \\
		$\Dsmax, \migrationdelay$             & continuous, max. service delay of a SFC $\sS$ and service downtime duration caused by a migration, respectively.                                      \\
		$\maxdelay$                           & continuous, maximum allowed processing delay for a VNF of type $t$.                                                                                   \\
		$\processtrafficdelay, \processdelay$ & continuous, delay of a VNF $v$ of type $t$ due to queues and processing, respectively.                                                                \\
		\toprule
		\textbf{Vars.}                        & \textbf{Meaning}                                                                                                                                      \\
		\midrule
		$\rsp$                                & binary, 1 if SFC $s$ uses path $\pPs$.                                                                                                                \\
		$\rspd$                               & binary, 1 if traffic demand $\lambda$ from SFC $s$ uses path $\pPs$.                                                                                  \\
		$\fx$                                 & binary, 1 if server $x$ is used, 0 otherwise.                                                                                                         \\
		$\fxsv$                               & binary, 1 if VNF $\vVs$ from SFC $s$ is allocated at server $\xX$, 0 otherwise.                                                                       \\
		$\fxsvd$                              & binary, 1 if VNF $\vVs$ from SFC $s$ is used at server $\xX$ by traffic demand $\lambda$, 0 otherwise.                                                \\
		$\hsvp$                               & binary, 1 if VNF $\vVs$ from SFC $s$ uses path $\pP$ for state synchronization, 0 otherwise.                                                          \\
		$\dsdp$                               & continuous, service delay of a traffic demand $\lambda$ in path $p$.                                                                                  \\
		$\uell$, $\ux$                        & continuous, utilization of a link $\lL$ and server $\xX$, respectively.                                                                               \\
		\bottomrule
	\end{tabular}
\end{table}

\section{Problem Formulation} \label{LP_models}

We model the network as $\mathbb{G}=(\mathbb{N} \cup \mathbb{X}, \mathbb{L})$
where $\mathbb{N} = \{1,...,N\}$ is a set of nodes, $\mathbb{X} = \{1,...,X\}$
is a set of servers and $\mathbb{L} = \{1,...,L\}$ is a set of directed links.
Specifically, $\mathbb{X}_n$ is a subset of servers $\xX$ attached to node
$\nN$. We denote the set of all SFCs as $\mathbb{S} = \{1,...,S\}$, where a
specific SFC $\sS$ is an ordered set of VNFs $\mathbb{V}_s = \{1,...,V_s\}$,
each VNF being of type $t$, $\tT$, $\mathbb{T} = \{1,...,T\}$, where $\vVs$ is
the $v$\textsuperscript{th} VNF in set $V_s$. Table \ref{notation} summarizes
the notations. It should be noted that the model is written such that it can be
efficiently used in optimization solvers. For instance, the big M method is
avoided when possible or its value is minimized in order to avoid numerical
issues with the solver.

\subsection{Objective Function}

We define the joint optimization problem as the minimization of the sum of the
number of migrations and replications, i.e.,
\begin{subequations} \label{obj_func}
	\begin{align}
		\text{\emph{minimize}}: \quad & \sum_{\sS} \sum_{\vVs} \bigg[ W_m \sum_{\xX} \Fxsv (1 - \fxsv) \\
		                              & + W_r \Big[ \big(\sum_{\xX} \fxsv \big) - 1 \Big] \bigg]       \\
		                              & + W_c \sum_{\xXC} \sum_{\sS} \sum_{\vVs} \fxsv
	\end{align}
\end{subequations}
, where the variable $\fxsv$ specifies if a VNF $v$ from service chain $s$ is
allocated in server $x$. Since the optimization process follows two different
phases, after the first placement we take the value of variables $\fxsv$ and
convert them into the input parameters $\Fxsv$ for the next placement step, i.e.
\begin{equation} \label{initialsolutionmapping}
	\forall \sS, \forall \vVs, \forall \xX:      \fxsv   \Rightarrow  \Fxsv
\end{equation}
The parameter $\Fxsv$ determines if a VNF $v$ of a service chain $s$ was placed
on server $x$ during the initial placement. In this way, the first term of the
equation (\ref{obj_func}) counts the number of migrations, the second term
counts the number of replications and the third term counts the number of
functions allocated in cloud servers (here only $\mathbb{X}_C$ subset is
considered). We next follow up with the definition of constraints.

\subsection{General Constraints}
The general constraints are related to the traffic routing, the VNF placement
and the mapping between VNFs and paths.

\subsubsection{Routing}
For a given network, the input set $\pPs$ is the set of all pre-calculated paths
for SFC $s$. The binary variable $\rspd=1$ indicates, that a traffic demand
$\dDs$ of the SFC $s$ is using path  $\pPs$. The first routing constraint
specifies that each traffic demand $\dDs$ from SFC $\sS$ has to use only one
path $\pPs$, i.e.:
\begin{equation} \label{onePathPerDemand}
	\forall \sS, \forall \dDs: \sum_{\pPs} \rspd = 1
\end{equation}
Then, the next constraint takes the activated paths from the variable $\rspd$
and activates the path for a certain SFC $s$:
\begin{equation} \label{activatePathForService}
	\forall \sS, \forall \pPs, \forall \dDs:  \rspd \leq \rsp \leq \sum_{\ddDs} \rspdd
\end{equation}
This forces  $\rsp$ to be 1 when at least one traffic demand is using path $p$,
whereas the right side forces to $\rsp$ to be 0 when no traffic demand $\lambda$
is using path $p$.

\subsubsection{VNF placement}
VNF placement is modeled using the binary variable $\fxsvd$, which has only
value 1, if VNF $v$ from SFC $s$ is allocated at server $\xX$ and used by
traffic demand $ \dDs$. Similar to (\ref{onePathPerDemand}), the next constraint
defines that each traffic demand $\dDs$ from SFC $\sS$ traverses every VNF
$\vVs$ in only one specific server $\xX$:
\begin{equation} \label{oneFunctionPerDemand}
	\forall \sS, \forall \vVs, \forall \dDs: \sum_{\xX} \fxsvd = 1
\end{equation}

Then, similarly to (\ref{activatePathForService}), the next constraint takes the
activated VNFs for each traffic demand from the variable $\fxsvd$ and activates
the VNF for a certain SFC $s$ as follows:
\begin{equation} \label{mappingFunctionsWithDemands}
	\begin{split}
		\forall \sS, \forall \vVs, \forall \xX, \forall \dDs: \\
		\fxsvd \leq \fxsv \leq  \!\!\!\! \sum_{\ddDs} \fxsvdd
	\end{split}
\end{equation}
, where the left side forces to $\fxsv$ to be 1 when at least one traffic demand
$\dDs$ is using VNF $\vVs$ at server $\xX$ and the right side forces to $\fxsv$
to be 0 when no traffic demand is using that specific VNF $v$ on server $x$.
Likewise, we determine if a server is being used or not by constraining the
variable $\fx$ as:
\begin{equation}  \label{used-server}
	\forall \xX: \frac{1}{|\mathbb{S}||\mathbb{V}_s|} \sum_{\sS} \sum_{\vVs} \fxsv \leq \fx \leq \sum_{\sS} \sum_{\vVs} \fxsv     \text{  ,}
\end{equation}
where $\fx$ is 1 if at least one VNF from any SFC is allocated at server $\xX$,
0 otherwise.

\subsubsection{Mapping VNFs to paths}
The next equation maps the activated VNF to the activated paths defined in the
previous constraints. The first one defines how many times a VNF can be
replicated:
\begin{equation} \label{pathsConstrainedByFunctions}
	\forall \sS, \forall \vVs:  \sum_{\xX} \fxsv  \leq \replicable \sum_{\pPs} \rsp + 1 - \replicable
\end{equation}
, where $\replicable$ specifies if a certain VNF $v$ of type $t$ is replicable.
When $\replicable$ is 0, the total number of activated VNFs $\vVs$ from SFC
$\sS$ is $\sum_{\xX} \fxsv \leq 1$. In case the VNF is replicable, then the
maximum number of replicas is limited by the total number of activated paths
$\sum_{\pPs} \rsp$ for that specific SFC $s$. The next constraint activates
the VNFs on the activated paths:
\begin{equation}  \label{functionPlacement}
	\forall \sS, \forall \pPs, \forall \dDs , \forall \vVs: \rspd \leq \sum_{\xXp} \fxsvd
\end{equation}
If the variable $\rspd$ is activated, then every VNF $\vVs$ from SFC $\sS$
has to be activated in some server $\xXp$ from the path $\pP$ for a specific
traffic demand $\lambda$. When $\rspd$ is deactivated, then no VNFs can be
placed for that specific traffic demand. The last general constraint controls
that all VNFs $V_s$ from a specific SFC $s$ are traversed by every traffic
demand $\dDs$ in the given order, i.e.:
\begin{equation}  \label{functionSequenceOrder}
	\begin{split}
		\forall \sS, \forall \dDs, \forall \pPs, \forall \vVs, \forall n, m \in \mathbb{N}: \\
		\Bigg( \sum_{\m = 1}^{n} \sum_{\yXm}  \!  \fysvd \!\! \Bigg) - \! \!\!  \sum_{\xXn} \!\!  \fxsvd \geq \rspd \! - 1
		\left\{
		\begin{array}{ll}
			1< v \leq |\mathbb{V}_s | \\
			n \neq m
		\end{array}
		\right.
	\end{split}
\end{equation}
, where the variable $\rspd$ activates the ordering constraint side (left side)
when is 1 and deactivates it, otherwise. Then, if path $\pP$ is activated, the
ordering is checked for every traffic demand $\dDs$ individually by using the
variable $\fxsvd$. Hence, for every traffic demand $\lambda$ of SFC $s$, the
$v$\textsuperscript{th} VNF is allocated at server $\xXn$ only if the previous
$(v - 1)$\textsuperscript{th} VNF is allocated at any server $\yXm$, where $m$
is the i\textsuperscript{th} node from 1 until $n$ traversed by path $p$. It
should be noted, that the correct sequence of VNFs relies on the correct
sequence of subset of servers, i.e.  $\xXn$. This assumes that the  correct
sequence of VNFs inside these subsets  is organized by the local routing, which
may be located at the node $n$ or at a local switch not modeled in detail.

\subsection{Traffic and Performance Constraints}

\subsubsection{Synchronization traffic}

When performing replications of VNFs, the stateful states between the original
and replicas has to be maintained in order to be reliable against VNF failures
and avoiding the loss of information. For this reason, we consider that when a
VNF is replicated, the generated synchronization traffic between replicas and
the original has to be also considered. The amount of the state synchronization
traffic depends on the state space and its time dynamic, where it is assumed,
that each VNF has full knowledge on the state of all its instances used to
implement the VNF $\vVs$. Let us assume, that this amount is proportional to the
total traffic offered to the SFC weighted by an synchronization ratio
$\syncload$, which depends on the type of VNF $t$. In summary, the directional
traffic from a VNF to its replica is given by $\syncload  | \Lambda_s | $, and
its routing should be optimized within the network.

In order to know if the same VNF $\vVs$ from SFC $s$ is placed in two
different servers $\xX$ and $\yX$, we define:
\begin{equation}  \label{sync-traffic}
	\begin{split}
		\forall \sS, \forall \vVs, \forall \xX, \forall \yX: \\
		\gxysv = \fxsv \fysv \text{, \quad for }  y \! \neq \! x
	\end{split}
\end{equation}
, where the variable $\gxysv$ is 1 only when both variables $\fxsv$ and $\fysv$
are also 1, and 0 otherwise. In this way, this variable is used to know if two
different servers have the same VNF placed, which means that model is allocating
one replica. We use the well-known linearization method when multiplying two
binary variables. In case $\gxysv = 1$, we need to carry the synchronization
traffic from server $x$ to $y$, by selecting only one predefined path  between
them, i.e.:
\begin{equation} \label{hpvs}
	\begin{split}
		\forall \sS, \forall \vVs,  \forall n, m \in \mathbb{N}, \forall \xXn, \forall \yXm:  \\
		\gxysv  \leq  \sum_{\pP} \hsvp \cdot \pnm \leq 1   \text {,\quad for } n \neq m
	\end{split}
\end{equation}
\begin{equation} \label{hpvs_2}
	\begin{split}
		\forall \sS, \forall \vVs,  \forall n, m \in \mathbb{N}:  \\
		\sum_{\pP} \hsvp \cdot \pnm  \leq  \sum_{\xXn} \sum_{\yXm}  \gxysv  \text {,\quad for } n \neq m
	\end{split}
\end{equation}
, where the constant $ \pnm = 1$ indicates, that the path $ \pP$ exists which
connects servers  $ \xXn$ and $ y \in X_m$ using the shortest path between nodes
$n$ and $m$. The right term of \eqref{hpvs} guarantees that only one path $\pP$
is selected by variable $\hsvp$. Moreover, \eqref{hpvs_2} guarantees that this
path is only used if at least one $\gxysv$ is 1. Note that $\hsvp$ is a binary
variable used for every VNF $v$ of SFC $s$.

\subsubsection{Link and server utilization}

The utilization of a link is calculated as follows:
\begin{equation} \label{linkutil}
	\begin{split}
		\forall \lL: \uell =   \frac{1}{\Cl}  \sum_{\sS} \sum_{\pPs} \sum_{\dDs} \lambda \cdot \pl \cdot \rspd +  \\
		\frac{1}{\Cl}  \sum_{\pP}  \pl \sum_{\sS} \sum_{\vVs}  \syncload \cdot  |\Lambda_s| \cdot \hsvp  \leq 1  \text{ ,}
	\end{split}
\end{equation}
where $\lambda \cdot \pl$ adds the traffic demands from SFC $\sS$ when a path
$\pPs$ traverses the link $\lL$. Then, the variable $\rspd$ specifies if the
traffic demand $\lambda$ from SFC $s$ is using path $p$. The second term is the
sum of the extra traffic generated due to the state synchronization between VNFs
$\vVs$ from SFC $s$, which is proportional to its total traffic $|\Lambda_s|$
multiplied by the synchronization traffic ratio $\syncload$ of the VNF of type
$t$. This traffic is only added, if the variable $\hsvp$ is 1, which indicates
that path $\pP$ is used for synchronization by a VNF $v$ from SFC $s$, and the
link  $ \lL$ belongs to this path. Both summation terms are divided by the
maximum link capacity $\Cl$ to restrict the utilization.

The processing load of a server is derived as
\begin{equation} \label{server_load}
	\gamma_x = \sum_{\sS} \sum_{\vVs} \Big(  \loadratio  \sum_{\dDs} \lambda \cdot \fxsvd + \overhead \cdot \fxsv \Big)
\end{equation}
, where the first term sums the traffic $\dDs$ that is using the VNF $\vVs$ from
SFC $\sS$ at server $\xX$, which is determined by the variable $\fxsvd$, and
multiplied by the processing load ratio $\loadratio$ of the VNF of type $t$. The
second term adds the overhead generated by the VM where the VNF is running and
is only added, when the variable $\fxsv$ determines that this VNF is placed in
server $x$. Then, the utilization follows to be given by
\begin{equation} \label{serverutil}
	\forall \xX:  \ux  =    \frac{\gamma_x}{\Cx}      \leq 1    \text{ ,}
\end{equation}
where  $\Cx$  is  the maximum processing capacity.

\subsubsection{Service delay} \label{service_delay} Since every service has a
maximum allowed delay $\Dsmax$ specified in the SLA agreement, in case of
exceeding it, some penalty costs are applied. In our model, and for simplicity,
we take into account the propagation delay due to the traversed links, the
processing delay that every VNF requires in the servers and, where applicable,
the downtime delays caused by the interruption of the service during the
migrations of VNFs.

\emph{Processing delay}: The processing delay $\dpro$ of a VNF $v$ in a server
$x$ depends, on the one side, on the amount of traffic being processed by a
specific VNF, described by $\dproq$, and on  $\dprox$, which is related to the
VNF type and  the total  server  load $\ux$, given as
\begin{subequations} \label{processing_delay_equations}
	\begin{equation}
		\forall \sS, \forall \vVs, \forall \xXp:  \dpro = \dproq + \dprox
	\end{equation}
	\begin{equation}
		\dproq = \processtrafficdelay \frac{ \loadratio \cdot \sum_{\dDs} \fxsvd \cdot \lambda}{\maxcapserver} \label{processing_delay_equations_B}
	\end{equation}
	\begin{equation}
		\dprox =  \minprocessdelay \cdot \fxsv + \processdelay \cdot \ux  \label{processing_delay_equations_C}
	\end{equation}
\end{subequations}

In \eqref{processing_delay_equations_B}, the numerator of $\dproq$  determines
the total processing load assigned to the VNF of type $t$, which is controlled
by the variables $\fxsvd$. Thus, if the assigned processing load is equal to
$\maxcapserver$, the VNF adds the processing delay $\processtrafficdelay$. The
second delay term, given in \eqref{processing_delay_equations_C}, adds the load
independent minimum delay associated to the usage of a type of this VNF, and a
delay part which increases with the server utilization. As a consequence the
processing delay $\dpro(\vec{ \lambda})$ depends on the server $x$, the used VNF
type and linearly increases with increasing traffic. Furthermore, the dependency
on all traffic demands is denoted by the vector $\vec{ \lambda}$, which is
omitted for simplicity in \eqref{processing_delay_equations}.

\emph{Downtime duration}: If a VNF $v$ of SFC $s$ has to be migrated, we assume
an interruption of the service with duration $\migrationdelay$. Thus, the total
service downtime will consider the migration of all VNFs in that SFC which
yields a constraint as follows:
\begin{equation}  \label{migration_delay_equations}
	\forall \sS: \dmgr = \migrationdelay \sum_{\xX} \sum_{\vVs} \Fxsv (1 - \fxsv)
\end{equation}
, where the parameter $\Fxsv$ determines if a VNF $v$ was placed on server $x$
during the first placement. Thus, if a VNF migrates to another server $y \neq
	x$, the variable $\fxsv$ is equal to zero and the service downtime
$\migrationdelay$ has to be taken into account.

\emph{Total delay}: Because the model allows that different traffic demands per
service can be assigned to different paths, we define individual end-to-end
delay $\dsdph$ for every traffic demand, as follows:
\begin{equation} \label{exact_service_delay}
	\begin{split}
		\forall \sS, \forall \dDs, \forall \pPs: \\
		\dsdph =  \sum_{\lL} \Dl \cdot \pl  + \sum_{\xXp} \sum_{\vVs}    \dpro(\vec{\lambda}) \cdot  \fxsvd   + \dmgr
	\end{split}
\end{equation}
The first term is the propagation delay, where $\Dl$ is the delay of the link
$\ell$, and $\pl$ specifies if the link $\ell$ is traversed by path $\pPs$. The
second term adds the processing delays caused by all VNFs from the SFC placed on
the servers $\xXp$, in which the variable $\fxsvd$ has to ensure that the demand
$\lambda$ is processed at a specific server $x$. Finally, the third term is the
total downtime duration due to the migrations of that service chain. It should
be noted that the second term of \eqref{exact_service_delay} includes a
nonlinear relation between the binary variable $f^{v,s}_{x, \lambda}$ and the
delay variable $\dpro$, which also depends on all decision variables
$f^{v{'},s{'}}_{x, \lambda{'}}$. To solve that, we introduce a new delay
variable $\daux$, which  is  bounded as follows:
\begin{equation} \label{new-variable}
	\dpro - \maxdelay(1 - \fxsvd)  \leq  \daux   \leq \maxdelay \cdot \fxsvd
\end{equation}
If the VNF is selected at server $x$ by $\fxsvd=1$, the variable is lower
bounded by the exact delay $\dpro$ and upper bounded by the maximum VNF delay
$\maxdelay$. Since $\dpro \leq \daux \leq \maxdelay$, the specific delay of a
VNF can be restricted. If the VNF is not selected, i.e., $\fxsvd=0$, the
variable has value $\daux=0$, since the constant $\maxdelay$ makes the left size
of \eqref{new-variable} to be negative. Hence, the end-to-end delay is mapped to
an upper and lower bounded variable $\dsdp$ given as
\begin{equation} \label{total_service_delay}
	\begin{split}
		\forall \sS, \forall \dDs, \forall \pPs: \\
		\dsdp =  \sum_{\lL} \Dl \cdot \pl  + \sum_{\xXp} \sum_{\vVs} \daux + \dmgr \quad \text{,}
	\end{split}
\end{equation}
in which the bounding feature is used in the optimization scenarios described
next.

\section{Online Heuristic Approaches}

Since the model presented is a MILP optimization problem and these models are
known to be NP-hard \cite{Bulut2015}, in this section we propose a greedy
algorithm to work as an online solution and, First-Fit and Random-Fit algorithms
for comparison purposes.

\subsection{First-Fit and Random-Fit algorithms}

Both \emph{First-Fit} (FF) and \emph{Random-Fit} (RF) algorithms are described
in Algorithm \ref{algorithm:ff_rf}. While both approaches share most of the
code, the \emph{FF\_RF} parameter specifies whether the code has to run FF or
RF. The process starts with a loop where every demand from every SFC is going to
be considered (line \ref{main_loop_FFRF}). The first step is to then retrieve
all the paths with enough link resources to assign traffic demand $\lambda$ and
that also connect both source and destination nodes (line
\ref{admissible_paths_FFRF}). These paths are saved into $\mathbb{P}_s'$, from
where one admissible path $p$, first one for FF or a random one for RF, is
selected (line \ref{choose_path_FFRF}). In this point, we make sure here that in
this path, there are enough server resources to allocate all the VNFs for SFC
$s$. Then, from that path, for every VNF $v$ from SFC $s$ (line
\ref{for_functions_FFRF}) we start with the process of selecting servers for
allocations. First, we retrieve all servers with enough free capacity to
allocate the VNF $v$ and to provide service to demand $\lambda$ (line
\ref{available_servers_FFRF}), and then we choose the first available server in
FF or a random one in RF (line \ref{choose_server_FFRF}). It is to be noted
here, that to satisfy VNF ordering (see equation \ref{functionSequenceOrder}),
the procedure \emph{chooseServer} will return a valid server from before/after
the previous/next VNF allocated. While for the FF case, we assure in line
\ref{choose_path_FFRF} that there will always be a server where to allocate the
next VNF in the chain, in RF case we make sure here (line
\ref{choose_server_FFRF}) that after the random server selected there is still
place to allocate all the rest of the VNFs from the chain in next servers in the
path, or we select another server instead. In line
\ref{add_function_to_server_FFRF}, we assign the demand and the VNF to the
server (i.e. equations (\ref{oneFunctionPerDemand}) and
(\ref{mappingFunctionsWithDemands})). After all the VNFs have been placed, the
next step is to route traffic demand $\lambda$ to path $p$ (line
\ref{route_demand_to_path_FFRF}), to finally add the synchronization traffic for
the service chain (line \ref{add_synch_traffic_FFRF}).

\begin{algorithm}[!t]
	\caption{First-Fit/Random-Fit: \textit{main(FF\_RF)}}
	\begin{algorithmic}[1]
		\For{$\sS$, $\dDs$} \label{main_loop_FFRF}
		\State $\mathbb{P}_s' \gets$ getAdmissiblePaths($s$, $\lambda$) \label{admissible_paths_FFRF}
		\State $p \gets$ choosePath(FF\_RF, $\mathbb{P}_s'$) \label{choose_path_FFRF}
		\For{$\vVs$} \label{for_functions_FFRF}
		\State $\mathbb{X}_p' \gets$ getAvailableServers($s$, $\lambda$, $v$, $p$) \label{available_servers_FFRF}
		\State $x \gets$ chooseServer(FF\_RF, $\mathbb{X}_p'$) \label{choose_server_FFRF}
		\State addVNFToServer($s$, $v$, $\lambda$, $x$) \label{add_function_to_server_FFRF}
		\EndFor
		\State routeDemandToPath($s$, $p$, $\lambda$) \label{route_demand_to_path_FFRF}
		\State addSynchronizationTraffic($s$)  \label{add_synch_traffic_FFRF}
		\EndFor
	\end{algorithmic}
	\label{algorithm:ff_rf}
\end{algorithm}

\subsection{Greedy algorithm}

The greedy algorithm main function is described in Algorithm
\ref{algorithm:greedy_main}. The procedure starts with the natural ordering of
SFCs by the total traffic demand value (line \ref{order_services_GRD}). This is
done in order to first allocate services with lower impact on the utilization
resources in order to avoid the creation of bottlenecks in servers and links
during the firsts phases of the allocation. Then, it starts iterating over each
service (line \ref{for_services_GRD}) and over each traffic demand for certain
service (line \ref{for_demands_GRD}). Then, for each traffic demand we first
retrieve all paths with enough free link resources in $\mathbb{P}_s'$ (line
\ref{admissible_paths_GRD}). Then, we choose a path $p$ inside of a loop from
all retrieved paths (line \ref{choose_path_GRD}, details explained later).This
is done to ensure that in case a path cannot be used for allocating all VNFs,
the algorithm tries with the next one. Once the path is selected, we start with
the placement of all VNFs on the selected path. First, all the available servers
for a specific VNF $v$ on path $p$ are retrieved in variable $\mathbb{X}_p'$
(line \ref{available_servers_GRD}), then we choose one server $x$ for that
specific VNF in line \ref{choose_server_GRD} (this procedure explained later)
and place the VNF (line \ref{map_function_to_server_GRD}). In case the VNF has
been already placed by another demand of the same service, the demand is
associated to that VNF, instead. Finally after all VNFs are placed, we map the
demand over path (line \ref{map_demand_to_path_GRD}). Finally, as in the
previous case, the synchronization traffic for that service is added (line
\ref{add_synch_traffic_GRD})

\begin{algorithm}[!t]
	\caption{Greedy: \textit{main()}}
	\begin{algorithmic}[1]
		\State $\mathbb{S}'$ = orderServicesByTotalDemandValue($\mathbb{S}$) \label{order_services_GRD}
		\For{$\sS'$}  \label{for_services_GRD}
		\For{$\lambda \in \Lambda_s$}  \label{for_demands_GRD}
		\State $\mathbb{P}_s' \gets$ getAdmissiblePaths($s$, $\lambda$) \label{admissible_paths_GRD}
		\For{$p \in \mathbb{P}_s'$}
		\State $p \gets$ choosePath($s$, $\lambda$, $\mathbb{P}_s'$) \label{choose_path_GRD} \Comment{go to Alg. \ref{algorithm:greedy_choosePath}}
		\For{$\vVs$}
		\State $\mathbb{X}_p' \gets$ getAvailableServers($s$, $\lambda$, $p$, $v$) \label{available_servers_GRD}
		\State $x \gets$ chooseServer($s$, $\lambda$, $p$, $v$, $\mathbb{X}_p'$) \label{choose_server_GRD} \Comment{go to Alg. \ref{algorithm:greedy_chooseServer}}
		\State mapVNFToServer($v$ , $x$) \label{map_function_to_server_GRD}
		\EndFor
		\State mapDemandToPath($s$, $p$, $\lambda$) \label{map_demand_to_path_GRD}
		\EndFor
		\EndFor
		\State addSynchronizationTraffic($s$)  \label{add_synch_traffic_GRD}
		\EndFor
	\end{algorithmic}
	\label{algorithm:greedy_main}
\end{algorithm}

\begin{algorithm}[!t]
	\caption{Greedy: \textit{choosePath($s$, $\lambda$, $\mathbb{P}_s'$)}}
	\begin{algorithmic}[1]
		\State $p \gets$ getUsedPathDemandInitPlacement($s$, $\lambda$, $\mathbb{P}_s'$) \label{get_used_path_for_demand_init_GRD}
		\If{$p$}
		\Return $p$
		\EndIf
		\State $p \gets$ getUsedPathInitialPlacement($s$, $\mathbb{P}_s'$) \label{get_used_path_init_GRD}
		\If{$p$}
		\Return $p$
		\EndIf
		\State $p \gets$ getUsedPathForSFC($s$, $\mathbb{P}_s'$) \label{get_used_path_for_service_GRD}
		\If{$p$}
		\Return $p$
		\EndIf
		\State \Return getPathWithShortestDelay($s$, $\lambda$, $\mathbb{P}_s'$) \label{get_path_shortest_delay_GRD}
	\end{algorithmic}
	\label{algorithm:greedy_choosePath}
\end{algorithm}

\begin{algorithm}[!t]
	\caption{Greedy: \textit{chooseServer($s$, $\lambda$, $p$, $v$, $\mathbb{X}_p'$, A)}}
	\begin{algorithmic}[1]
		\State $\mathbb{X}_p' \gets$ removeServersPreviousVNFs($\mathbb{X}_p'$) \label{remove_servers_previous_vnfs_GRD}
		\State $\mathbb{X}_p' \gets$ removeServersNextVNFs($\mathbb{X}_p'$) \label{remove_servers_next_vnfs_GRD}
		\State $c \gets$ getCloudServer($\mathbb{X}_p'$) \label{get_cloud_server_GRD}
		\State $x \gets$ getUsedServerDemandInitialPlace($s$, $v$, $\lambda$, $\mathbb{X}_p'$) \label{get_used_server_demand_init_GRD}
		\State checkPosition(x, c, A) \Comment{go to line \ref{check_position}} \label{check_position_1_GRD}
		\State $x \gets$ getUsedServerInitialPlacement($s$, $v$, $\mathbb{X}_p'$) \label{get_used_server_init_GRD}
		\State checkPosition(x, c, A) \Comment{go to line \ref{check_position}} \label{check_position_2_GRD}
		\State $x \gets$ getUsedServerForSFC($s$, $v$, $\mathbb{X}_p'$) \label{get_used_server_GRD}
		\State checkPosition(x, c, A) \Comment{go to line \ref{check_position}} \label{check_position_3_GRD}
		\If{!A} \label{check_last_try_GRD}
		\Return null
		\Else \ \Return $\mathbb{X}_p'$[0] \label{get_first_server_GRD}
		\EndIf

		\Procedure{checkPosition}{$x$, $c$, A} \label{check_position}
		\If{$x$ != null} \label{check_x_valid_GRD}
		\If{$A$ \textbf{\&} $c$ \textbf{\&} indexOf($x$) < indexOf($c$)} \label{check_x_c_1_GRD}
		\State \Return x
		\ElsIf{$A$ \textbf{\&} $c$ \textbf{\&} indexOf($x$) > indexOf($c$)} \label{check_x_c_2_GRD}
		\State \Return c
		\Else \ \Return $x$ \label{check_x_c_3_GRD}
		\EndIf
		\EndIf
		\EndProcedure
	\end{algorithmic}
	\label{algorithm:greedy_chooseServer}
\end{algorithm}

When selecting a path for a specific traffic demand in line
\ref{choose_path_GRD}, the procedure described in Algorithm
\ref{algorithm:greedy_choosePath} is executed. This procedure execute the
following methods in this specific order: return an already used path for the
same demand $\lambda$ during the initial placement (line
\ref{get_used_path_for_demand_init_GRD}), return any used path for SFC $s$
during the initial placement (line \ref{get_used_path_init_GRD}), return any
used path for SFC $s$ (line \ref{get_used_path_for_service_GRD}) or return the
path with shortest path delay (line \ref{get_path_shortest_delay_GRD}). If one
method does not return a path, then the next one is executed.

Going back to Algorithm \ref{algorithm:greedy_main}, when choosing a server for
a specific VNF in line \ref{choose_server_GRD}, the procedure described in
Algorithm \ref{algorithm:greedy_chooseServer} is executed. In this point, we
first remove servers from the set $\mathbb{X}_p'$ that have already allocated
VNFs before/after the current VNF in the path (lines
\ref{remove_servers_previous_vnfs_GRD} and \ref{remove_servers_next_vnfs_GRD}),
in order to satisfy with sequence order equation (\ref{functionSequenceOrder}).
Then, we follow up with the selection of a server from the remaining ones. Here,
in case it exists, we first retrieve the cloud server $c$ in the path (line
\ref{get_cloud_server_GRD}). Then, we retrieve a server already used for VNF $v$
and demand $\lambda$ during the initial placement (line
\ref{get_used_server_demand_init_GRD}) into server $x$. In line
\ref{check_position_1_GRD}, we check the position in the path of that server,
where the procedure is specified in line \ref{check_position}. This procedure
receives the server $x$, the cloud server $c$ in case it exists and the boolean
variable $A$ which specifies whether this is the last attempt in terms of
remaining available paths. The objective here is to first check if $x$ is valid
(line \ref{check_x_valid_GRD}), otherwise it finishes. In case is valid, then we
return $x$ if it is the last attempt $A$, if it exists a cloud server in the
path and if the index of $x$ is lower than the index of $c$ in the array. In
case this condition does not apply, we continue with the next condition in line
\ref{check_x_c_2_GRD} with the difference of checking whether $x$ is after the
cloud server in the array. In that case, the cloud server is returned. If none
of the previous applies, then $x$ is returned in line \ref{check_x_c_3_GRD}.
This procedure is basically performed to make sure that in all cases there will
be a location where to place VNFs which is in the cloud server, but choosing it
as the last option. Continuing with line \ref{get_used_server_init_GRD},
similarly here we try to retrieve a server used during the initial placement for
service $s$ regardless for which traffic demand and perform the same procedure
like in the previous case (line \ref{check_position_2_GRD}). While the first
method tries to reuse the same exact server like in the initial placement in
order to avoid a migration, here we try to use a server already used by some
other demand during the initial placement for the same service in order to avoid
a replication. Similarly, the next case in line \ref{get_used_server_GRD}
retrieves an already used server by the same service regarless it is from
initial placement or allocated during the current placement. Here again we are
trying to avoid an unnecessary replication and we check again like before the
position of the returned server in line \ref{check_position_3_GRD}. If none of
the previous methods returned a valid server, then we return null in line
\ref{check_last_try_GRD} in order to later try with the next available path in
case this is not the latest path. If it is the latest path, then we just return
the first available server in the set (line \ref{get_first_server_GRD}).

\subsubsection{Computational complexity} 
In terms of complexity from bottom to top, for the Algorithm
\ref{algorithm:greedy_chooseServer} considering $V_{L}$ as the length of the
longest SFC, it is in the order of $\Theta = O(V_{L} \cdot |\mathbb{X}|)$. The
Algorithm \ref{algorithm:greedy_choosePath} is in the order of $\Theta' =
	O(P_{S})$ where $P_{S}$ is the number of paths per SFC. The Algorithm
\ref{algorithm:greedy_main} is calculated based on the complexity of Algorithm
\ref{algorithm:greedy_choosePath} and \ref{algorithm:greedy_chooseServer}, and
the complexity of the synchronization traffic (line \ref{add_synch_traffic_GRD})
which is in the order of $\Theta'' = O(V_{L} \cdot |\mathbb{X}^2| \cdot
	|\mathbb{P}|)$. Considering $L_{P}$ as the length of the longest path, then the
complexity of the entire Algorithm \ref{algorithm:greedy_main} is in the order
of $O(|\mathbb{S}^2| + |\Lambda| \cdot L_{P} \cdot P_{S} \cdot [\Theta' + V_{L}
		\cdot \Theta] \cdot \Theta'')$, which can be simplified as $O(|\mathbb{S}^2| +
	|\Lambda| \cdot L_{P} \cdot V_{L} \cdot |\mathbb{X}^2| \cdot |\mathbb{P}| \cdot
	[P_{S}^2 + V_{L}^2 \cdot  |\mathbb{X}| \cdot P_{S}])$.


\begin{figure}[!t]
	\centering
	\subfloat[N7 network]{\includegraphics[width=0.40\columnwidth]{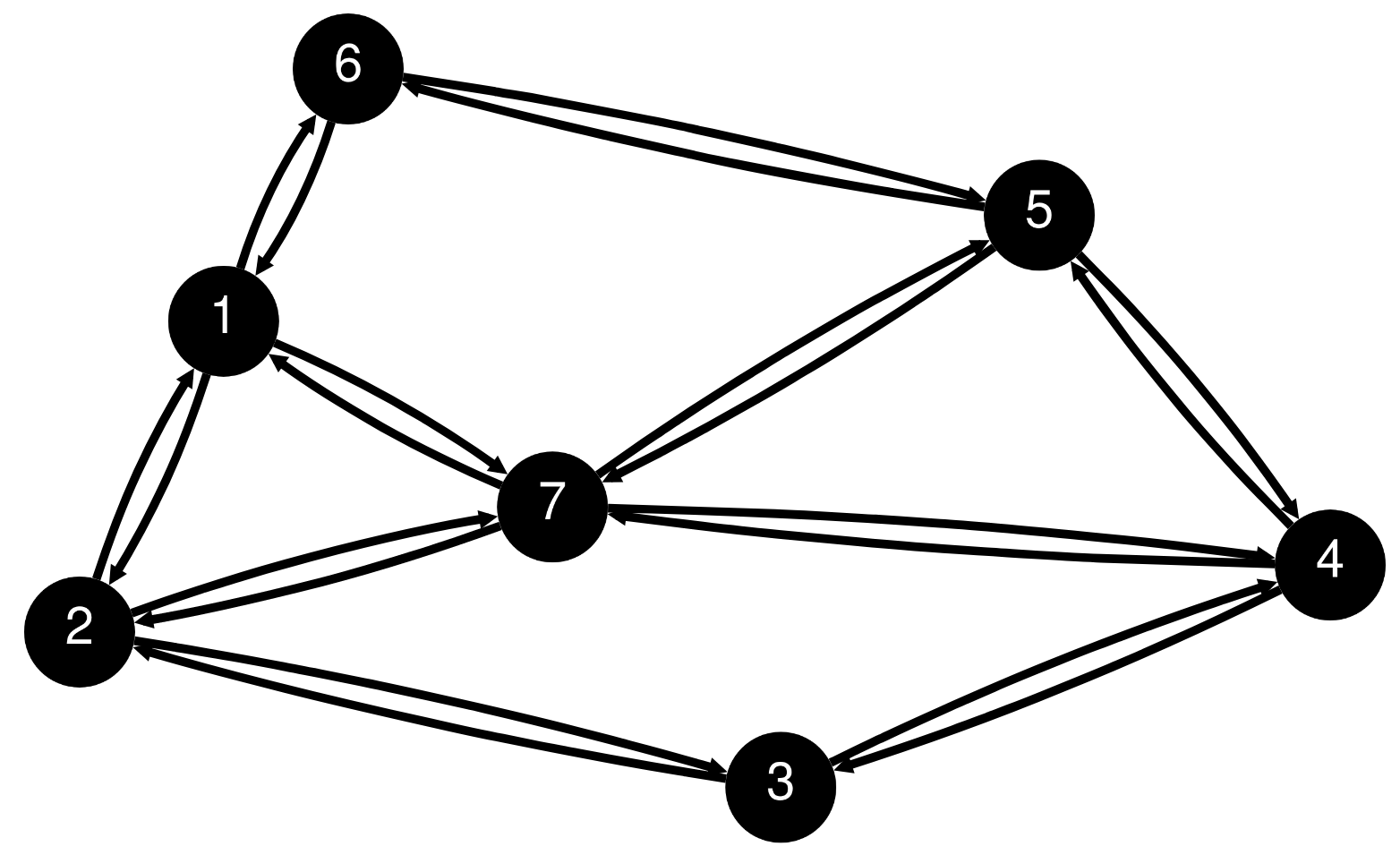}%
		\label{fig:N7}}
	\hfil
	\subfloat[N45 network]{\includegraphics[width=0.60\columnwidth]{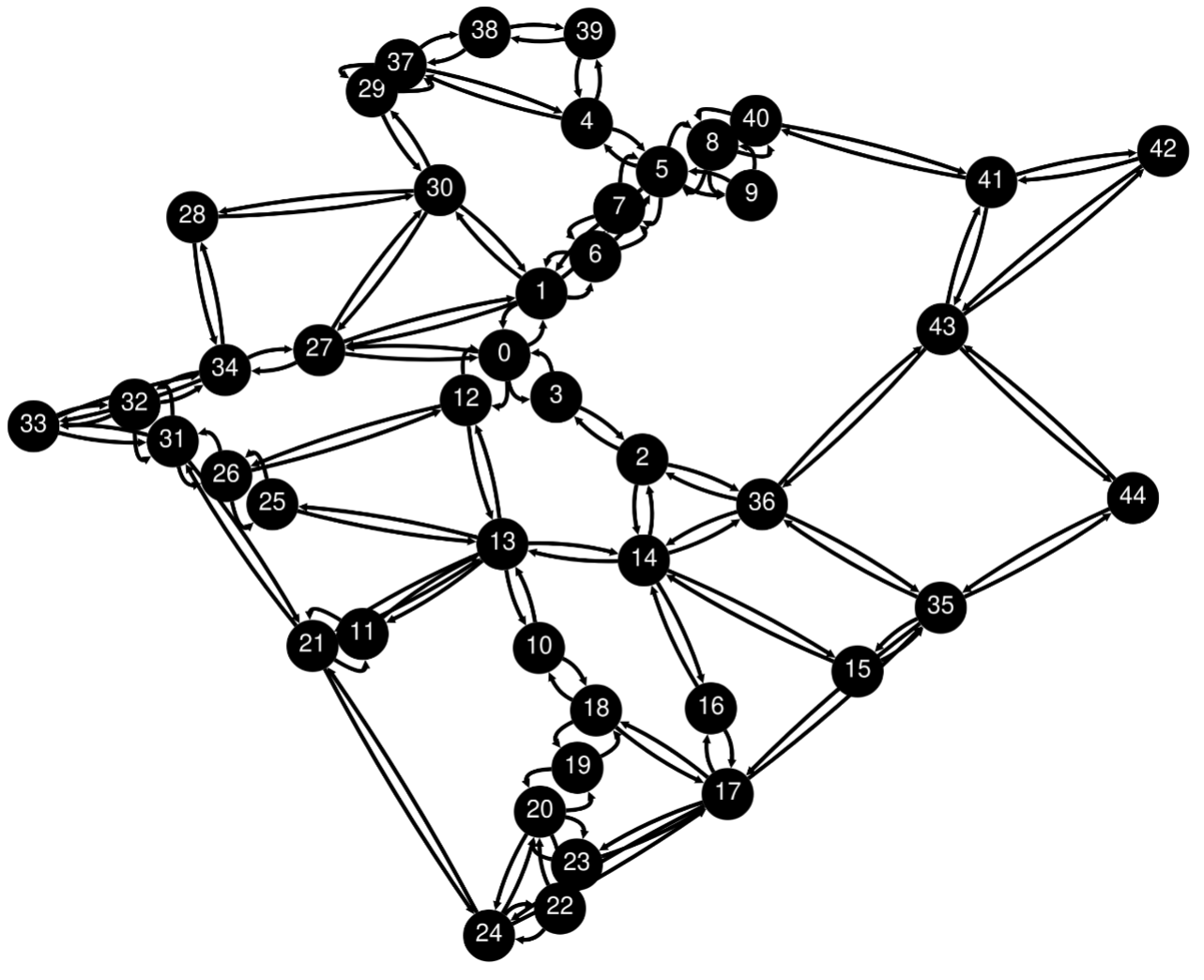}%
		\label{fig:N45}}
	\caption{Network topologies used in the performance evaluation.}
\end{figure}

\begin{figure}[!t]
	\centering
	\subfloat[RMSE]{\includegraphics[width=0.24\textwidth]{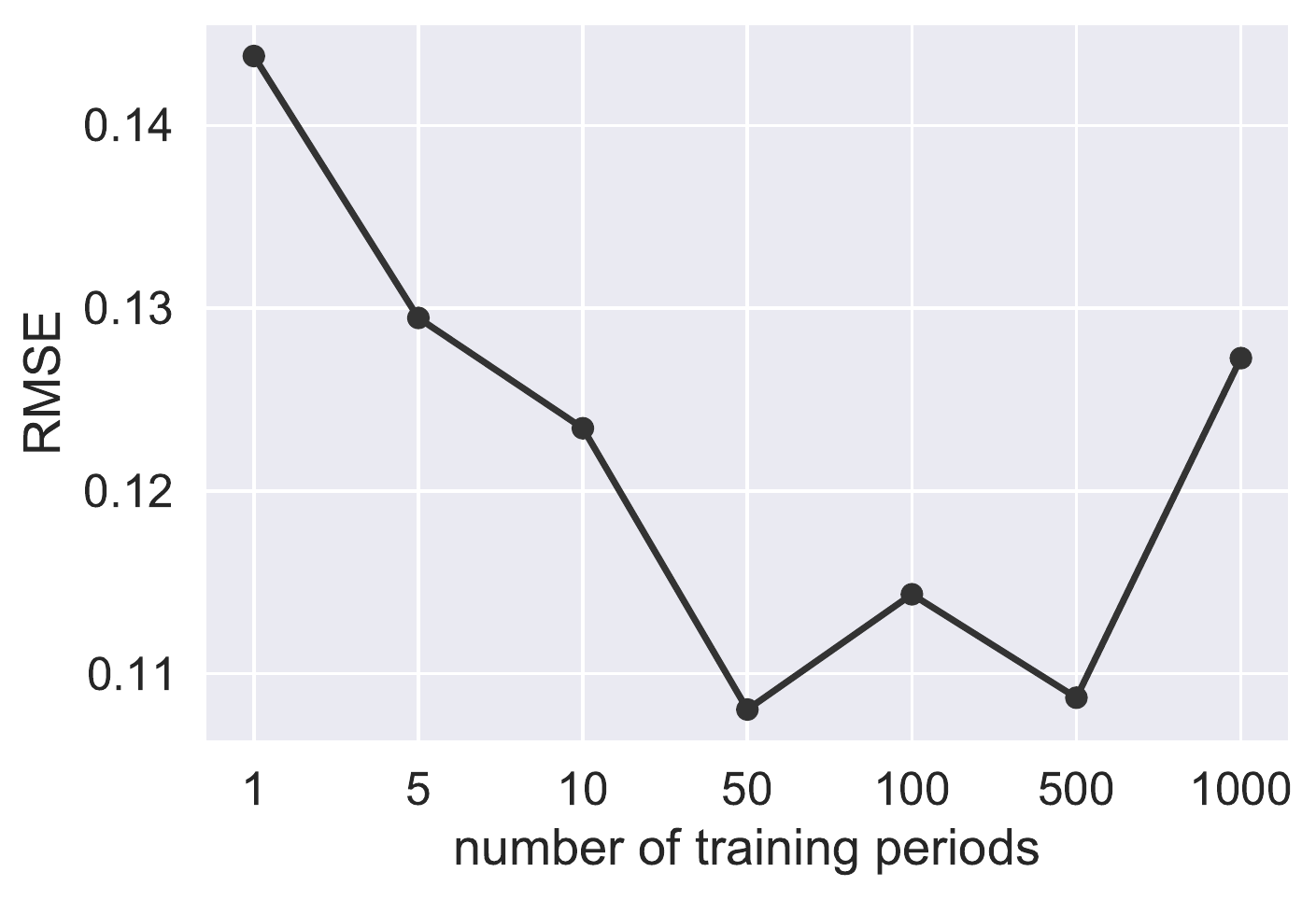}%
		\label{fig:rmse}}
	\hfil
	\subfloat[Training time]{\includegraphics[width=0.236\textwidth]{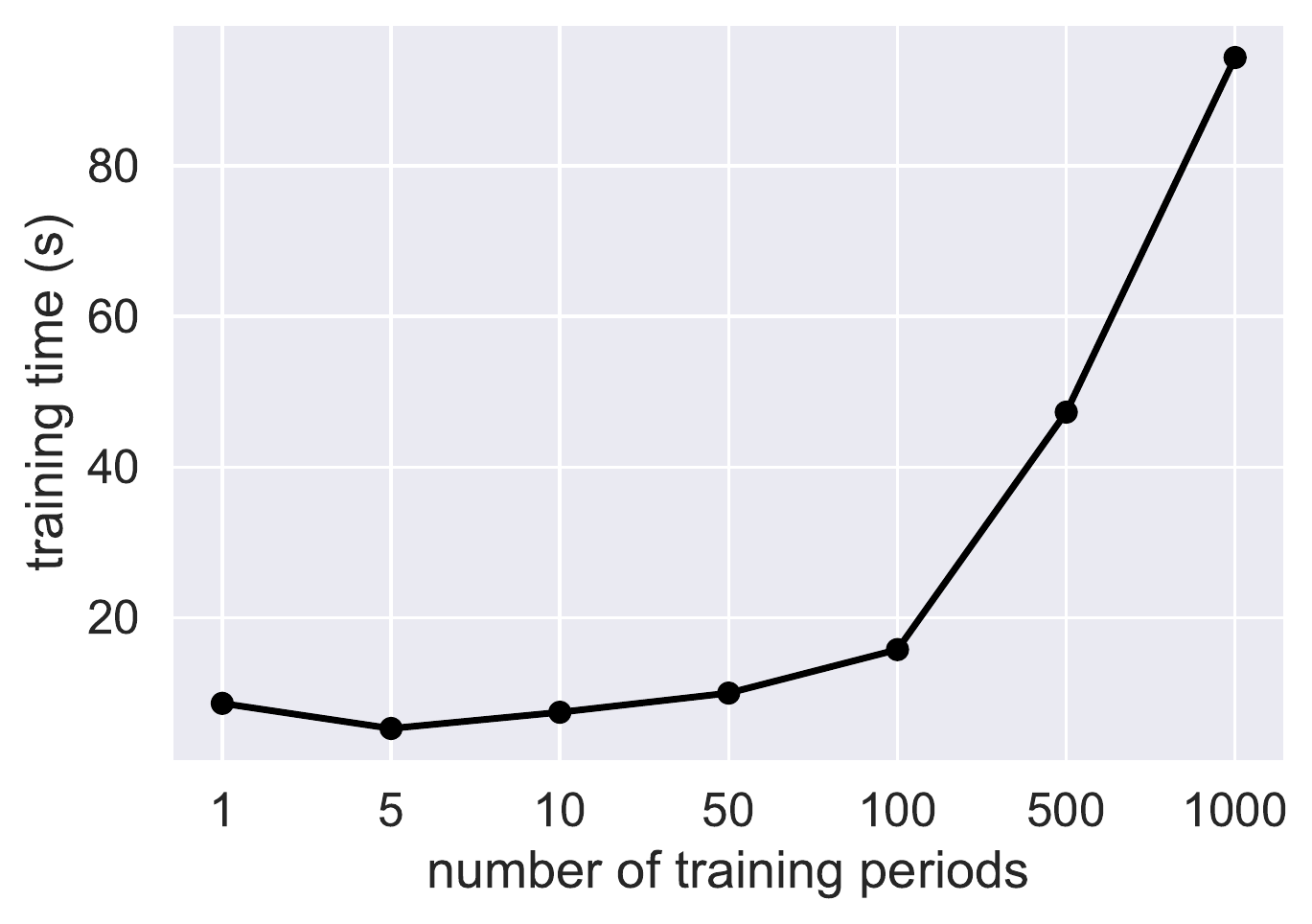}%
		\label{fig:time}}
	\hfil
	\subfloat[Validation for 1 period]{\includegraphics[width=0.237\textwidth]{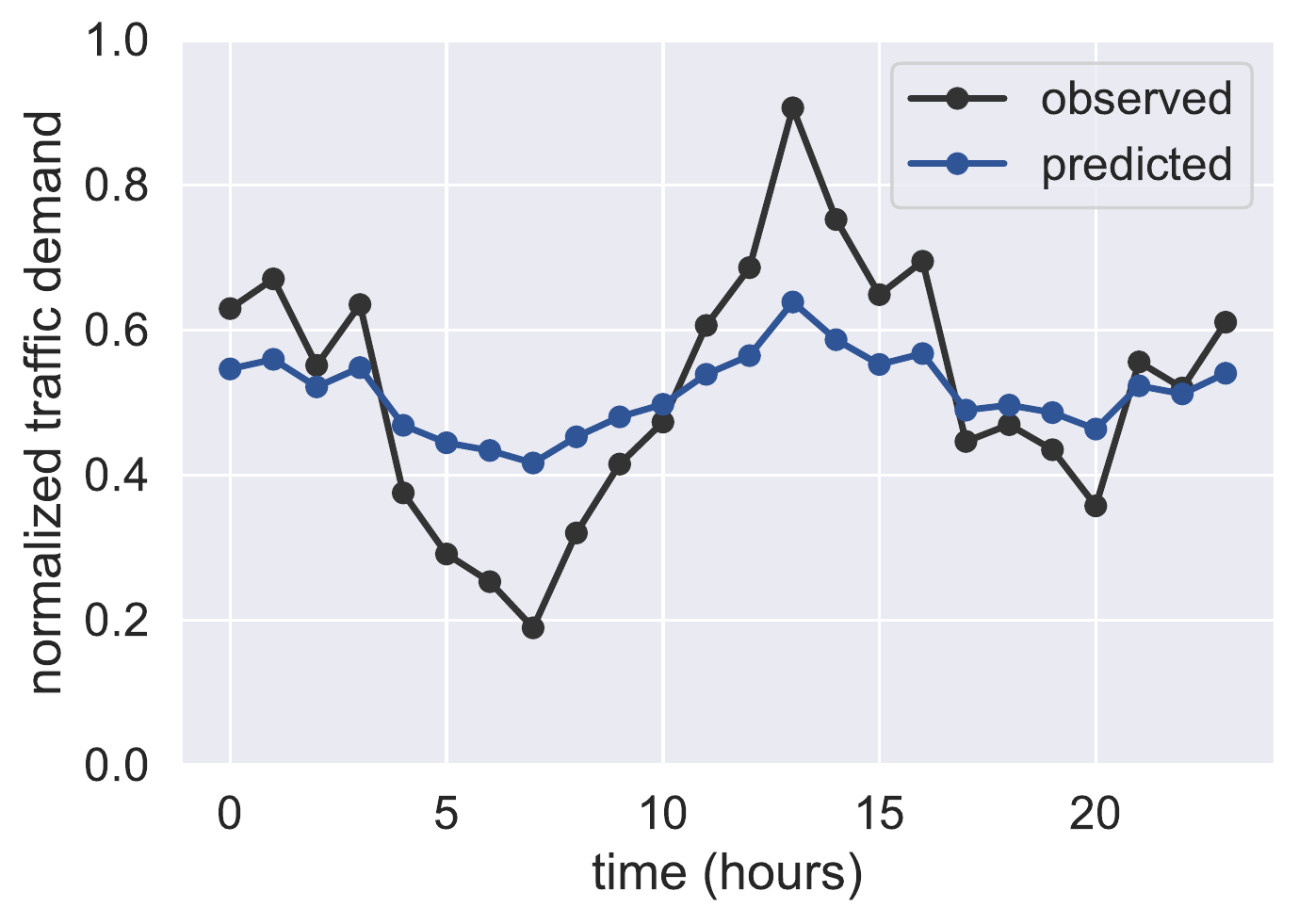}%
		\label{fig:test_1-days}}
	\hfil
	\subfloat[Validation for 50 periods]{\includegraphics[width=0.237\textwidth]{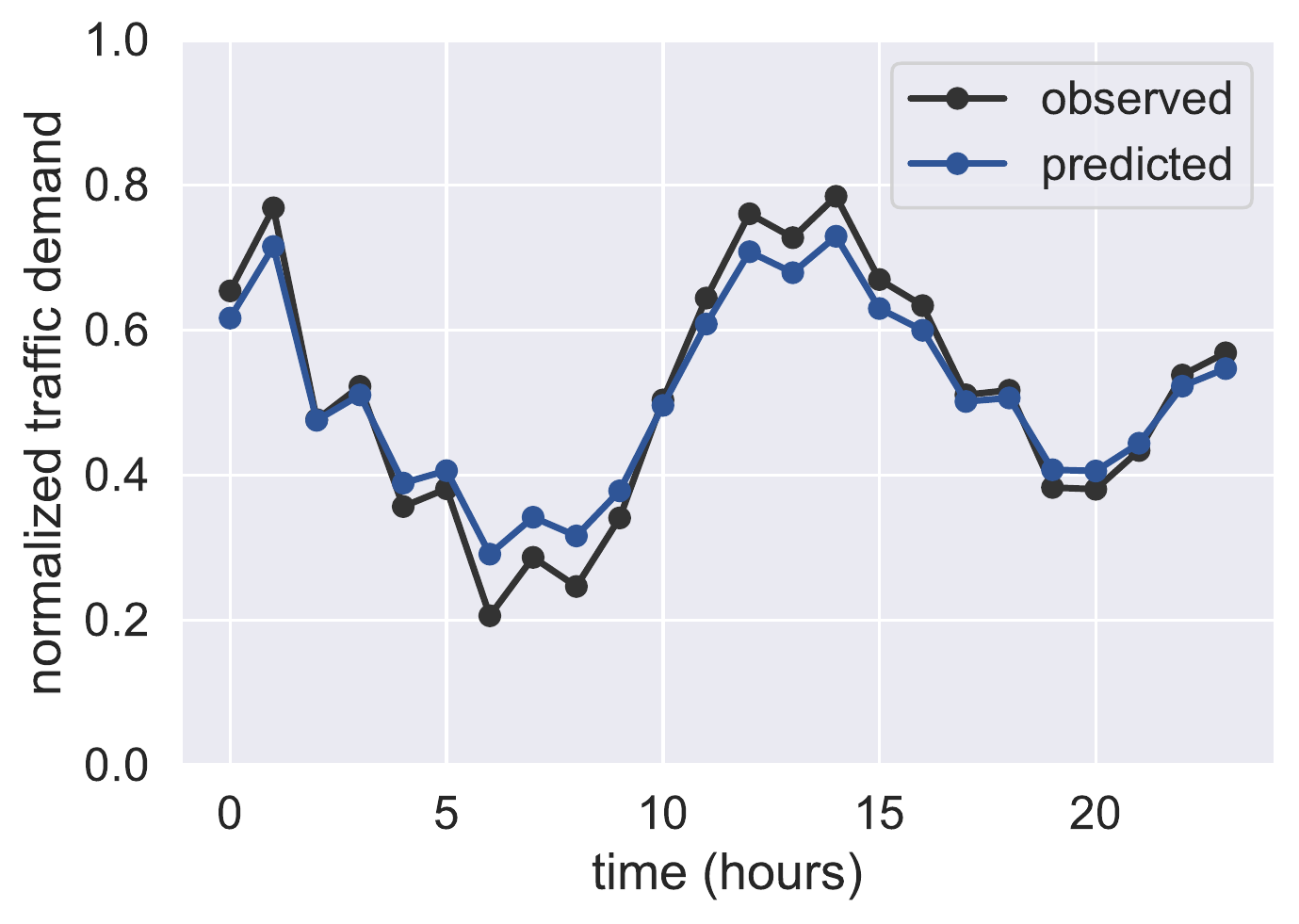}
		\label{fig:test_50-days}}
	\caption{Traffic prediction model results}
	\label{fig:traffic_prediction}
\end{figure}

\section{Performance evaluation}

We use MILP model implemented with Gurobi Optimizer tool to evaluate a smaller
size network N7 (7 nodes, 20 directed links with 500 units of capacity each, see
Fig. \ref{fig:N7}) and heuristics for a larger-size network N45 (45 nodes, 140,
directed links with 1000 units of capacity, Fig. \ref{fig:N45}). In N7, every
node is equipped with one server, whereas in N45 there are 8 servers per node.
In both networks, we assume that all nodes can establish on-demand connectivity
to a third-party cloud server of which the geographic location is determined
based on the closest common locations used by cloud providers. Thus, for the
7-nodes network, in N7 the geographic locations are based regionally, such as
the area of Braunschweig (Germany) for the network and the area of Frankfurt,
for the cloud server, respectively. For N45, we use a modified version of
Palmetto network in South Carolina, USA and the cloud server in North Virginia,
USA. The propagation delay is correspondingly calculated considering the
distance between nodes from their latitude and longitude using the Haversine
method using 2/3 of the speed of light. We thereby assume the links used to
connect to the third-party cloud have sufficient capacity for any demand, and
therefore do not impact the analysis of server utilization.

For each source-destination pair of nodes, 3 paths are pre-computed that do not
traverse the cloud node and 1 additional path that does. Also 2 additional paths
per node are computed for the synchronization traffic between possible VNFs
allocated in the cloud and in the network. The path computation is carried out
in this way to make sure the model has enough freedom to allocate all SFCs in
the network and at least there is one admissible path per SFC to allocate VNFs
in the cloud. We assume that every source-destination pair of nodes (except the
cloud node) instantiates independent SFCs with variable length from 1 to 10 VNFs
depending depending on the scenario. The processing load of a certain VNF is
calculated from the total amount of processed traffic in the VNF multiplied by a
random load ratio ($\loadratio$) between 1\% and 100\%. Additionally, an
overhead ($\overhead$) is calculated as a random percentage between 1\% and 10\%
of the processing load \cite{Reddy2014}. The synchronization traffic between
VNFs ($\syncload$) is calculated as 10\% of the processing load of the VNF. The
delay parameters per VNF, already explained is section \ref{service_delay}, are
specified using typical values as follows: $\processtrafficdelay = 3~ms$,
$\processdelay =5~ms$, $\minprocessdelay = 2~ms$ and $\maxdelay = 10~ms$. In
the networks studied, for all SFCs the service delay is constrained to $\Dsmax =
	400~ms$. The round trip time is, for both networks, always shorter than 5 ms
which leads to a service downtime of duration $\migrationdelay = 27.5~ms$ when
performing a migration, in the worst case scenario \cite{Taleb2019}.

Two types of results are produced: (i) one setting all SFCs with a certain
length while all servers have the same capacity and (ii) one setting all servers
to the certain capacity while all SFCs have a random length. The reason for that
is to independently see the effects that SFCs lengths and server capacities have
on the network. In case i), the server capacities are set to 1000 for N7 and
2000 units for N45, and all SFC lengths are chosen in increments from 1 to 10.
In case ii), the server capacities vary from 250 to 3000 units and every SFC is
of random length between 1 and 10.

\begin{figure}[!t]
	\centering
	\includegraphics[width=1.0\columnwidth]{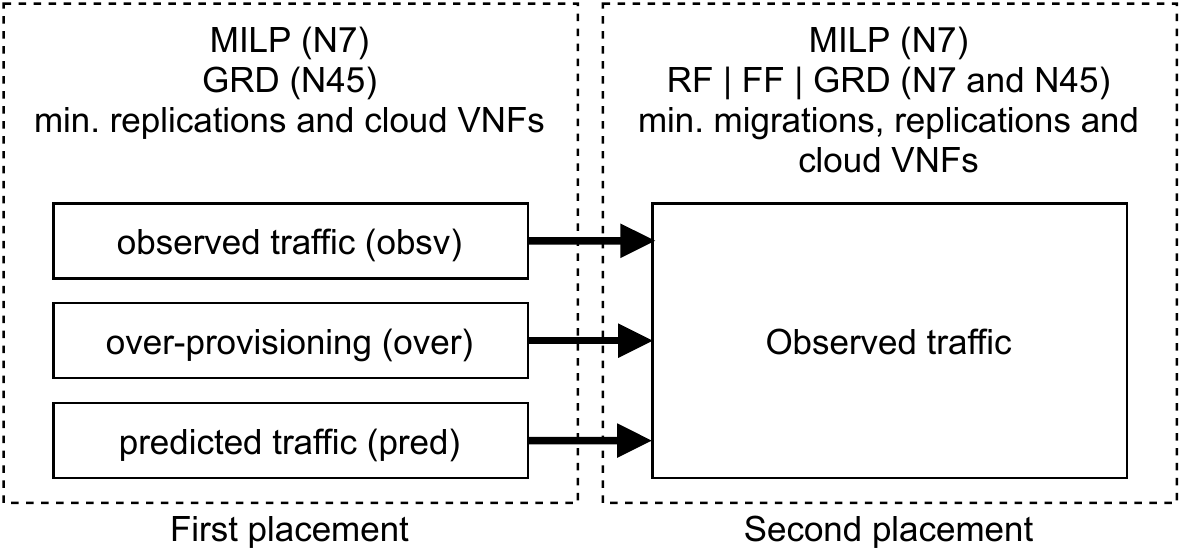}%
	\caption{Optimization scenarios}
	\label{fig:diagram}
\end{figure}

\subsection{Optimization scenarios}

We assume that every source-destination pair of nodes generates between 1 and 3
traffic flows, with traffic demand per flow set to a random value between 1 and
100 traffic units. For each traffic demand, 24 values are generated in one time
period following a lognormal distribution with time-varying mean and variance,
as explained in Section \ref{traffic_model}. For the time series forecasting,
one LSTM network is created and trained per each traffic flow for a certain
number of periods, and then evaluated for 1 time period.

To determine the optimum number of required training periods, the model has been
tested using from 1 to 1000 periods for training. The resulting RMSE is shown in
\ref{fig:rmse} where it shows that above 50 periods of time, the performance is
not improving anymore. However, the training time continues to increase with the
number of training periods as expected, see Fig. \ref{fig:time}. Taking 1 period
as the worst case and 50 as the best case, Fig. \ref{fig:test_1-days} and Fig.
\ref{fig:test_50-days} show the predicted and observed normalized traffic demand
values over time during the evaluation period, respectively. Here, we can see
how the number of training periods impacts the accuracy of the model.

To illustrate the issues of computation time, we show the results obtained by
using the CPU of a machine with an Intel Core i7-6700 and 32 GB of RAM. The
total computation time considering all traffic demands for takes $\approx$7
minutes in N7, when training for 1 period and $\approx$12 minutes when training
for 50 periods. For N45, it takes in total $\approx$13 hours for 50 training
periods. While the specific total computational time can be improved by using
GPUs or by training models in parallel, it should be noted that the network size
needs to be considered when using predictions.

\begin{figure}[!t]
	\centering
	\subfloat[Variable SFC lenght]{\includegraphics[width=0.50\columnwidth]{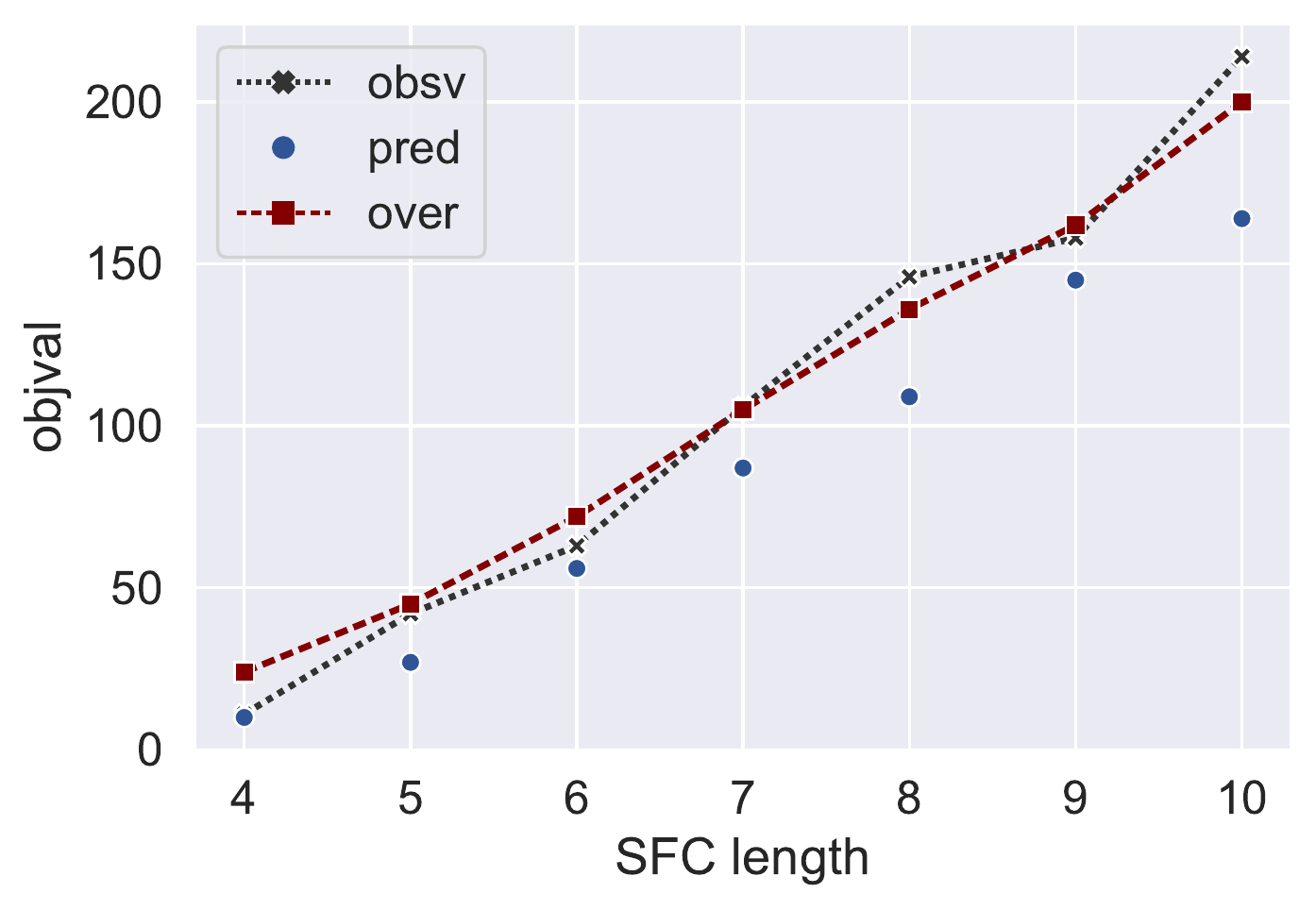}%
		\label{fig:7nodes_sfclen_MGR_REP_CLOUD_objval_LP}}
	\subfloat[Variable server capacity]{\includegraphics[width=0.50\columnwidth]{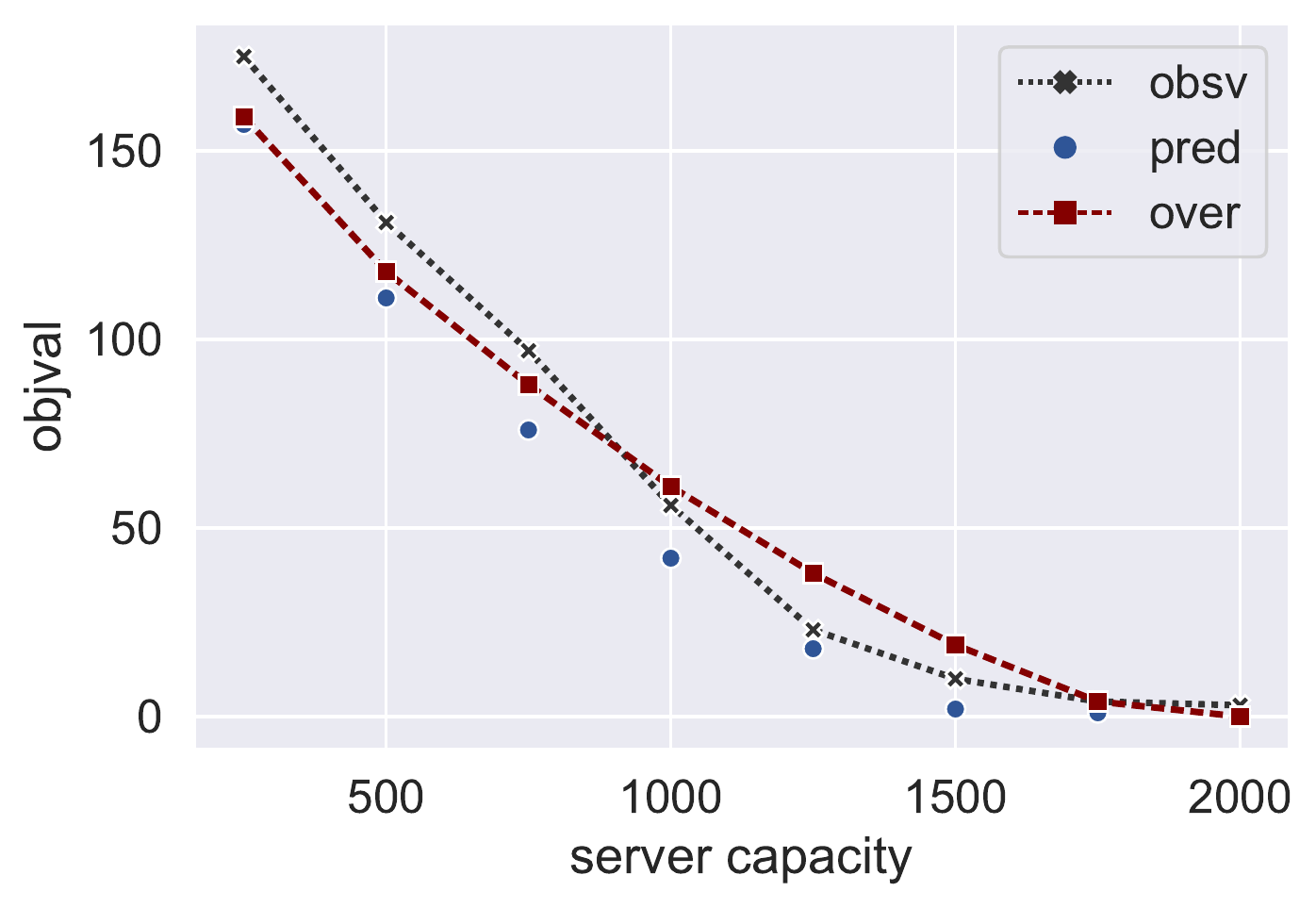}%
		\label{fig:7nodes_servercap_MGR_REP_CLOUD_objval_LP}} \caption{Objective
		function value for \texttt{obsv}, \texttt{over} and \texttt{pred} scenarios
		in the N7 network using the \texttt{MILP} model.}
	\label{fig:7nodes_MGR_REP_CLOUD_objval_LP}
\end{figure}
\begin{figure}[!t]
	\centering
	\subfloat[Variable SFC length]{\includegraphics[width=0.50\columnwidth]{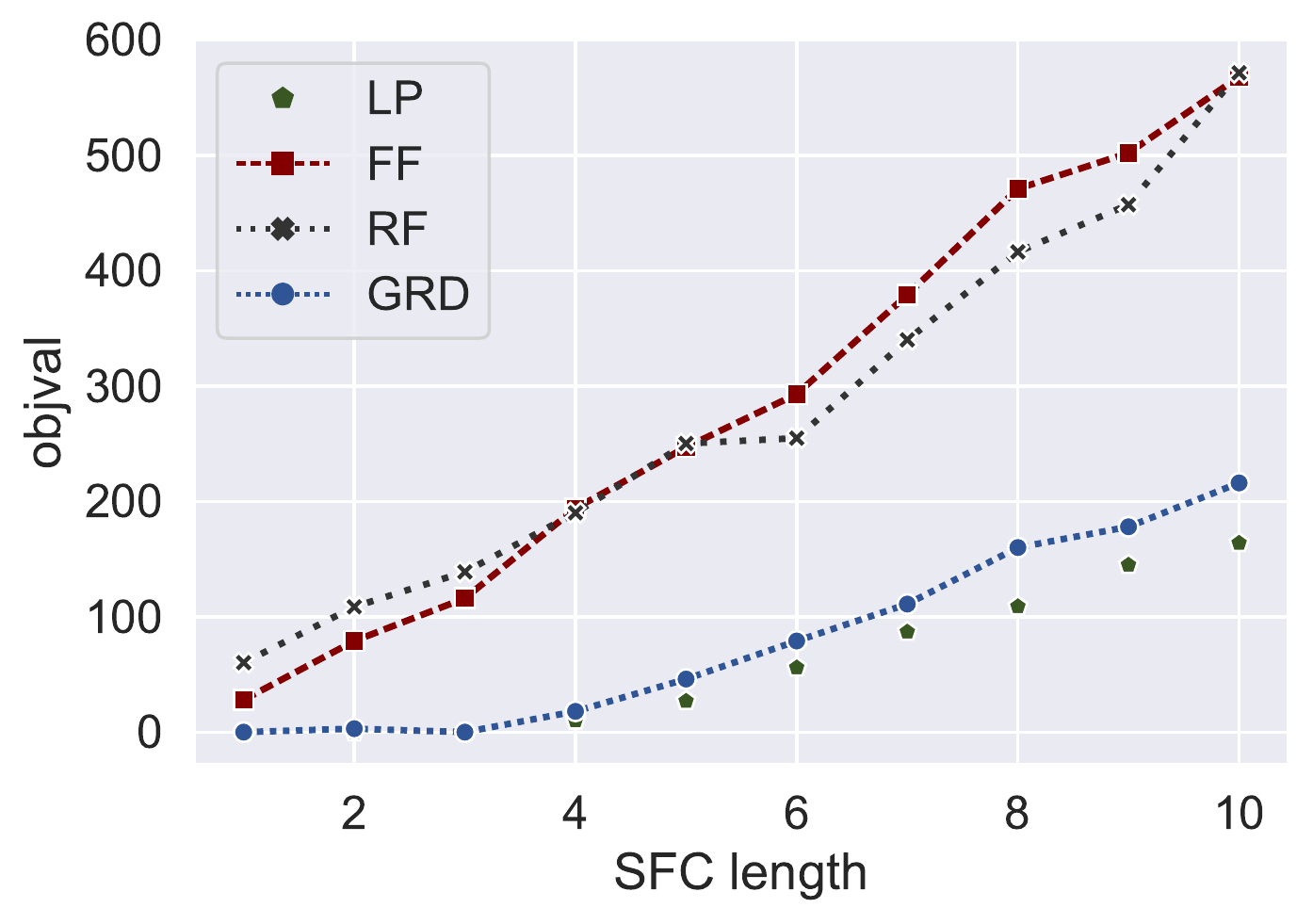}%
		\label{fig:7nodes_sfclen_MGR_REP_CLOUD_pred_objval}}
	\hfil
	\subfloat[Variable server capacity]{\includegraphics[width=0.50\columnwidth]{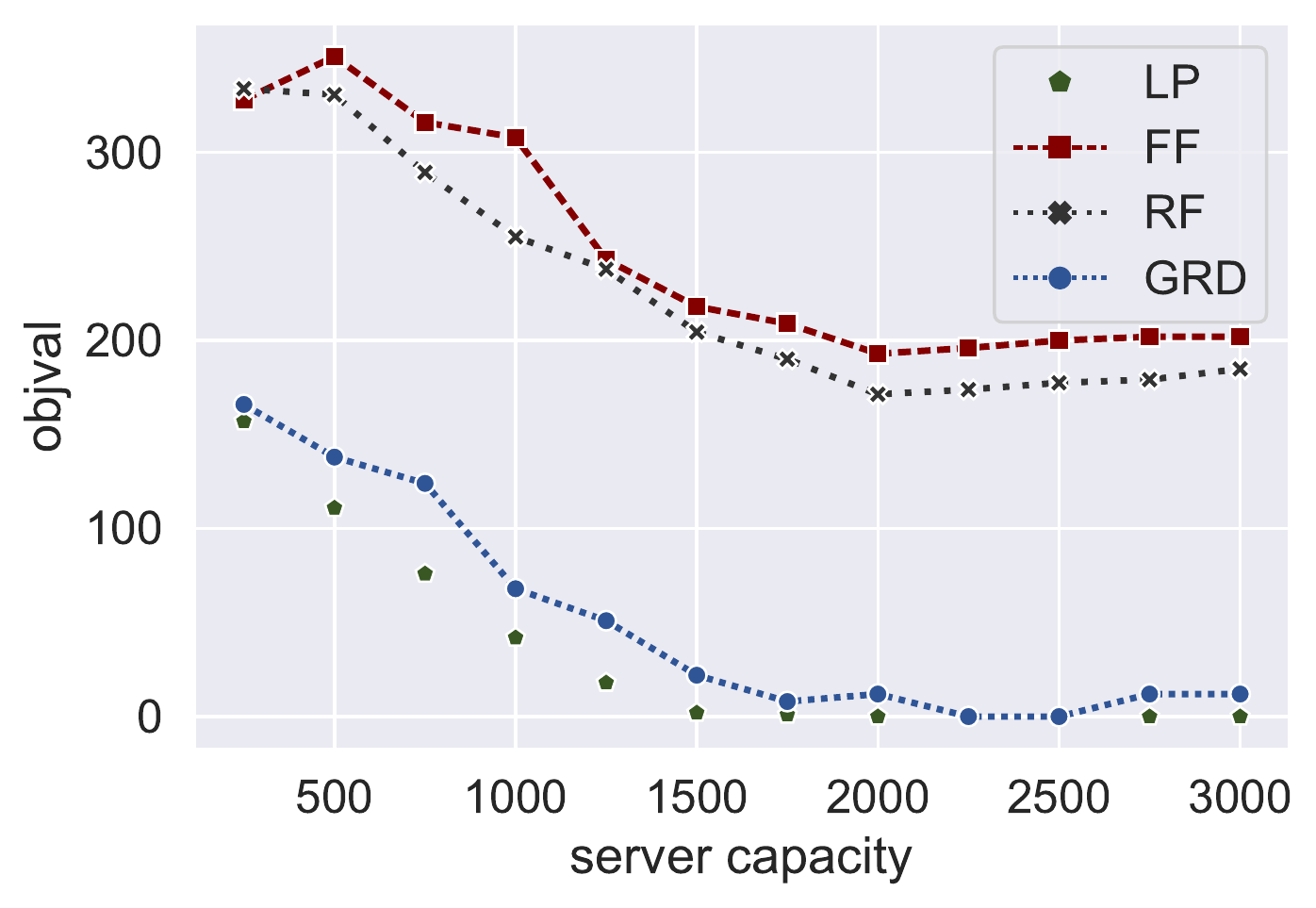}%
		\label{fig:7nodes_servercap_MGR_REP_CLOUD_pred_objval}}
	\caption{Objective function values of \texttt{RF}, \texttt{FF}, \texttt{GRD}
		and \texttt{MILP} for the \texttt{pred} scenario in the N7 network.}
	\label{fig:7nodes_MGR_REP_CLOUD_pred_objval}
\end{figure}

From the generated traffic demand values produced for the evaluation period,
three optimization scenarios are derived based on which values are considered
during the first placement: i) observed values (\texttt{obsv}), ii) 80\% of the
maximum individual traffic demand values, which corresponds to overprovisioning
(\texttt{over}) and iii) predicted values (\texttt{pred}). After the first
placement, the second placement is carried out considering the location of the
VNFs during the first placement, as explained in equation
\ref{initialsolutionmapping} and considering the new traffic demand values after
a $\Delta t$ time shift from the set of traffic demand values (see Fig.
\ref{fig:test_50-days}). In our case, the first time step for the first
placement is taken randomly from the first 18 time values and $\Delta t$ is set
to 6 time periods. Hence, for the first scenario \texttt{obsv}, only the current
observed values at time $t$ are considered for the placement of VNFs. In the
second scenario \texttt{over}, the observed values are ignored, and instead, the
VNFs are placed assuming the traffic is always at the 80\% of the maximum
traffic demand value. The third scenario places VNFs considering the predicted
traffic values after $\Delta t$.

Fig. \ref{fig:diagram}) illustrates the optimization process. The second
placement uses the first placement as input, and it   optimizes the placement
again by considering the real monitored and observed traffic demand values. The
first placement is carried out using either the \texttt{MILP} model in N7, or
the greedy algorithm (\texttt{GRD}) in N45. In all cases, the objective is to
allocate VNF while minimizing the number of replications and the number of
virtual functions placed in the cloud. In the first placement, the are no
migrations from any previous step to consider. In the second placement, the
\texttt{MILP} model in N7, and all heuristics for both networks, all by
considering the same objectives which to minimizing the number of migrations,
replications and cloud VNFs. Finally, for the reminder of the paper we show the
results obtained from the second placement, while using the three scenarios
during the first placement, as described.

\begin{figure}[!t]
	\centering
	\subfloat[Variable SFC lenght]{\includegraphics[width=0.50\columnwidth]{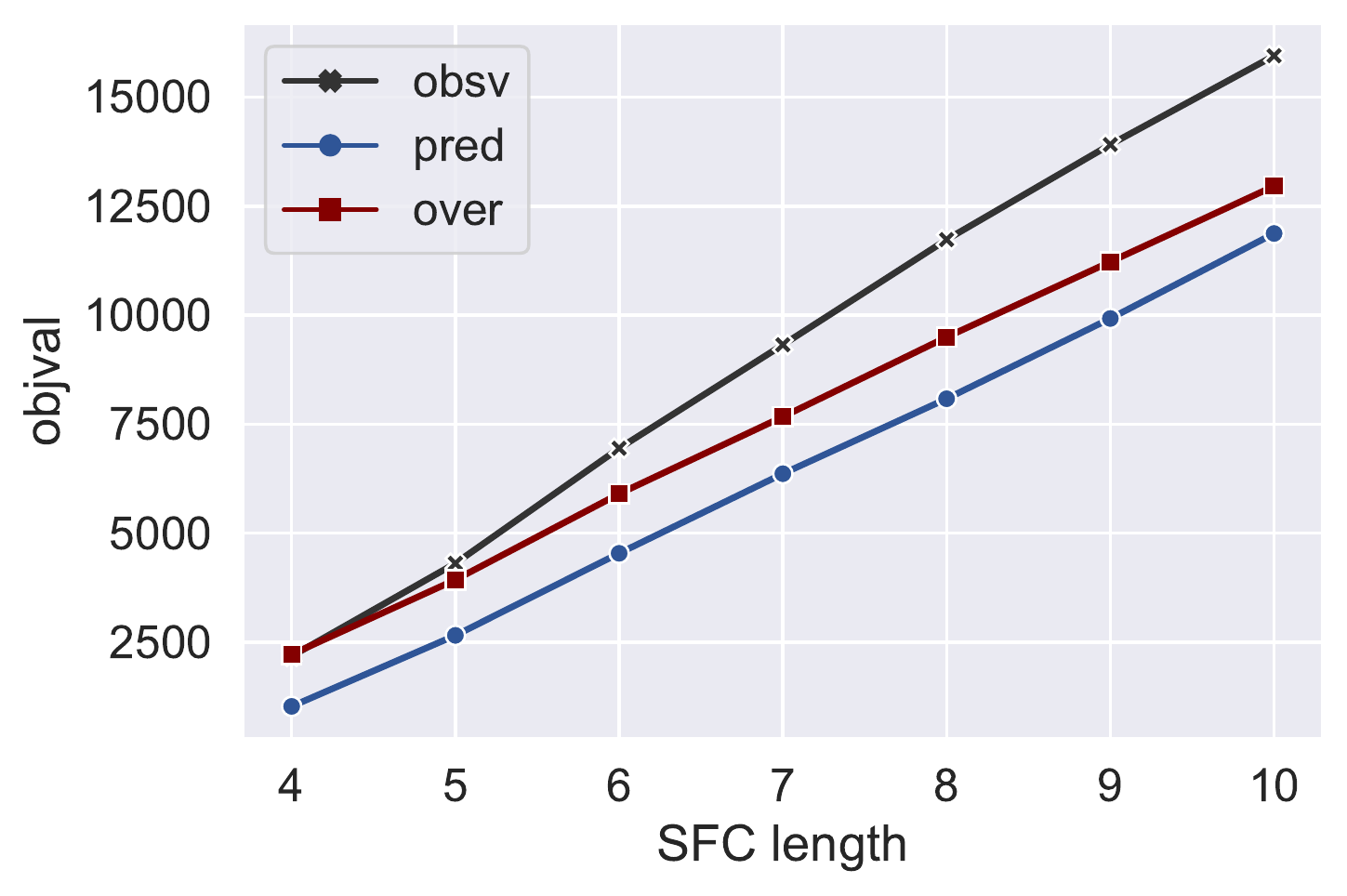}%
		\label{fig:palmetto_sfclen_MGR_REP_CLOUD_objval_GRD}}
	\subfloat[Variable server capacity]{\includegraphics[width=0.50\columnwidth]{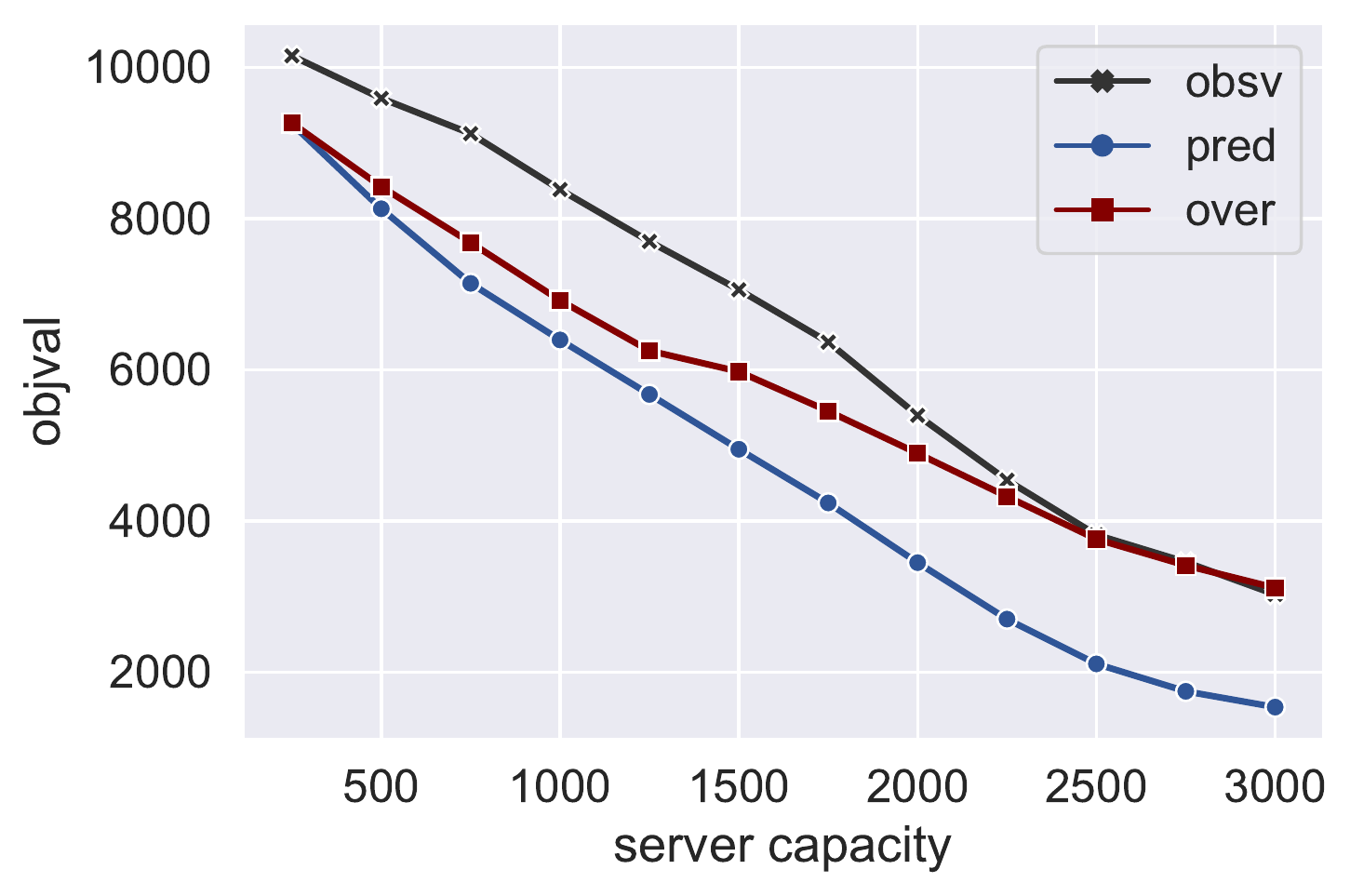}%
		\label{fig:palmetto_servercap_MGR_REP_CLOUD_objval_GRD}}
	\caption{Objective function value for \texttt{obsv}, \texttt{over} and \texttt{pred} scenarios in the N45 network using the \texttt{GRD} algorithm.}
	\label{fig:palmetto_MGR_REP_CLOUD_objval_GRD}
\end{figure}

\begin{figure}[!t]
	\centering
	\subfloat[Variable SFC length]{\includegraphics[width=0.50\columnwidth]{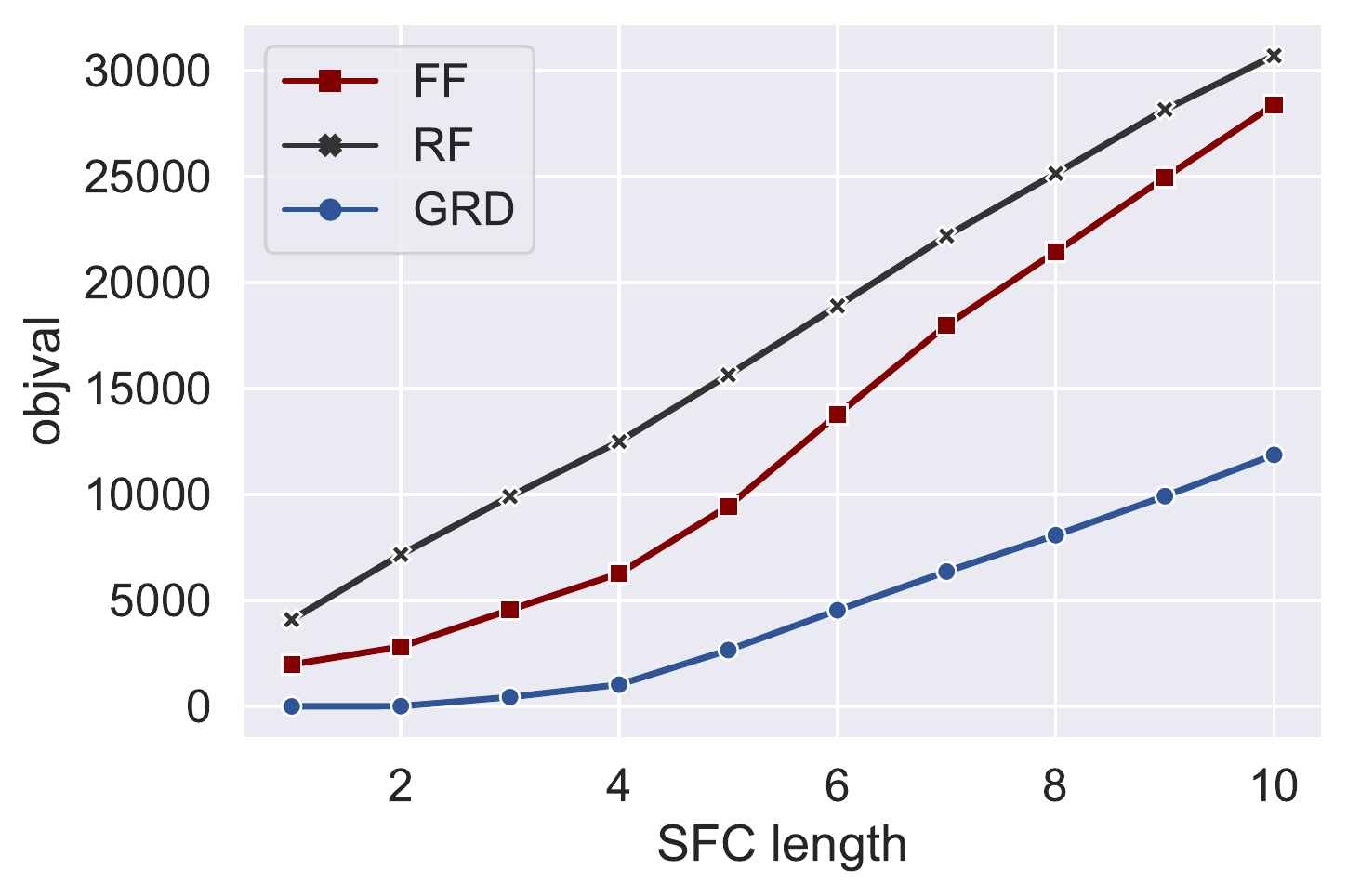}%
		\label{fig:palmetto_sfclen_MGR_REP_CLOUD_pred_objval}}
	\hfil
	\subfloat[Variable server capacity]{\includegraphics[width=0.50\columnwidth]{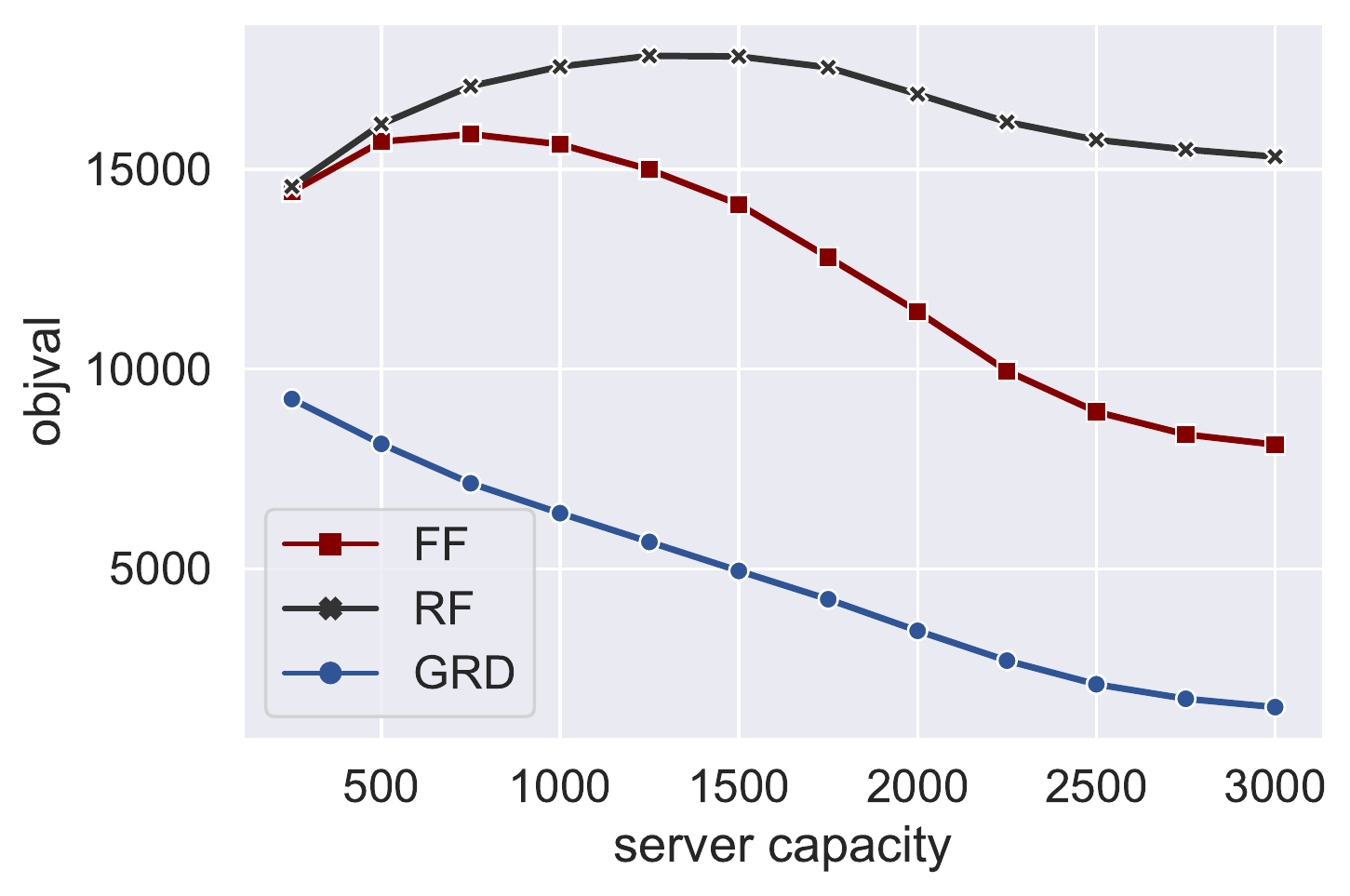}%
		\label{fig:palmetto_servercap_MGR_REP_CLOUD_pred_objval}}
	\caption{Objective function values for \texttt{RF}, \texttt{FF} and
		\texttt{GRD} for the \texttt{pred} scenario in the N45 network.}
	\label{fig:palmetto_MGR_REP_CLOUD_pred_objval}
\end{figure}

\subsection{Objective function}

Since the objective function (equation \eqref{obj_func}) is a joint optimization
from three different weighted terms, we first show the results when minimizing
all terms, so all three weights $W_m$, $W_r$ and $W_c$ are equal to 1. Fig.
\ref{fig:7nodes_MGR_REP_CLOUD_objval_LP} shows the objective value for the three
scenarios \texttt{obsv}, \texttt{over} and \texttt{pred} when varying the SFC
lengths and when varying the server capacities in N7. It should be noted that
some zero values for certain SFC lengths or server capacities are omitted in the
plots due to clarity. We can observe that \texttt{pred} overperforms the other
two cases. Between \texttt{over} and \texttt{obsv}, when the servers are
overloaded the \texttt{over} case performs slightly better than \texttt{obsv} as
expected, due to the overprovisioning factor.

Before analyzing the three scenarios in large network N45, let us first compare
how heuristics compare to \texttt{MILP} model in N7. Fig.
\ref{fig:7nodes_MGR_REP_CLOUD_pred_objval} shows again the objective values for
\texttt{pred} scenario, but now comparing the \texttt{MILP} model with the
heuristic algorithms \texttt{RF}, \texttt{FF} and \texttt{GRD}. Here we can see
that both \texttt{RF} and \texttt{FF} are far from the optimal solution, being
\texttt{RF} slightly better than \texttt{FF} in most cases.

When using the greedy algorithm for the N45 network, we compare again the three
scenarios \texttt{obsv}, \texttt{over} and \texttt{pred} in Fig.
\ref{fig:palmetto_MGR_REP_CLOUD_objval_GRD}. Here, we can see a more clear
difference between the three cases, being again the \texttt{pred} scenario the
one with a clear advantage compared to the other two. This case also better
illustrates how \texttt{over} case overperforms \texttt{obsv} case mostly when
the servers are overloaded confirming what we could slightly see with the N7
network. From Fig. \ref{fig:palmetto_MGR_REP_CLOUD_pred_objval} we can compare
\texttt{RF}, \texttt{FF} and \texttt{GRD}, in this case for the N45 network.
Different from N7, here we can see how \texttt{FF} outperforms \texttt{RF} in
all cases. We see here a trend on \texttt{FF} working better the more free the
network and servers are, but in any case the achieved values are comparable to
the \texttt{GRD} algorithm which performs always better.

\begin{figure}[!t]
	\centering
	\subfloat[Minimizing migrations]{\includegraphics[width=0.50\columnwidth]{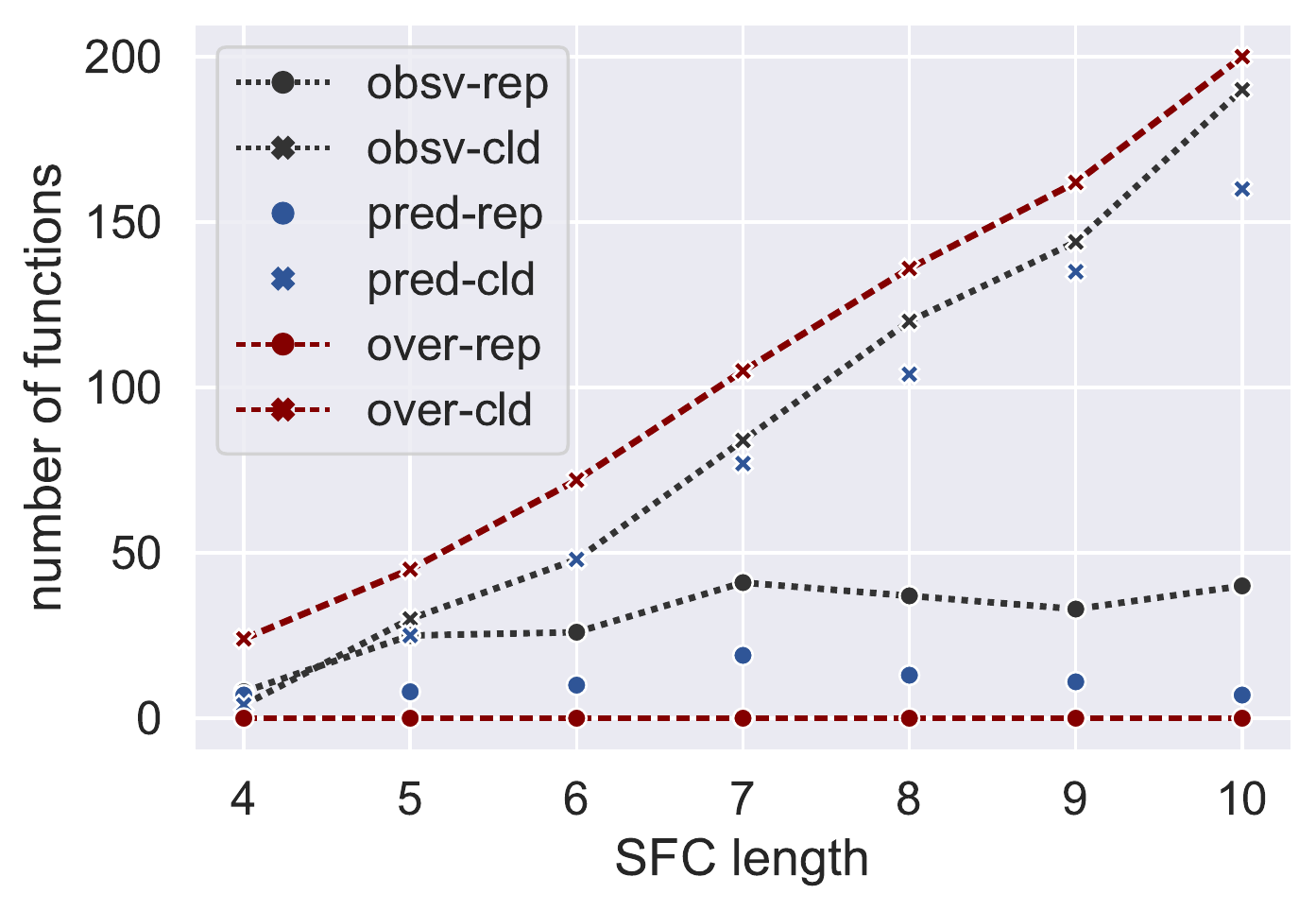}%
		\label{fig:7nodes_sfclen_LP_MGR_repcld}}
	\hfil
	\subfloat[Minimizing replications]{\includegraphics[width=0.50\columnwidth]{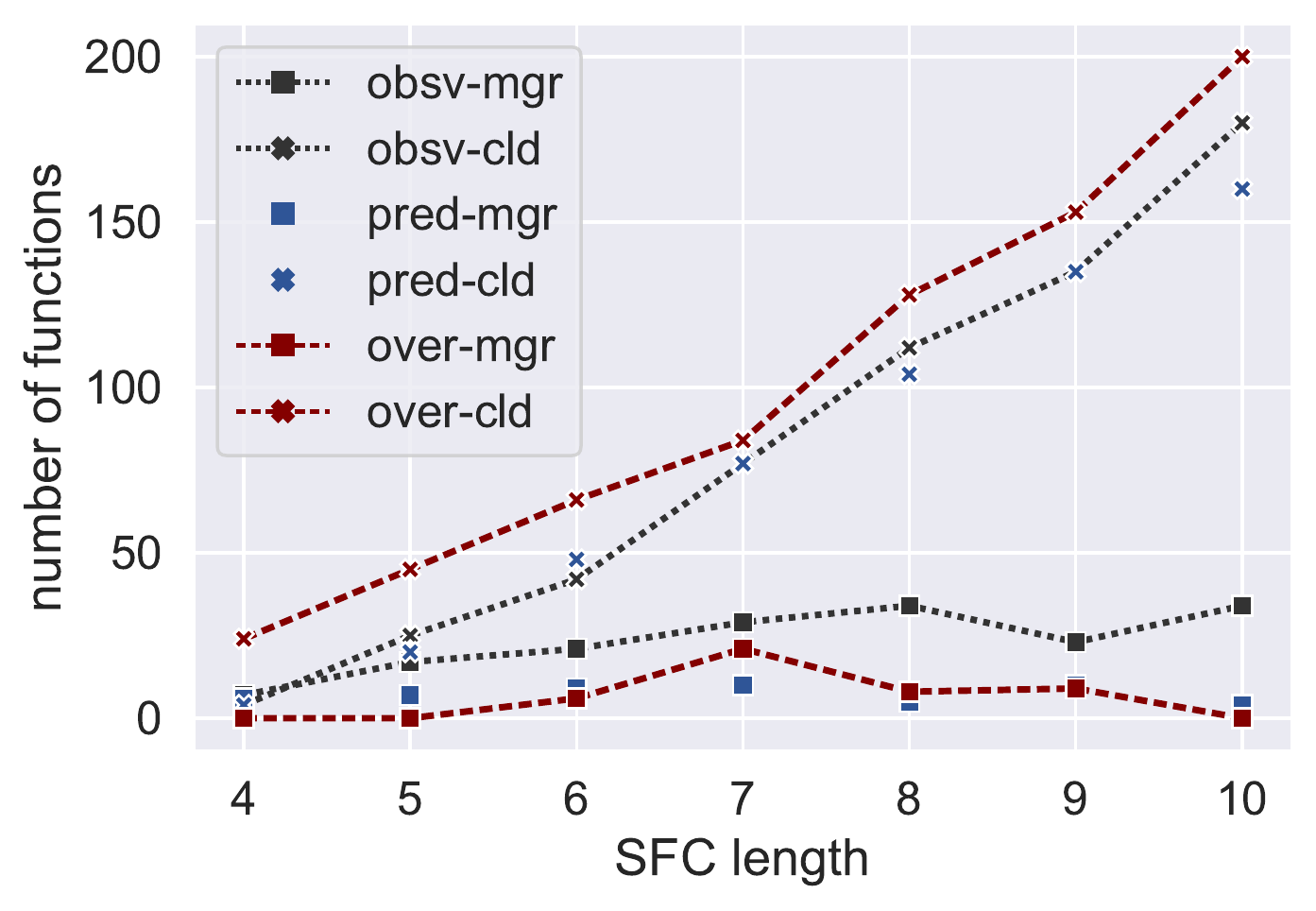}%
		\label{fig:7nodes_sfclen_LP_REP_mgrcld}}
	\hfil
	\subfloat[Minimizing cloud VNFs]{\includegraphics[width=0.50\columnwidth]{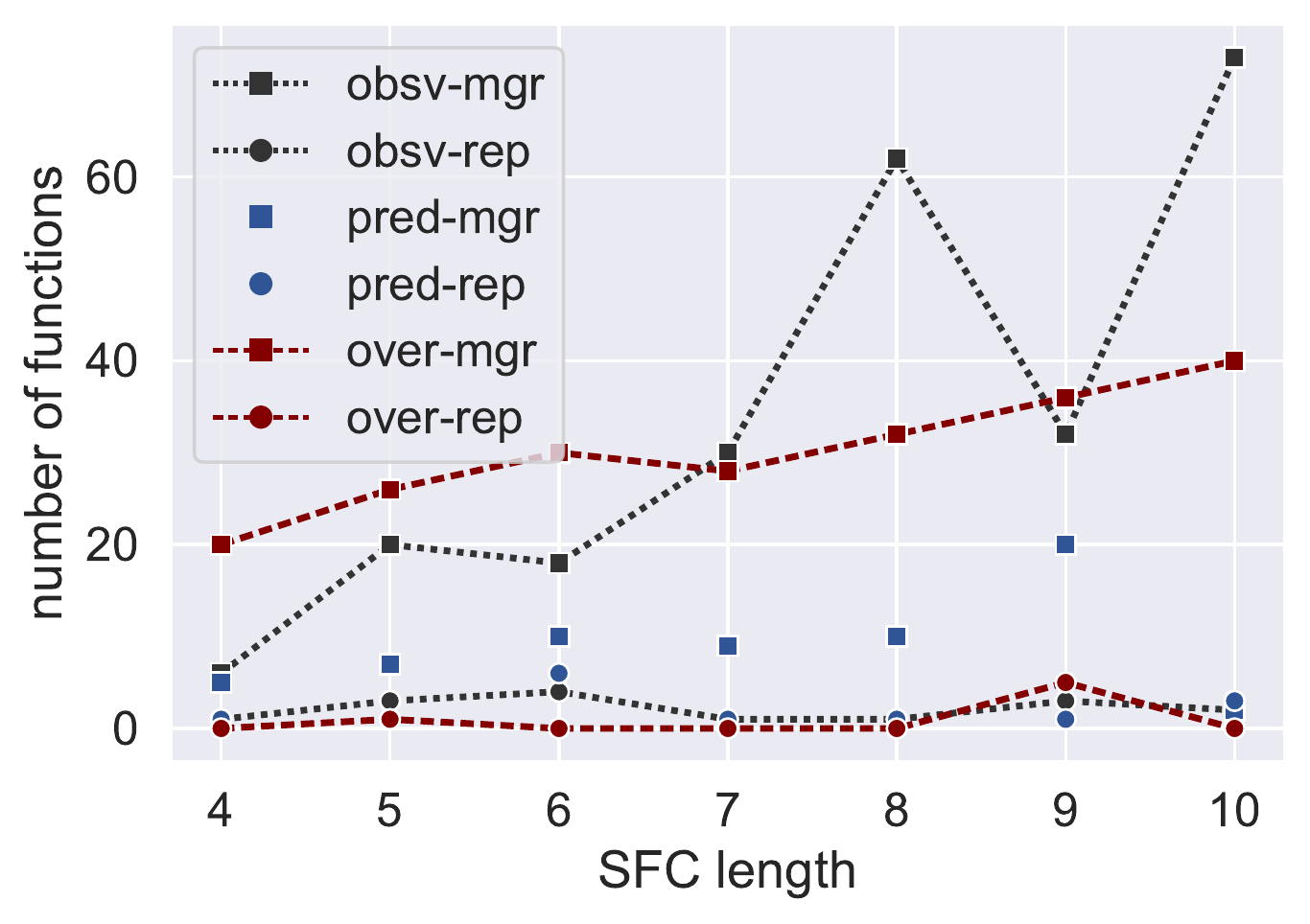}%
		\label{fig:7nodes_sfclen_LP_CLOUD_mgrrep}} \caption{Number of
		migrations, replications and cloud VNFs for different SFC lengths in the N7
		network using \texttt{MILP} model.}
	\label{fig:7nodes_sfclen_objval_LP}
\end{figure}

\begin{figure}[!t]
	\centering
	\subfloat[Number of migrations]{\includegraphics[width=0.50\columnwidth]{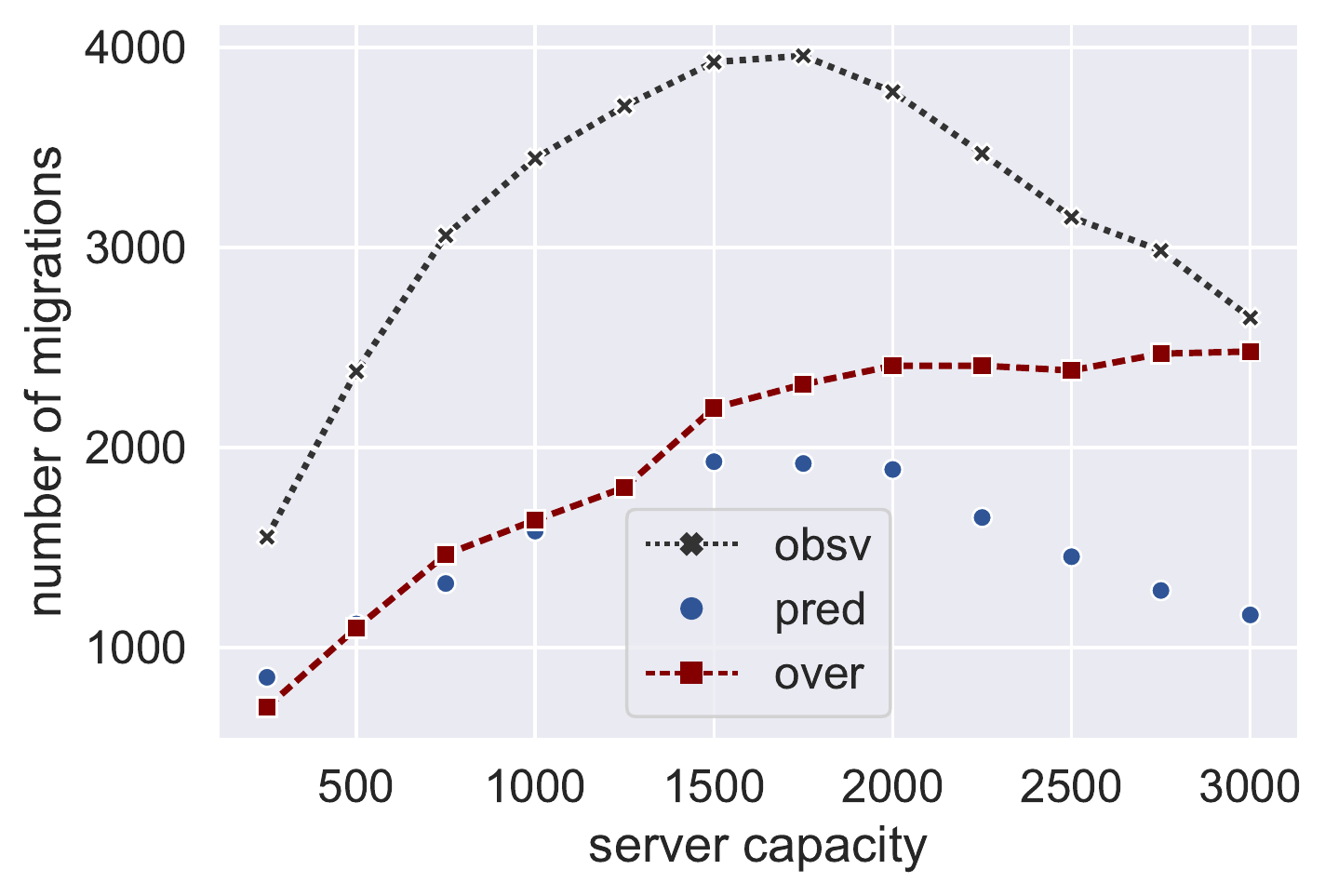}%
		\label{fig:palmetto_servercap_GRD_MGR_REP_CLOUD_mgr}}
	\hfil
	\subfloat[Number of replications]{\includegraphics[width=0.50\columnwidth]{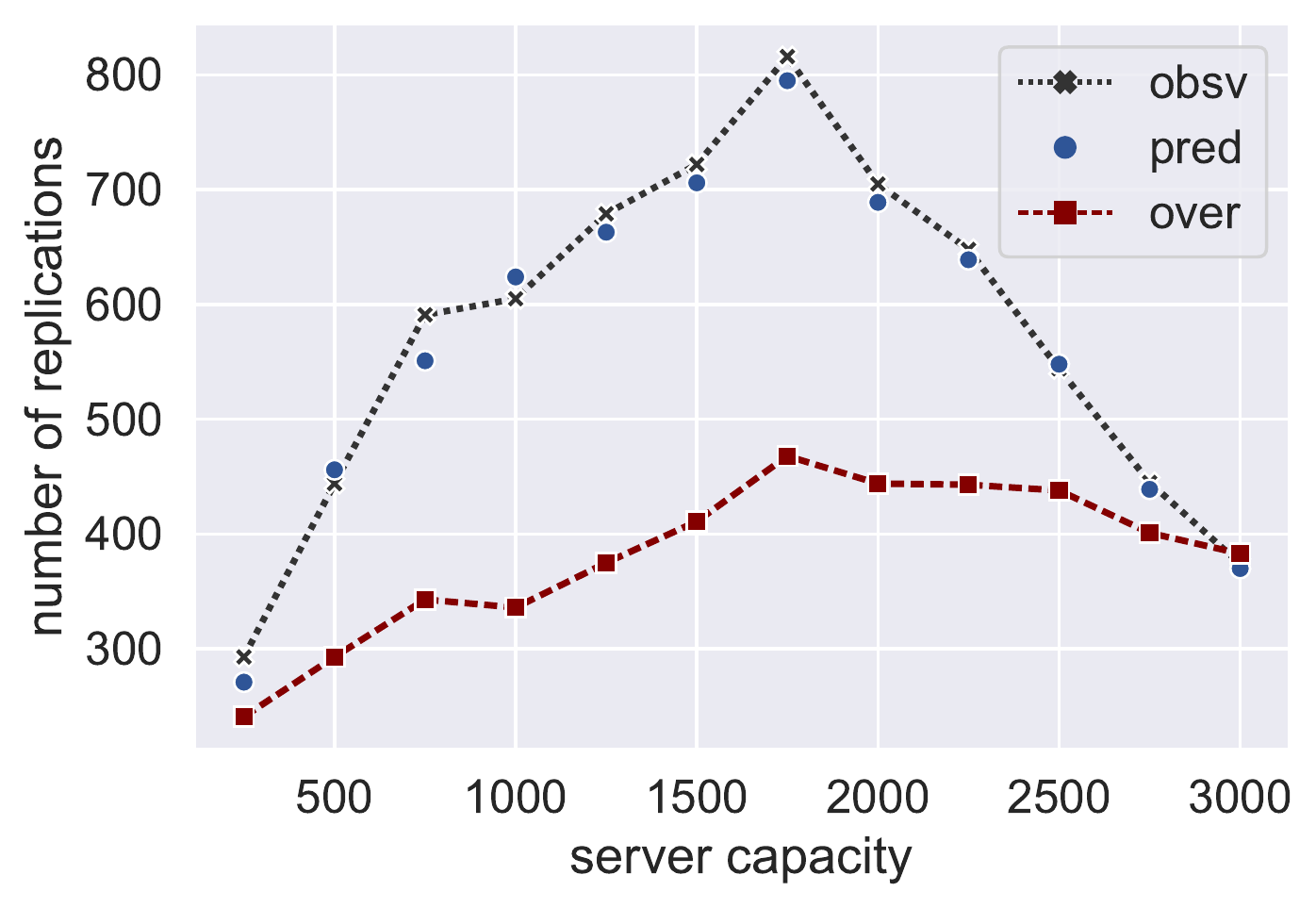}%
		\label{fig:palmetto_servercap_GRD_MGR_REP_CLOUD_rep}}
	\hfil
	\subfloat[Number of cloud VNFs]{\includegraphics[width=0.50\columnwidth]{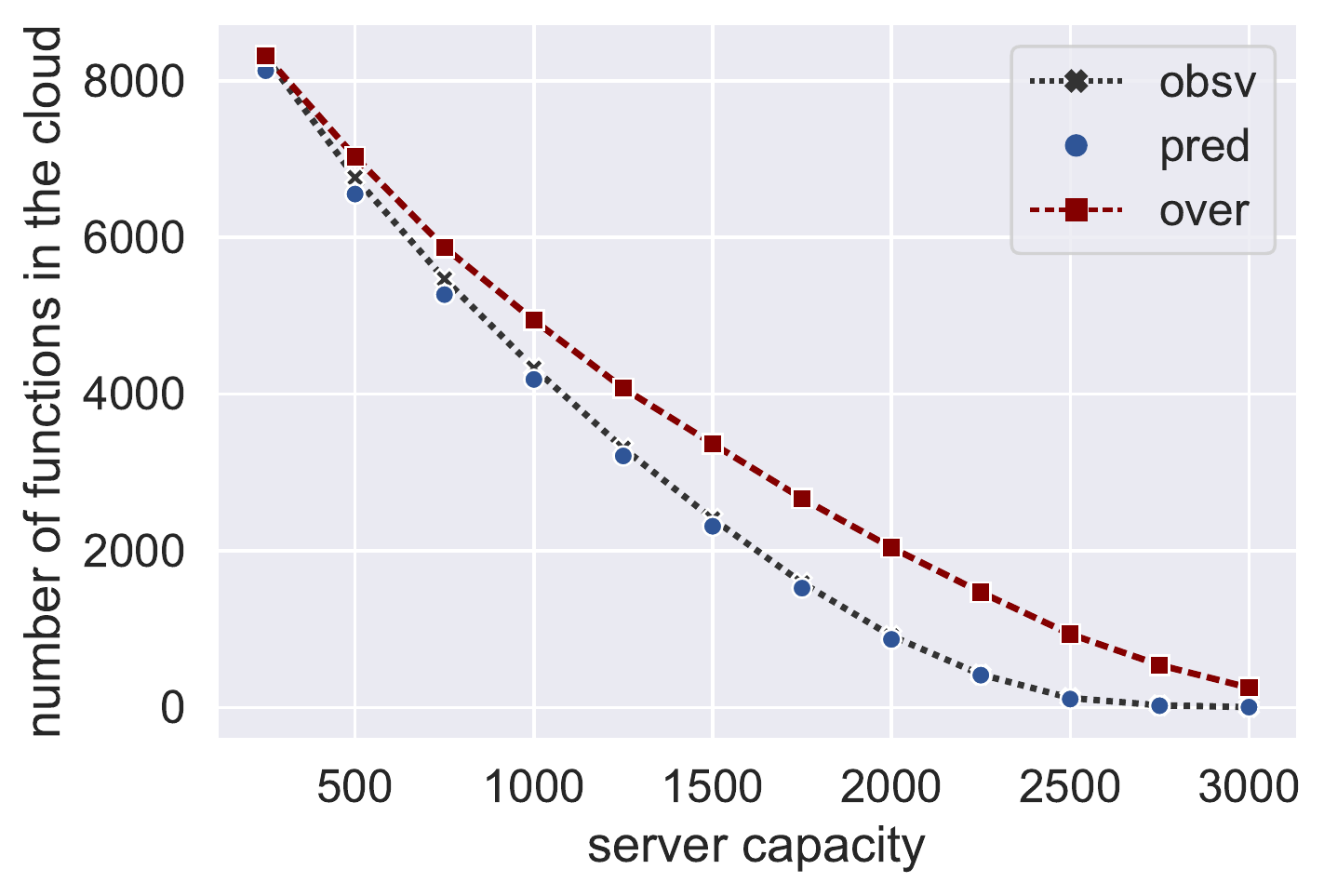}%
		\label{fig:palmetto_servercap_GRD_MGR_REP_CLOUD_cld}} \caption{Number of
		migrations, replications and cloud VNFs for different server capacity in
		N45 network using \texttt{GRD} algorithm.}
	\label{fig:palmetto_servercap_GRD_MGR_REP_CLOUD}
\end{figure}

\subsection{Migrations, Replications and Cloud VNFs}

In order to better see how the model behaves individually when minimizing only
one of the terms, we set a certain weight (i.e. $W_m$, $W_r$ or $W_c$) equal to
1, and the others close to 0 in such a way that the sum of all secondary terms
is within interval $[0,1)$. By doing that, we limit the freedom of the model
while, at the same time, we ensure there is no impact on the main term which
value is always going to be a positive integer. In this regard, Fig.
\ref{fig:7nodes_sfclen_LP_MGR_repcld} shows the results in terms of number of
replications (\texttt{rep}) and number of cloud VNFs (\texttt{cld}), when
minimizing the number of migrations for the three scenarios \texttt{obsv},
\texttt{over} and \texttt{pred} and different SFC lengths in N7. By looking at
\texttt{over-rep} and \texttt{over-cld}, we see how overprovisioning does not
allocate replicas and places more functions in the cloud compared to other
cases. In comparison, the \texttt{obsv} case allocates less functions in the
cloud at expenses of deploying a considerable number of replicas. The
\texttt{pred} case can be seen as a trade-off solution, as it allocates
considerably less VNFs in the cloud compared to \texttt{over}, independently
from the SFC length, and less than \texttt{obsv} mostly when the servers are
overloaded with long SFCs. In terms of replicas, the \texttt{pred} requires much
less resources in almost all cases compared to \texttt{obsv} case. When
minimizing the number of replications, see Fig.
\ref{fig:7nodes_sfclen_LP_REP_mgrcld}, the difference between \texttt{pred} and
\texttt{obsv} in terms of allocations in the cloud is much smaller, but still
reduces the number of migrations independently from the SFC length. Here the
\texttt{over} case behaves quite similar to \texttt{pred} in number of
migrations, but instead requires to allocate more cloud VNFs. When minimizing
the number of functions in the cloud, see Fig.
\ref{fig:7nodes_sfclen_LP_CLOUD_mgrrep}, we see how \texttt{pred} requires much
less migrations compared to the other two cases, but no remarkable difference
regarding replications.

To individually see the number of migrations, replications and cloud VNFs with
no influence from the weights (i.e. all terms the same weight), we now study N45
network.  Fig. \ref{fig:palmetto_servercap_GRD_MGR_REP_CLOUD_mgr} shows how
\texttt{obsv} case requires much more migrations compared to the other cases
except when the servers are either too overloaded or too underloaded where the
values become closer to \texttt{over} case. On the other hand, \texttt{pred}
case requires the same number of migrations than \texttt{over} when the servers
are overloaded and improves when there is enough free available resources. In
Fig. \ref{fig:palmetto_servercap_GRD_MGR_REP_CLOUD_rep}, regarding the number of
replications we see that there is no much difference between \texttt{pred} and
\texttt{obsv}, but \texttt{over} case is the one requiring significantly less
replications, except for the cases where the servers are either too overloaded
or too underloaded. This effect can be explained by the fact that when there are
no available resources in the servers, the model cannot perform replications,
and when there are more than enough available resources, the model avoids
replications that are not essential. When looking at Fig.
\ref{fig:palmetto_servercap_GRD_MGR_REP_CLOUD_cld}, we see that there is almost
no difference between \texttt{obsv} and \texttt{pred}, but the \texttt{over}
case allocates considerably more cloud VNFs than the other two cases.

\begin{figure}[!t]
	\centering
	\subfloat[Average link utilization]{\includegraphics[width=0.50\columnwidth]{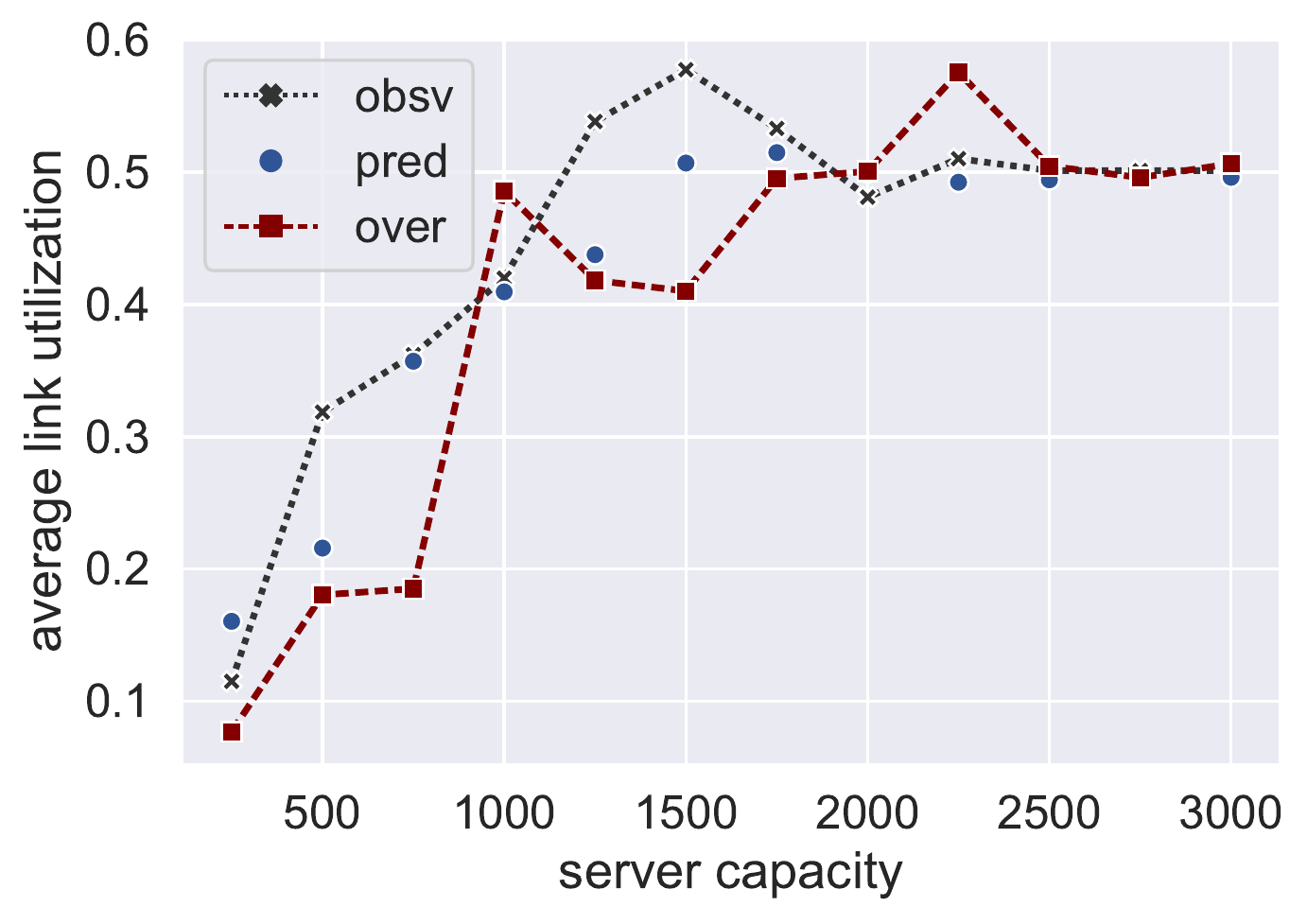}%
		\label{fig:7nodes_servercap_LP_MGR_REP_CLOUD_lu}}
	\hfil
	\subfloat[Average server utilization]{\includegraphics[width=0.50\columnwidth]{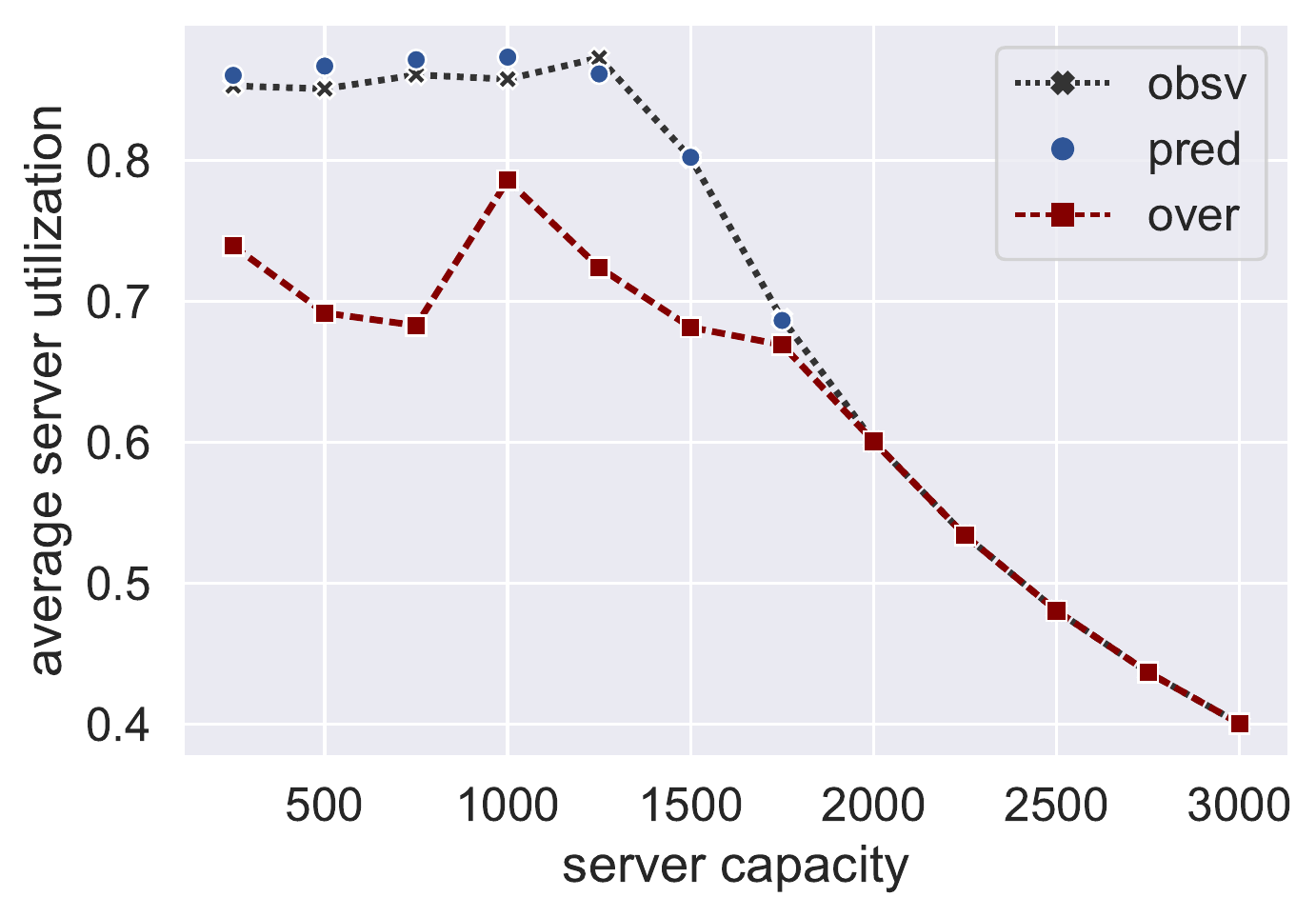}%
		\label{fig:7nodes_servercap_LP_MGR_REP_CLOUD_xu}}
	\hfil
	\subfloat[Average service delay]{\includegraphics[width=0.50\columnwidth]{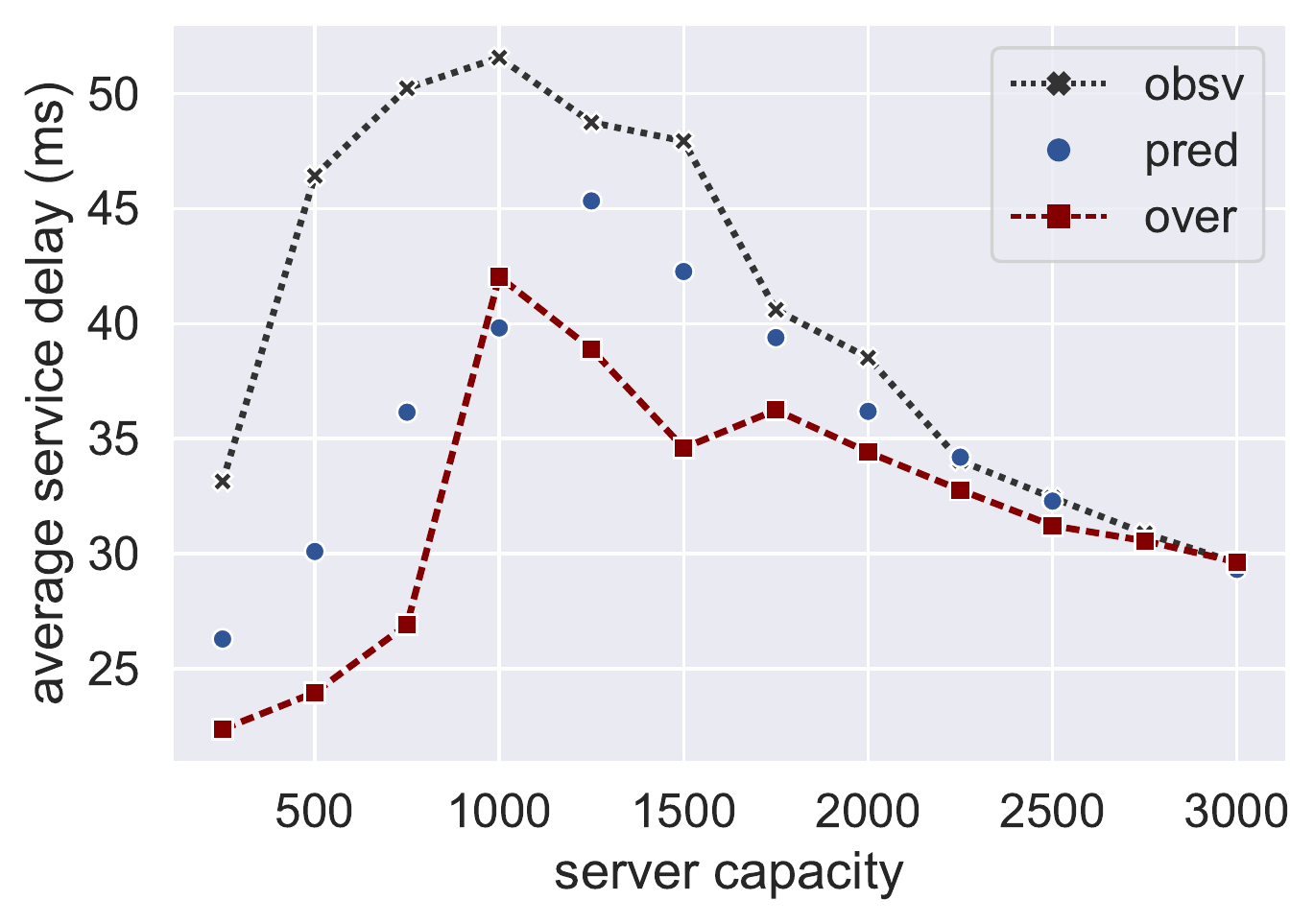}%
		\label{fig:7nodes_servercap_LP_MGR_REP_CLOUD_sd}}
	\caption{Resource utilization and service delays for different server capacities in the N7 network}
	\label{fig:7nodes_servercap_resources}
\end{figure}

\subsection{Resource Utilization and Service Delay}

To show the difference between the three scenarios, Fig.
\ref{fig:7nodes_servercap_LP_MGR_REP_CLOUD_lu}, Fig.
\ref{fig:7nodes_servercap_LP_MGR_REP_CLOUD_xu} and Fig.
\ref{fig:7nodes_servercap_LP_MGR_REP_CLOUD_sd} show the average link
utilization, server utilization and service delay, respectively, versus a
varying server capacity for N7. For both link and server utilization, the link
capacity connecting to the cloud and the cloud servers are not considered. Here,
in most cases when the network is not overloaded, the \texttt{over} case has
slightly lower link utilization compared to the other cases, since this case
allocates more cloud VNFs, so the edge network is less utilized and less
replicas are used, so less synchronization traffic is added to the network.
Between \texttt{pred} and \texttt{obsv} cases, the first one has slightly lower
link utilization in some specific cases. This difference is inexistent when
looking at the server utilization, and here only \texttt{over} case has lower
utilization for the same reason as before. When comparing the three cases for
the average service delay, we notice how \texttt{over} has the lowest delay,
even though it allocates generally more cloud VNFs as we have seen before, so
the propagation delay is larger. However, this case performs less migrations
compared to the other cases, and therefore, there is less penalty due to service
interruptions. When comparing \texttt{pred} with \texttt{obsv}, we see how
\texttt{pred} has less service delay, so less migrations are required.

\begin{figure}[!t]
	\centering
	\subfloat[Average link utilization]{\includegraphics[width=0.50\columnwidth]{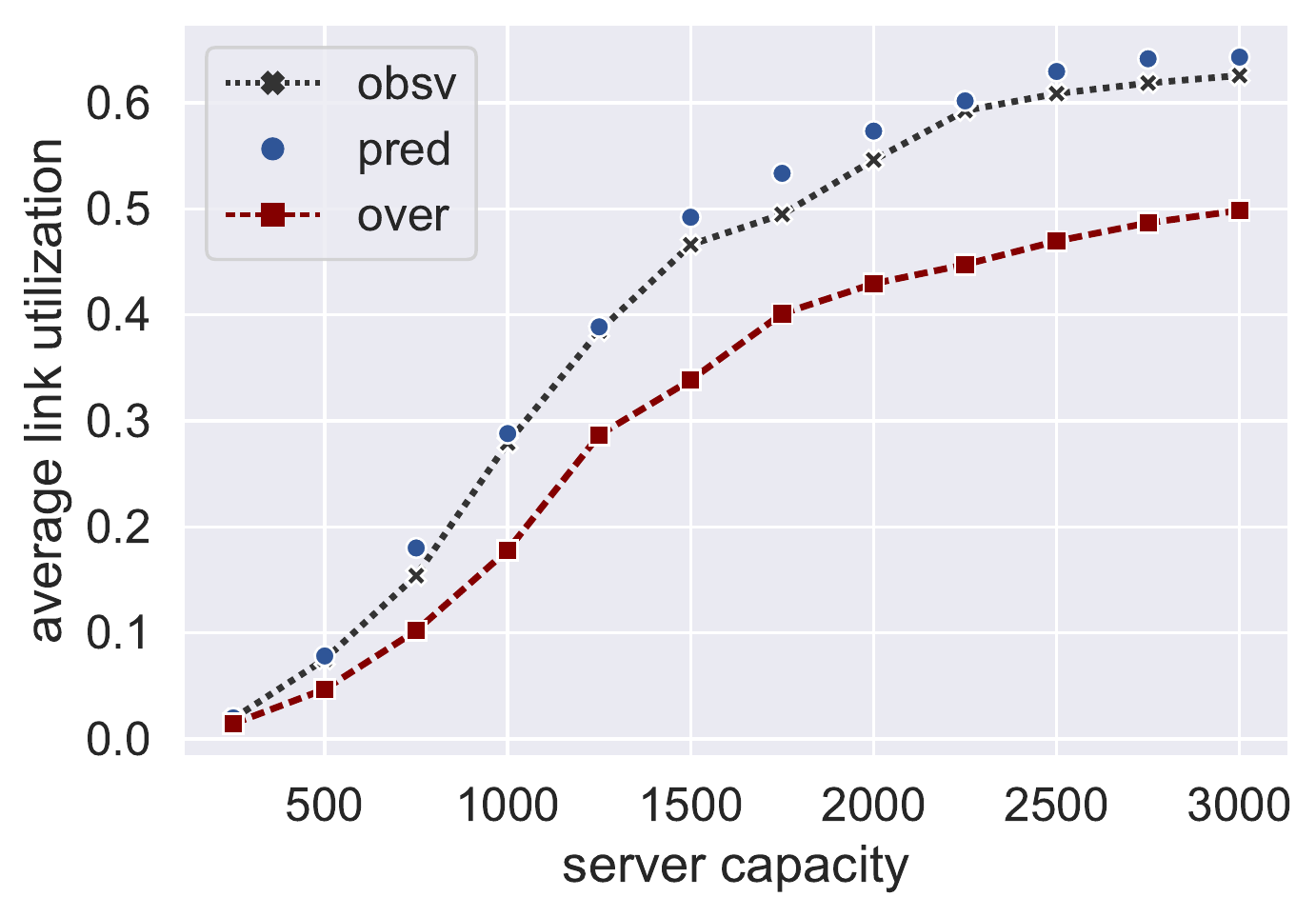}%
		\label{fig:palmetto_servercap_GRD_MGR_REP_CLOUD_lu}}
	\hfil
	\subfloat[Average server utilization]{\includegraphics[width=0.50\columnwidth]{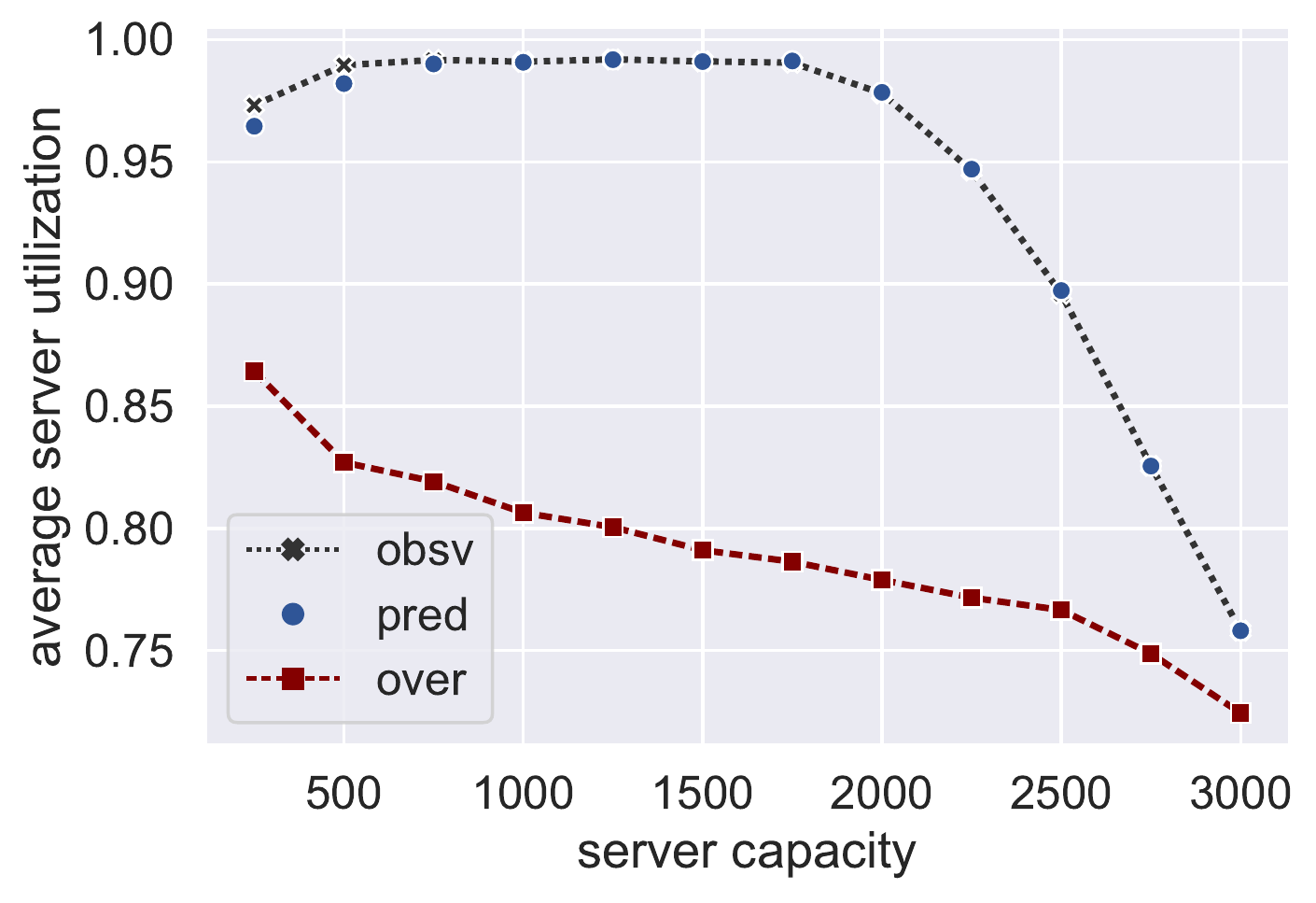}%
		\label{fig:palmetto_servercap_GRD_MGR_REP_CLOUD_xu}}
	\hfil
	\subfloat[Average service delay]{\includegraphics[width=0.50\columnwidth]{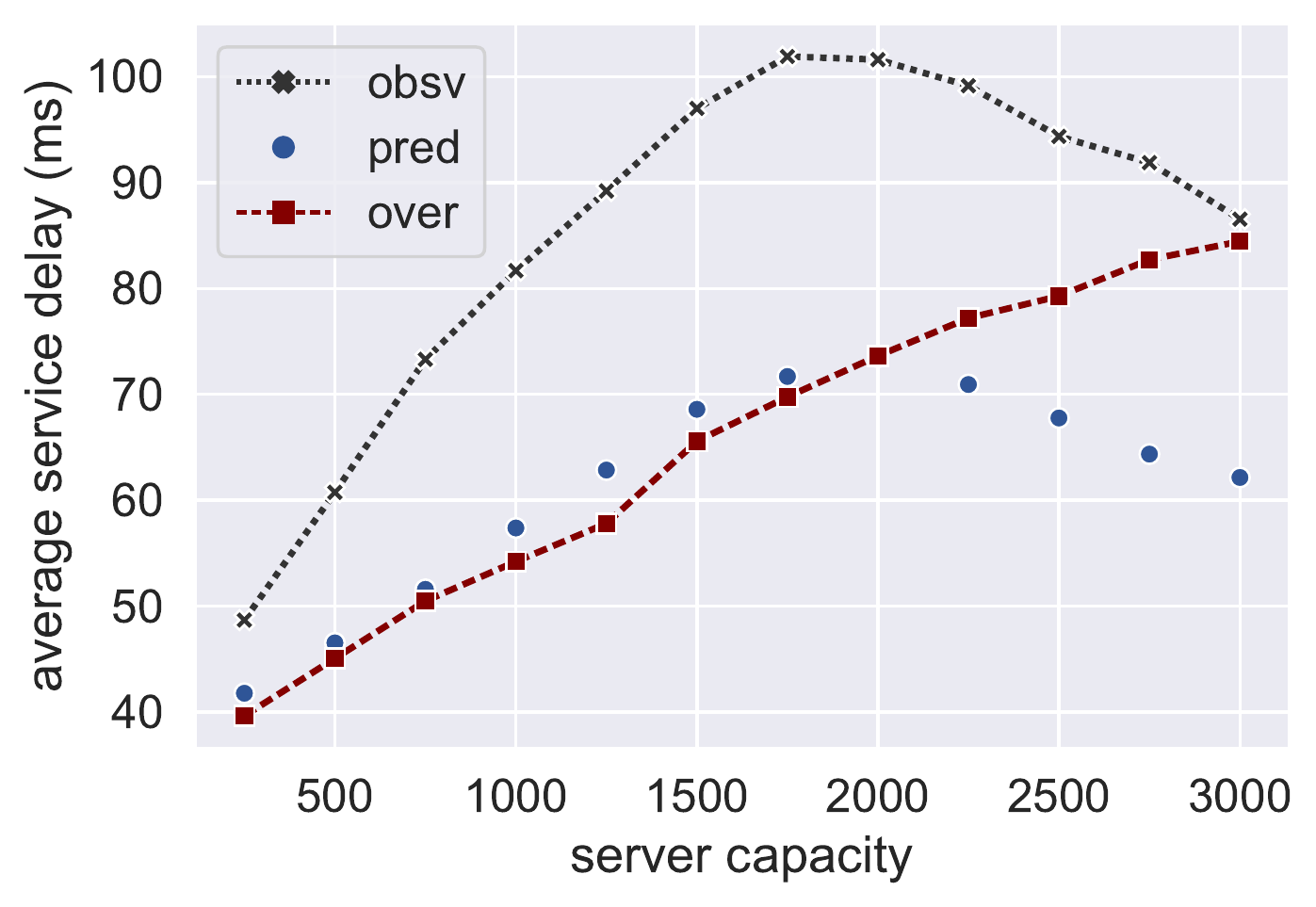}%
		\label{fig:palmetto_servercap_GRD_MGR_REP_CLOUD_sd}}
	\caption{Resource utilization and service delays for different server capacities in the N45 network}
	\label{fig:palmetto_servercap_resources}
\end{figure}

Fig. \ref{fig:palmetto_servercap_resources} shows again the same results, but
this time for the N45 network. Here, we can better see the difference in the
lower utilization of links of the \texttt{over} case compared with the other
two. This is again due to the fact that overprovisioning results into a higher
usage of the cloud, so the network is less utilized. This is also confirmed when
looking at the average server utilization where \texttt{pred} and \texttt{obsv}
cases make full usage of all server resources at the edge before using the
cloud, contrary to the \texttt{over} case. The most interesting case is with
regard the service delay, where we can see how the \texttt{pred} case is able to
outperform \texttt{over} when the servers are not overloaded since the number of
migrations are much lower as we could see from Fig.
\ref{fig:palmetto_servercap_GRD_MGR_REP_CLOUD_mgr}.

\subsection{Discussion and remarks}

From all three scenarios analyzed, we see observe that in all cases, predicting
the traffic demands helps to reduce the overall number of migrations,
replications and usage of the cloud. More specifically, the overprovisioning
case requires in general less replications compared to the other two cases, but
requires as many migrations as with the prediction case, when the network is
overloaded and considerably more when the network is underloaded. Because
overprovisioning does not consider the fluctuations of traffic, it can, in the
best case, match the real traffic and, in the worst case, to provision excessive
resources in advance, which results in using more the cloud compared to the
other two cases. Placing VNFs only considering the observed traffic results in
using a similar total amount of resources as with prediction, since there is no
much difference in the number of replications and usage of the cloud, but it
requires significantly more migrations to be able to accommodate future
demands. In summary, we can say that when using traffic prediction, the number
of migrations can be reduced up to 45\% when there is enough available resources
to allocate replicas, compared to other cases studied. This is true at expenses
of using replications and cloud placements, as much as in the observed traffic
case. When comparing it to the overprovisioning case, that statement remains
true, but also the usage of the cloud is reduced by allocating almost up to
double number of replications. However, for traffic prediction to successfully
help on this problem, it requires certain amount of training periods per
independent traffic demand in the network, which can result in high
computational resources and computational time for larger networks.

\section{Conclusions}

We studied the problem of optimal placement of VNFs from an ISP point of view,
when minimizing migrations and replications. We proposed a traffic forecasting
model using LSTM networks and used it to place VNFs accordingly to the predicted
traffic demands. We proposed an offline MILP model as well as an online greedy
algorithm for the placement optimization problem. We compared three scenarios by
either considering: (i) the current observed traffic demands only, (ii)
overprovisioning of the 80\% of every specific maximum traffic demand value had
in the past, or (iii) the predicted traffic values based on history. We showed
that with traffic prediction, the number of migrations can be reduced up to 45\%
when there is enough available resources to allocate replicas. This also results
in less usage of the third-party clouds as compared to capacity
overprovisioning. While overprovisioning can be valid a solution when unexpected
traffic peaks appear resulting in higher usage of the cloud temporarily, traffic
prediction can minimize the need for the same that by anticipating a proper
placement and replication inside the network. The usage of LSTM networks,
however, requires non-negligible training time and computational resources which
is also something that needs to be taken into consideration.



\printcredits

\bibliographystyle{cas-model2-names}

\bibliography{refs}

\begin{thebibliography}{33}
\expandafter\ifx\csname natexlab\endcsname\relax\def\natexlab#1{#1}\fi
\providecommand{\url}[1]{\texttt{#1}}
\providecommand{\href}[2]{#2}
\providecommand{\path}[1]{#1}
\providecommand{\DOIprefix}{doi:}
\providecommand{\ArXivprefix}{arXiv:}
\providecommand{\URLprefix}{URL: }
\providecommand{\Pubmedprefix}{pmid:}
\providecommand{\doi}[1]{\href{http://dx.doi.org/#1}{\path{#1}}}
\providecommand{\Pubmed}[1]{\href{pmid:#1}{\path{#1}}}
\providecommand{\bibinfo}[2]{#2}
\ifx\xfnm\relax \def\xfnm[#1]{\unskip,\space#1}\fi
\bibitem[{Alawe et~al.(2018)Alawe, Ksentini, Hadjadj-Aoul and
  Bertin}]{Alawe_2018}
\bibinfo{author}{Alawe, I.}, \bibinfo{author}{Ksentini, A.},
  \bibinfo{author}{Hadjadj-Aoul, Y.}, \bibinfo{author}{Bertin, P.},
  \bibinfo{year}{2018}.
\newblock \bibinfo{title}{Improving traffic forecasting for 5g core network
  scalability: A machine learning approach}.
\newblock \bibinfo{journal}{{IEEE} Network} \bibinfo{volume}{32},
  \bibinfo{pages}{42--49}.
\newblock \DOIprefix\doi{10.1109/MNET.2018.1800104}.
\bibitem[{Alharbi et~al.(2019)Alharbi, Elgorashi, Lawey and
  Elmirghani}]{Alharbi2019}
\bibinfo{author}{Alharbi, H.A.}, \bibinfo{author}{Elgorashi, T.E.},
  \bibinfo{author}{Lawey, A.Q.}, \bibinfo{author}{Elmirghani, J.M.},
  \bibinfo{year}{2019}.
\newblock \bibinfo{title}{{The Impact of Inter-Virtual Machine Traffic on
  Energy Efficient Virtual Machines Placement}}.
\newblock \bibinfo{journal}{2019 IEEE Sustainability through ICT Summit, StICT
  2019} \DOIprefix\doi{10.1109/STICT.2019.8789381}.
\bibitem[{Basta et~al.(2017)Basta, Blenk, Hoffmann, Morper, Hoffmann and
  Kellerer}]{Basta2017}
\bibinfo{author}{Basta, A.}, \bibinfo{author}{Blenk, A.},
  \bibinfo{author}{Hoffmann, K.}, \bibinfo{author}{Morper, H.J.},
  \bibinfo{author}{Hoffmann, M.}, \bibinfo{author}{Kellerer, W.},
  \bibinfo{year}{2017}.
\newblock \bibinfo{title}{{Towards a Cost Optimal Design for a 5G Mobile Core
  Network based on SDN and NFV}}.
\newblock \bibinfo{journal}{IEEE Transactions on Network and Service
  Management} \bibinfo{volume}{4537}, \bibinfo{pages}{1--14}.
\newblock \DOIprefix\doi{10.1109/TNSM.2017.2732505}.
\bibitem[{Bulut and Ralphs(2015)}]{Bulut2015}
\bibinfo{author}{Bulut, A.}, \bibinfo{author}{Ralphs, T.},
  \bibinfo{year}{2015}.
\newblock \bibinfo{title}{{On the Complexity of Inverse Mixed Integer Linear
  Optimization}}.
\newblock \URLprefix
  \url{https://coral.ise.lehigh.edu/~ted/files/papers/InverseMILP15.pdf}.
\bibitem[{Carpio et~al.(2017a)Carpio, Bziuk and Jukan}]{Carpio2017b}
\bibinfo{author}{Carpio, F.}, \bibinfo{author}{Bziuk, W.},
  \bibinfo{author}{Jukan, A.}, \bibinfo{year}{2017}a.
\newblock \bibinfo{title}{{Replication of Virtual Network Functions: Optimizing
  link utilization and resource costs}}, in: \bibinfo{booktitle}{2017 40th
  International Convention on Information and Communication Technology,
  Electronics and Microelectronics ({MIPRO})}, \bibinfo{publisher}{Croatian
  Society MIPRO}.
\newblock \DOIprefix\doi{10.23919/MIPRO.2017.7973481}.
\bibitem[{Carpio et~al.(2017b)Carpio, Dhahri and Jukan}]{Carpio2017a}
\bibinfo{author}{Carpio, F.}, \bibinfo{author}{Dhahri, S.},
  \bibinfo{author}{Jukan, A.}, \bibinfo{year}{2017}b.
\newblock \bibinfo{title}{{VNF placement with replication for Load balancing in
  NFV networks}}, in: \bibinfo{booktitle}{IEEE International Conference on
  Communications}, \bibinfo{publisher}{{IEEE}}.
\newblock \DOIprefix\doi{10.1109/ICC.2017.7996515}.
\bibitem[{Carpio et~al.(2018)Carpio, Jukan and Pries}]{Carpio2018}
\bibinfo{author}{Carpio, F.}, \bibinfo{author}{Jukan, A.},
  \bibinfo{author}{Pries, R.}, \bibinfo{year}{2018}.
\newblock \bibinfo{title}{{Balancing the Migration of Virtual Network Functions
  with Replications in Data Centers}}, in: \bibinfo{booktitle}{{NOMS} 2018 -
  2018 {IEEE}/{IFIP} Network Operations and Management Symposium},
  \bibinfo{publisher}{IEEE}.
\newblock \DOIprefix\doi{10.1109/NOMS.2018.8406275}.
\bibitem[{Cziva et~al.(2018)Cziva, Anagnostopoulos and Pezaros}]{Cziva2018}
\bibinfo{author}{Cziva, R.}, \bibinfo{author}{Anagnostopoulos, C.},
  \bibinfo{author}{Pezaros, D.P.}, \bibinfo{year}{2018}.
\newblock \bibinfo{title}{{Dynamic, Latency-Optimal vNF Placement at the
  Network Edge}}.
\newblock \bibinfo{journal}{Proceedings - IEEE INFOCOM}
  \bibinfo{volume}{2018-April}, \bibinfo{pages}{693--701}.
\newblock \DOIprefix\doi{10.1109/INFOCOM.2018.8486021}.
\bibitem[{Ding et~al.(2017)Ding, Yu and Luo}]{Ding2017}
\bibinfo{author}{Ding, W.}, \bibinfo{author}{Yu, H.}, \bibinfo{author}{Luo,
  S.}, \bibinfo{year}{2017}.
\newblock \bibinfo{title}{{Enhancing the reliability of services in NFV with
  the cost-efficient redundancy scheme}}.
\newblock \bibinfo{journal}{IEEE International Conference on Communications}
  \bibinfo{volume}{1}.
\newblock \DOIprefix\doi{10.1109/ICC.2017.7996840}.
\bibitem[{Engelmann and Jukan(2018)}]{Engelmann2018}
\bibinfo{author}{Engelmann, A.}, \bibinfo{author}{Jukan, A.},
  \bibinfo{year}{2018}.
\newblock \bibinfo{title}{{A Reliability Study of Parallelized VNF Chaining}},
  in: \bibinfo{booktitle}{2018 {IEEE} International Conference on
  Communications ({ICC})}, \bibinfo{publisher}{IEEE}. pp.
  \bibinfo{pages}{2--7}.
\newblock \DOIprefix\doi{10.1109/ICC.2018.8422595}.
\bibitem[{Eramo et~al.(2017)Eramo, Miucci, Ammar and Lavacca}]{Eramo2017}
\bibinfo{author}{Eramo, V.}, \bibinfo{author}{Miucci, E.},
  \bibinfo{author}{Ammar, M.}, \bibinfo{author}{Lavacca, F.G.},
  \bibinfo{year}{2017}.
\newblock \bibinfo{title}{{An Approach for Service Function Chain Routing and
  Virtual Function Network Instance Migration in Network Function
  Virtualization Architectures}}.
\newblock \bibinfo{journal}{IEEE/ACM Transactions on Networking}
  \DOIprefix\doi{10.1109/TNET.2017.2668470}.
\bibitem[{Gember-Jacobson et~al.(2014)Gember-Jacobson, Viswanathan, Prakash,
  Grandl, Khalid, Das and Akella}]{Gember-Jacobson2014}
\bibinfo{author}{Gember-Jacobson, A.}, \bibinfo{author}{Viswanathan, R.},
  \bibinfo{author}{Prakash, C.}, \bibinfo{author}{Grandl, R.},
  \bibinfo{author}{Khalid, J.}, \bibinfo{author}{Das, S.},
  \bibinfo{author}{Akella, A.}, \bibinfo{year}{2014}.
\newblock \bibinfo{title}{{OpenNF: Enabling Innovation in Network Function
  Control}}.
\newblock \bibinfo{journal}{SIGCOMM Comput. Commun. Rev.} \bibinfo{volume}{44},
  \bibinfo{pages}{163--174}.
\newblock \DOIprefix\doi{10.1145/2740070.2626313}.
\bibitem[{Golkarifard et~al.(2021)Golkarifard, Chiasserini, Malandrino and
  Movaghar}]{Golkarifard_2021}
\bibinfo{author}{Golkarifard, M.}, \bibinfo{author}{Chiasserini, C.F.},
  \bibinfo{author}{Malandrino, F.}, \bibinfo{author}{Movaghar, A.},
  \bibinfo{year}{2021}.
\newblock \bibinfo{title}{Dynamic {VNF} placement, resource allocation and
  traffic routing in 5g}.
\newblock \bibinfo{journal}{Computer Networks} \bibinfo{volume}{188},
  \bibinfo{pages}{107830}.
\newblock \DOIprefix\doi{https://doi.org/10.1016/j.comnet.2021.107830}.
\bibitem[{Huang et~al.(2018)Huang, Liang, Ma and Guo}]{Huang2018}
\bibinfo{author}{Huang, M.}, \bibinfo{author}{Liang, W.}, \bibinfo{author}{Ma,
  Y.}, \bibinfo{author}{Guo, S.}, \bibinfo{year}{2018}.
\newblock \bibinfo{title}{{Throughput maximization of delay-sensitive request
  admissions via virtualized network function placements and migrations}}.
\newblock \bibinfo{journal}{IEEE International Conference on Communications}
  \bibinfo{volume}{2018-May}.
\newblock \DOIprefix\doi{10.1109/ICC.2018.8422337}.
\bibitem[{Kim et~al.(2019)Kim, Lee, Jeong, Choi, Yoo and Hong}]{Kim_2019}
\bibinfo{author}{Kim, H.G.}, \bibinfo{author}{Lee, D.Y.},
  \bibinfo{author}{Jeong, S.Y.}, \bibinfo{author}{Choi, H.},
  \bibinfo{author}{Yoo, J.H.}, \bibinfo{author}{Hong, J.W.K.},
  \bibinfo{year}{2019}.
\newblock \bibinfo{title}{{Machine Learning-Based Method for Prediction of
  Virtual Network Function Resource Demands}}, in: \bibinfo{booktitle}{2019
  {IEEE} Conference on Network Softwarization ({NetSoft})},
  \bibinfo{publisher}{{IEEE}}.
\newblock \URLprefix \url{https://ieeexplore.ieee.org/document/8806687},
  \DOIprefix\doi{10.1109/NETSOFT.2019.8806687}.
\bibitem[{Laghrissi and Taleb(2019)}]{Laghrissi2019}
\bibinfo{author}{Laghrissi, A.}, \bibinfo{author}{Taleb, T.},
  \bibinfo{year}{2019}.
\newblock \bibinfo{title}{{A Survey on the Placement of Virtual Resources and
  Virtual Network Functions}}.
\newblock \bibinfo{journal}{IEEE Communications Surveys and Tutorials}
  \bibinfo{volume}{21}, \bibinfo{pages}{1409--1434}.
\newblock \DOIprefix\doi{10.1109/COMST.2018.2884835}.
\bibitem[{{Michael Till Beck, Juan Felipe Botero} et~al.(2016){Michael Till
  Beck, Juan Felipe Botero}, Beck, Botero and {Michael Till Beck, Juan Felipe
  Botero}}]{michael2016}
\bibinfo{author}{{Michael Till Beck, Juan Felipe Botero}, K.S.},
  \bibinfo{author}{Beck, M.T.}, \bibinfo{author}{Botero, J.F.},
  \bibinfo{author}{{Michael Till Beck, Juan Felipe Botero}, K.S.},
  \bibinfo{year}{2016}.
\newblock \bibinfo{title}{{Resilient Allocation of Service Function Chains}},
  in: \bibinfo{booktitle}{IEEE Conference on Network Function Virtualization
  and Software Defined Networks (NFV-SDN)}, \bibinfo{publisher}{{IEEE}}.
\newblock \DOIprefix\doi{10.1109/NFV-SDN.2016.7919487}.
\bibitem[{Mijumbi et~al.(2016)Mijumbi, Hasija, Davy, Davy, Jennings and
  Boutaba}]{Mijumbi_2016}
\bibinfo{author}{Mijumbi, R.}, \bibinfo{author}{Hasija, S.},
  \bibinfo{author}{Davy, S.}, \bibinfo{author}{Davy, A.},
  \bibinfo{author}{Jennings, B.}, \bibinfo{author}{Boutaba, R.},
  \bibinfo{year}{2016}.
\newblock \bibinfo{title}{A connectionist approach to dynamic resource
  management for virtualised network functions}
  \DOIprefix\doi{10.1109/CNSM.2016.7818394}.
\bibitem[{Mijumbi et~al.(2017)Mijumbi, Hasija, Davy, Davy, Jennings and
  Boutaba}]{Mijumbi2017a}
\bibinfo{author}{Mijumbi, R.}, \bibinfo{author}{Hasija, S.},
  \bibinfo{author}{Davy, S.}, \bibinfo{author}{Davy, A.},
  \bibinfo{author}{Jennings, B.}, \bibinfo{author}{Boutaba, R.},
  \bibinfo{year}{2017}.
\newblock \bibinfo{title}{{Topology-Aware Prediction of Virtual Network
  Function Resource Requirements}}.
\newblock \bibinfo{journal}{IEEE Transactions on Network and Service
  Management} \bibinfo{volume}{14}, \bibinfo{pages}{106--120}.
\newblock \URLprefix \url{https://ieeexplore.ieee.org/document/7849149},
  \DOIprefix\doi{10.1109/TNSM.2017.2666781}.
\bibitem[{Qu et~al.(2020)Qu, Zhuang, Shen, Li and Rao}]{Qu_2020}
\bibinfo{author}{Qu, K.}, \bibinfo{author}{Zhuang, W.}, \bibinfo{author}{Shen,
  X.}, \bibinfo{author}{Li, X.}, \bibinfo{author}{Rao, J.},
  \bibinfo{year}{2020}.
\newblock \bibinfo{title}{Dynamic resource scaling for {VNF} over nonstationary
  traffic: A learning approach}.
\newblock \bibinfo{journal}{{IEEE} Transactions on Cognitive Communications and
  Networking} , \bibinfo{pages}{1--1}\DOIprefix\doi{10.1109/TCCN.2020.3018157}.
\bibitem[{Qu et~al.(2017)Qu, Assi, Shaban and Khabbaz}]{Qu_2017}
\bibinfo{author}{Qu, L.}, \bibinfo{author}{Assi, C.}, \bibinfo{author}{Shaban,
  K.}, \bibinfo{author}{Khabbaz, M.J.}, \bibinfo{year}{2017}.
\newblock \bibinfo{title}{A reliability-aware network service chain
  provisioning with delay guarantees in {NFV}-enabled enterprise datacenter
  networks}.
\newblock \bibinfo{journal}{{IEEE} Transactions on Network and Service
  Management} \bibinfo{volume}{14}, \bibinfo{pages}{554--568}.
\newblock \DOIprefix\doi{10.1109/TNSM.2017.2723090}.
\bibitem[{Rahman et~al.(2018)Rahman, Ahmed, Huynh, Tornatore and
  Mukherjee}]{Rahman_2018}
\bibinfo{author}{Rahman, S.}, \bibinfo{author}{Ahmed, T.},
  \bibinfo{author}{Huynh, M.}, \bibinfo{author}{Tornatore, M.},
  \bibinfo{author}{Mukherjee, B.}, \bibinfo{year}{2018}.
\newblock \bibinfo{title}{Auto-scaling {VNFs} using machine learning to improve
  {QoS} and reduce cost} \DOIprefix\doi{10.1109/ICC.2018.8422788}.
\bibitem[{Reddy and Rajamani(2014)}]{Reddy2014}
\bibinfo{author}{Reddy, P.V.V.}, \bibinfo{author}{Rajamani, L.},
  \bibinfo{year}{2014}.
\newblock \bibinfo{title}{{Virtualization overhead findings of four hypervisors
  in the CloudStack with SIGAR}}.
\newblock \bibinfo{journal}{2014 4th World Congress on Information and
  Communication Technologies, WICT 2014} ,
  \bibinfo{pages}{140--145}\DOIprefix\doi{10.1109/WICT.2014.7077318}.
\bibitem[{Shi et~al.(2015)Shi, Zhang, Chu, Bao, Jin, Gong, Zhu, Yu and
  Rosenberg}]{Shi2015}
\bibinfo{author}{Shi, R.}, \bibinfo{author}{Zhang, J.}, \bibinfo{author}{Chu,
  W.}, \bibinfo{author}{Bao, Q.}, \bibinfo{author}{Jin, X.},
  \bibinfo{author}{Gong, C.}, \bibinfo{author}{Zhu, Q.}, \bibinfo{author}{Yu,
  C.}, \bibinfo{author}{Rosenberg, S.}, \bibinfo{year}{2015}.
\newblock \bibinfo{title}{{MDP and Machine Learning-Based Cost-Optimization of
  Dynamic Resource Allocation for Network Function Virtualization}}, in:
  \bibinfo{booktitle}{2015 {IEEE} International Conference on Services
  Computing}, \bibinfo{publisher}{{IEEE}}. pp. \bibinfo{pages}{65--73}.
\newblock \URLprefix \url{https://ieeexplore.ieee.org/document/7207337},
  \DOIprefix\doi{10.1109/SCC.2015.19}.
\bibitem[{Subramanya and Riggio(2019)}]{Subramanya_2019}
\bibinfo{author}{Subramanya, T.}, \bibinfo{author}{Riggio, R.},
  \bibinfo{year}{2019}.
\newblock \bibinfo{title}{Machine learning-driven scaling and placement of
  virtual network functions at the network edges}
  \DOIprefix\doi{10.1109/NETSOFT.2019.8806631}.
\bibitem[{Sun et~al.(2016)Sun, Lu, Lu and Zhu}]{Sun_2016}
\bibinfo{author}{Sun, Q.}, \bibinfo{author}{Lu, P.}, \bibinfo{author}{Lu, W.},
  \bibinfo{author}{Zhu, Z.}, \bibinfo{year}{2016}.
\newblock \bibinfo{title}{Forecast-assisted {NFV} service chain deployment
  based on affiliation-aware {vNF} placement}
  \DOIprefix\doi{10.1109/GLOCOM.2016.7841846}.
\bibitem[{Tajiki et~al.(2017)Tajiki, Salsano, Chiaraviglio, Shojafar and
  Akbari}]{Tajiki2017}
\bibinfo{author}{Tajiki, M.M.}, \bibinfo{author}{Salsano, S.},
  \bibinfo{author}{Chiaraviglio, L.}, \bibinfo{author}{Shojafar, M.},
  \bibinfo{author}{Akbari, B.}, \bibinfo{year}{2017}.
\newblock \bibinfo{title}{{Joint Energy Efficient and QoS-aware Path Allocation
  and VNF Placement for Service Function Chaining}}.
\newblock \bibinfo{journal}{IEEE Transactions on Network and Service
  Management} \bibinfo{volume}{PP}, \bibinfo{pages}{1}.
\newblock \DOIprefix\doi{10.1109/TNSM.2018.2873225}.
\bibitem[{Taleb et~al.(2019)Taleb, Ksentini and Frangoudis}]{Taleb2019}
\bibinfo{author}{Taleb, T.}, \bibinfo{author}{Ksentini, A.},
  \bibinfo{author}{Frangoudis, P.A.}, \bibinfo{year}{2019}.
\newblock \bibinfo{title}{{Follow-me cloud: When cloud services follow mobile
  users}}.
\newblock \bibinfo{journal}{IEEE Transactions on Cloud Computing}
  \bibinfo{volume}{7}, \bibinfo{pages}{369--382}.
\newblock \DOIprefix\doi{10.1109/TCC.2016.2525987}.
\bibitem[{Tang et~al.(2019)Tang, Zhou and Chen}]{Tang_2019}
\bibinfo{author}{Tang, H.}, \bibinfo{author}{Zhou, D.}, \bibinfo{author}{Chen,
  D.}, \bibinfo{year}{2019}.
\newblock \bibinfo{title}{Dynamic network function instance scaling based on
  traffic forecasting and {VNF} placement in operator data centers}.
\newblock \bibinfo{journal}{{IEEE} Transactions on Parallel and Distributed
  Systems} \bibinfo{volume}{30}, \bibinfo{pages}{530--543}.
\newblock \DOIprefix\doi{10.1109/TPDS.2018.2867587}.
\bibitem[{Xia et~al.(2016a)Xia, Cai and Xu}]{Xia2016a}
\bibinfo{author}{Xia, J.}, \bibinfo{author}{Cai, Z.}, \bibinfo{author}{Xu, M.},
  \bibinfo{year}{2016}a.
\newblock \bibinfo{title}{{Optimized Virtual Network Functions Migration for
  NFV}}.
\newblock \bibinfo{journal}{IEEE 22nd International Conference on Parallel and
  Distributed Systems Optimized} \DOIprefix\doi{10.1109/ICPADS.2016.0053}.
\bibitem[{Xia et~al.(2016b)Xia, Pang, Cai, Xu and Hu}]{Xia2016}
\bibinfo{author}{Xia, J.}, \bibinfo{author}{Pang, D.}, \bibinfo{author}{Cai,
  Z.}, \bibinfo{author}{Xu, M.}, \bibinfo{author}{Hu, G.},
  \bibinfo{year}{2016}b.
\newblock \bibinfo{title}{{Reasonably Migrating Virtual Machine in NFV-Featured
  Networks}}.
\newblock \bibinfo{journal}{IEEE International Conference on Computer and
  Information Technology (CIT)} \DOIprefix\doi{10.1109/CIT.2016.96}.
\bibitem[{Yao et~al.(2020)Yao, Guo, Li, Liu and Zeng}]{Yao_2020}
\bibinfo{author}{Yao, Y.}, \bibinfo{author}{Guo, S.}, \bibinfo{author}{Li, P.},
  \bibinfo{author}{Liu, G.}, \bibinfo{author}{Zeng, Y.}, \bibinfo{year}{2020}.
\newblock \bibinfo{title}{Forecasting assisted {VNF} scaling in {NFV}-enabled
  networks}.
\newblock \bibinfo{journal}{Computer Networks} \bibinfo{volume}{168},
  \bibinfo{pages}{107040}.
\newblock \DOIprefix\doi{10.1016/j.comnet.2019.107040}.
\bibitem[{Yuan et~al.(2020)Yuan, Ji, Tang and You}]{Yuan2020}
\bibinfo{author}{Yuan, Q.}, \bibinfo{author}{Ji, X.}, \bibinfo{author}{Tang,
  H.}, \bibinfo{author}{You, W.}, \bibinfo{year}{2020}.
\newblock \bibinfo{title}{{Toward Latency-Optimal Placement and Autoscaling of
  Monitoring Functions in MEC}}.
\newblock \bibinfo{journal}{IEEE Access} \bibinfo{volume}{8},
  \bibinfo{pages}{41649--41658}.
\newblock \DOIprefix\doi{10.1109/ACCESS.2020.2976858}.

\end{thebibliography}

\end{document}